\documentclass[PAPER, atlasdraft=false, texlive=2025, cernpreprint=true, UKenglish, texmf, orcidlogo]{atlasdoc}
\usepackage{atlaspackage}[subfig=true,siunitx=true]
\usepackage{atlasbiblatex}

\usepackage{atlasphysics}
\makeatletter
\@ifundefined{qty}{
\newcommand{\qty}[2]{#1\,#2} %
}{}
\@ifundefined{qtylist}{
}{}
\@ifundefined{mm}{
\@ifpackageloaded{siunitx}{
\DeclareSIUnit{\mm}{\milli\metre}
}{
\newcommand{\mm}{mm}
}
}{}
\makeatother

\graphicspath{{logos/}{figures/}}

\usepackage{ANA-IDTR-2021-01-PAPER-defs}
\usepackage{subfig}

\AtlasTitle{Track and Vertex Reconstruction with the ATLAS Inner Detector}

\AtlasAbstract{%
Charged-particle reconstruction is a fundamental part of the event reconstruction in modern
multi-purpose high-energy physics detectors. This paper describes the algorithms used
to reconstruct charged particles and primary vertices with the ATLAS Inner Detector. The most recent software configuration deployed for data-taking is described, and the performance obtained when this software is used to process Run 2 (2015–2018) data, a subset (from 2022) of Run 3 (2022-2026) data, and corresponding simulated data is presented. The ATLAS track and vertex reconstruction performance is shown for up to 80 simultaneous proton--proton interaction. It maintains a high efficiency, good resolution for key parameters, and low rates of mis-reconstructed candidates for Run 2 and Run 3 conditions. }
\PreprintIdNumber{CERN-EP-2026-096}

\author{The ATLAS Collaboration}

\AtlasRefCode{IDTR-2021-01}

\AtlasJournal{Journal of Instrumentation}

\hypersetup{pdftitle={ATLAS document},pdfauthor={The ATLAS Collaboration}}

\begin{document}

\maketitle

\tableofcontents

\section{Introduction}
%

The reconstruction of charged particle trajectories is a crucial aspect of many high-energy physics experiments, like the ATLAS experiment~\cite{PERF-2007-01} at the Large Hadron Collider (LHC).
These trajectories, or \emph{tracks} are used both as physics objects\footnote{Here, \emph{object} refers to any item that can be selected for based on its reconstructed properties, e.g.\ electrons, muons, jets, etc, as well as generic charged particles for which tracks can be used directly.} for analysis in their own right~\cite{STDM-2015-02} and are fundamental inputs to the reconstruction of all other physics objects in ATLAS. Algorithms such as electron~\cite{EGAM-2021-01,EGAM-2021-02} and muon~\cite{MUON-2018-03,MUON-2022-01} reconstruction; the reconstruction of hadronic objects such as jets~\cite{PERF-2015-09,JETM-2018-06} and hadronically decaying $\tau$ leptons~\cite{TAUP-2023-02}; jet classification algorithms~\cite{ATL-PHYS-PUB-2023-020,JETM-2023-06,JETM-2024-06}, including \emph{Flavour Tagging}~\cite{FTAG-2023-05,ATL-PHYS-PUB-2023-021}; and pile-up suppression techniques~\cite{PERF-2014-03,JETM-2018-06}, are critically dependent on the performance of track reconstruction. Therefore, track reconstruction performance directly or indirectly affects the ability to perform a wide range of Standard Model (SM) measurements, including measurements of the $W$ boson mass~\cite{STDM-2014-18} and the properties of the Higgs boson~\cite{HIGG-2021-23}, as well as searches for physics beyond the Standard Model (BSM), such as long-lived particles decaying within the Inner Detector~\cite{SUSY-2018-13,EXOT-2021-32,EXOT-2022-31}, among others.
Charged-particle reconstruction is essential not only for offline analysis but also for online event selection, where tracking is available in the High-Level Trigger System~\cite{TRIG-2019-03,TRIG-2019-04,TRIG-2022-01}.
The efficient and precise reconstruction of tracks is therefore vital for the overall performance of the ATLAS experiment.

Track reconstruction relies on relatively few discrete measurements and is therefore highly impacted by material effects such as multiple scattering. Thus, the relevant detector systems, in this case the ATLAS Inner Detector (ID), are typically located as close as possible to the interaction point where the particles are produced. Consequently, the ID is exposed to a very high particle flux, which is further increased under typical operating conditions by the presence of multiple proton--proton (\pp) interactions within a single bunch crossing, referred to as \emph{pile-up}.

Pile-up is quantified by the parameter $\mu$, defined as the mean of the Poisson distribution of inelastic \pp\ interactions per bunch crossing for a given colliding bunch pair. In practice, $\mu$ varies across bunch pairs and over time, and the quantity $\langle\mu\rangle$ is therefore commonly used to represent the average over all colliding bunch pairs, either at a given moment (to describe the instantaneous luminosity of the LHC) or over a dataset. During Run~2 (2015--2018), typical values of $\langle\mu\rangle$ were around 34, while in Run~3 (2022--2026) the most probable value exceeds 60.

The high particle multiplicity associated with $\langle\mu\rangle$ values over 60 poses a formidable challenge for ID track reconstruction. The Inner Detector was designed for LHC design-luminosity conditions, corresponding to an average pile-up of order $\langle\mu\rangle\sim$20--25~\cite{ATLAS-TDR-04}, while present-day operation reaches substantially higher pile-up. The higher particle density in a single event leads to a substantially increased number of plausible combinations of measurements, which in turn increases the CPU time required for pattern‑recognition algorithms and raises the probability of spurious track candidates. These effects necessitate robust track‑finding strategies in the reconstruction software.

The reconstruction of \emph{primary vertices}, representing the estimated positions of the \pp collisions from which the associated charged particles originate, is a key application of tracking performance.
Reconstructing primary vertices and associating tracks to their most likely origin enables the identification of the vertex of interest and its separation from \pileup collisions within the same recorded event.

This paper describes the algorithmic design and performance of the offline track and primary vertex reconstruction in the most recent software configuration used for data taking with the ATLAS Inner Detector. This configuration was optimised ahead of Run~3 operation and subsequently applied in a full reprocessing of the Run~2 dataset.

Comparisons with the reconstruction configuration originally used during Run~2 data taking are discussed separately in Ref.~\cite{IDTR-2022-04}. A particular focus of the latest reconstruction strategy was on reducing the number of erroneous and low-quality tracks processed by rejecting them as early as possible. This strategy is crucial for reducing CPU usage for processing events with high instantaneous luminosity within the available resources, and for providing a pure collection of high-quality tracks to downstream algorithms. This must be carefully balanced against maintaining as high an efficiency as possible for tracks in the key phase space regions~\cite{IDTR-2022-04}. These developments also allowed for the inclusion of dedicated reconstruction strategies for specific topologies, such as tracks from long-lived particles decaying within the Inner Detector, which are reconstructed using a specialised chain running alongside the primary sequence (see Section~\ref{sec:trk:overview}) on all events~\cite{IDTR-2021-03}. In addition, a specific treatment for tracks within high-transverse-momentum (high-\pT) jets, where the particle density is extremely high (usually referred to as \emph{Dense Environments}) was refined. These targeted optimisations are particularly beneficial in searches for new physics~\cite{SUSY-2018-13,EXOT-2021-32,EXOT-2022-31,EXOT-2018-60,EXOT-2019-04,HIGG-2021-11}.

The paper presents performance results on the ATLAS track and primary vertex reconstruction obtained primarily with Run 2 data, complemented by dedicated simulation studies covering a wide range of \pileup conditions. It does not supersede the developments discussed in Ref.~\cite{IDTR-2022-04} nor the performance results for long-lived particle reconstruction presented in Ref.~\cite{IDTR-2021-03}, but instead provides a consolidated performance overview of the current ATLAS track and primary vertex reconstruction configuration.

The structure of the paper is as follows. In Section~\ref{sec:detector} the ATLAS detector, with particular focus on the Inner Detector, is outlined. This is followed by a discussion of the various data and Monte Carlo simulation samples used for the studies presented in Section~\ref{sec:samples}. In Section~\ref{sec:tracking} a comprehensive overview of the track reconstruction is provided, including the developments targeted at long-lived particle tracks and dense environments. The primary vertex reconstruction procedure is described in Section~\ref{sec:vertex}. The performance of the track and primary vertex reconstruction is shown in Section~\ref{sec:perf} for a number of key quantities, such as efficiencies and resolutions, both inclusively and for specific selections and phase space regions. These are performed in Monte Carlo samples and where possible also using data. The CPU consumption of the reconstruction is also discussed, before a summary and outlook are given in Section~\ref{sec:conclusions}.


\section{ATLAS detector}
\label{sec:detector}

The ATLAS detector~\cite{PERF-2007-01,GENR-2019-02} is a multipurpose particle physics
apparatus with forward--backward symmetric cylindrical geometry, covering nearly
the entire solid angle around the interaction point (IP).
ATLAS uses a right-handed coordinate system with its origin at the nominal IP
in the centre of the detector and the \(z\)-axis along the beam pipe.
The \(x\)-axis points from the IP to the centre of the LHC ring,
and the \(y\)-axis points upwards.
Polar coordinates \((r,\phi)\) are used in the transverse plane,
\(\phi\) being the azimuthal angle around the \(z\)-axis.
The pseudorapidity is defined in terms of the polar angle \(\theta\) as \(\eta = -\ln \tan(\theta/2)\).
Angular distance is measured in units of \(\Delta R \equiv \sqrt{(\Delta\eta)^{2} + (\Delta\phi)^{2}}\).

\begin{figure}[htbp]
\centering
\includegraphics[width=0.80\textwidth]{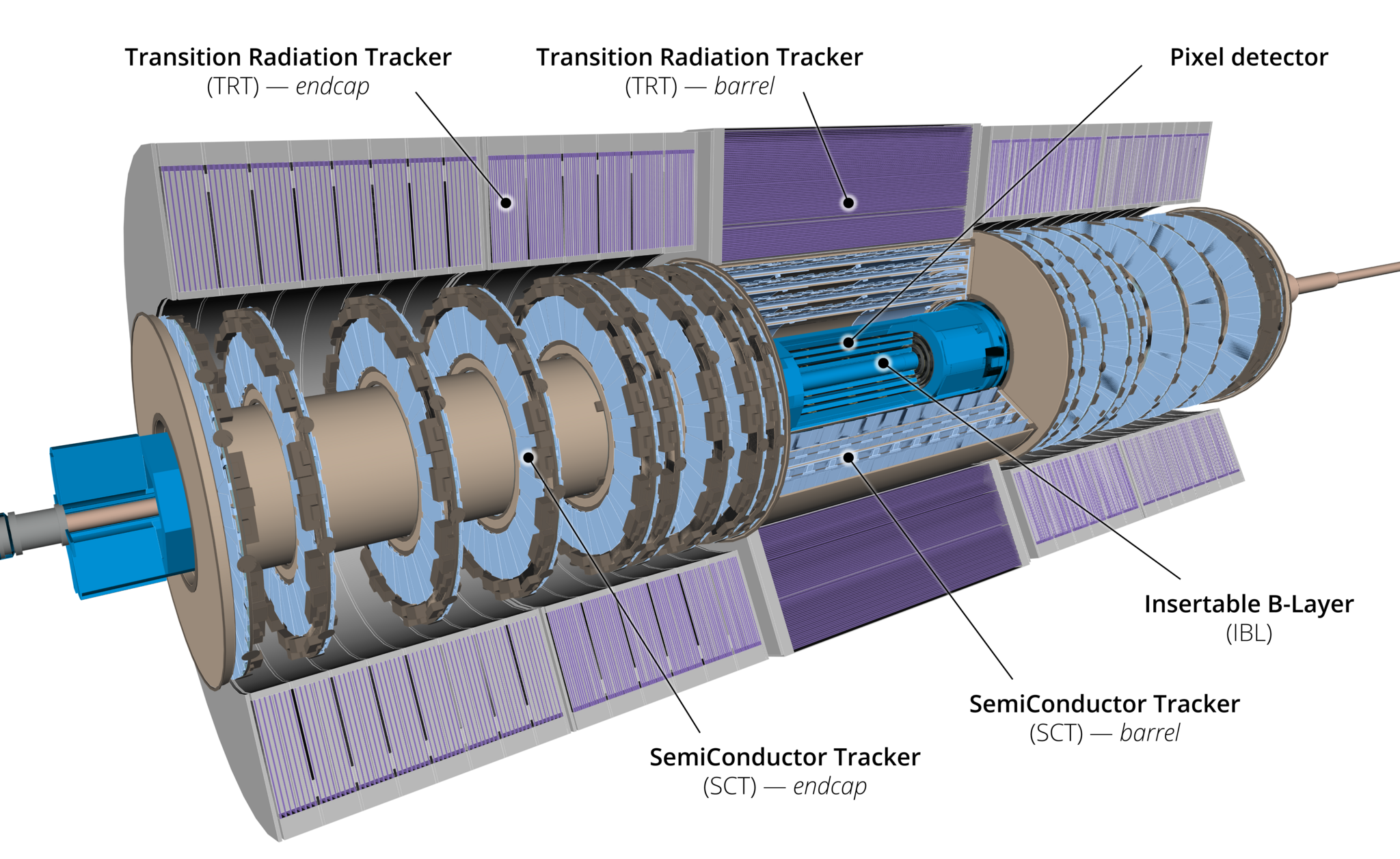}
\begin{center}
(a)
\end{center}
\includegraphics[width=0.80\textwidth]{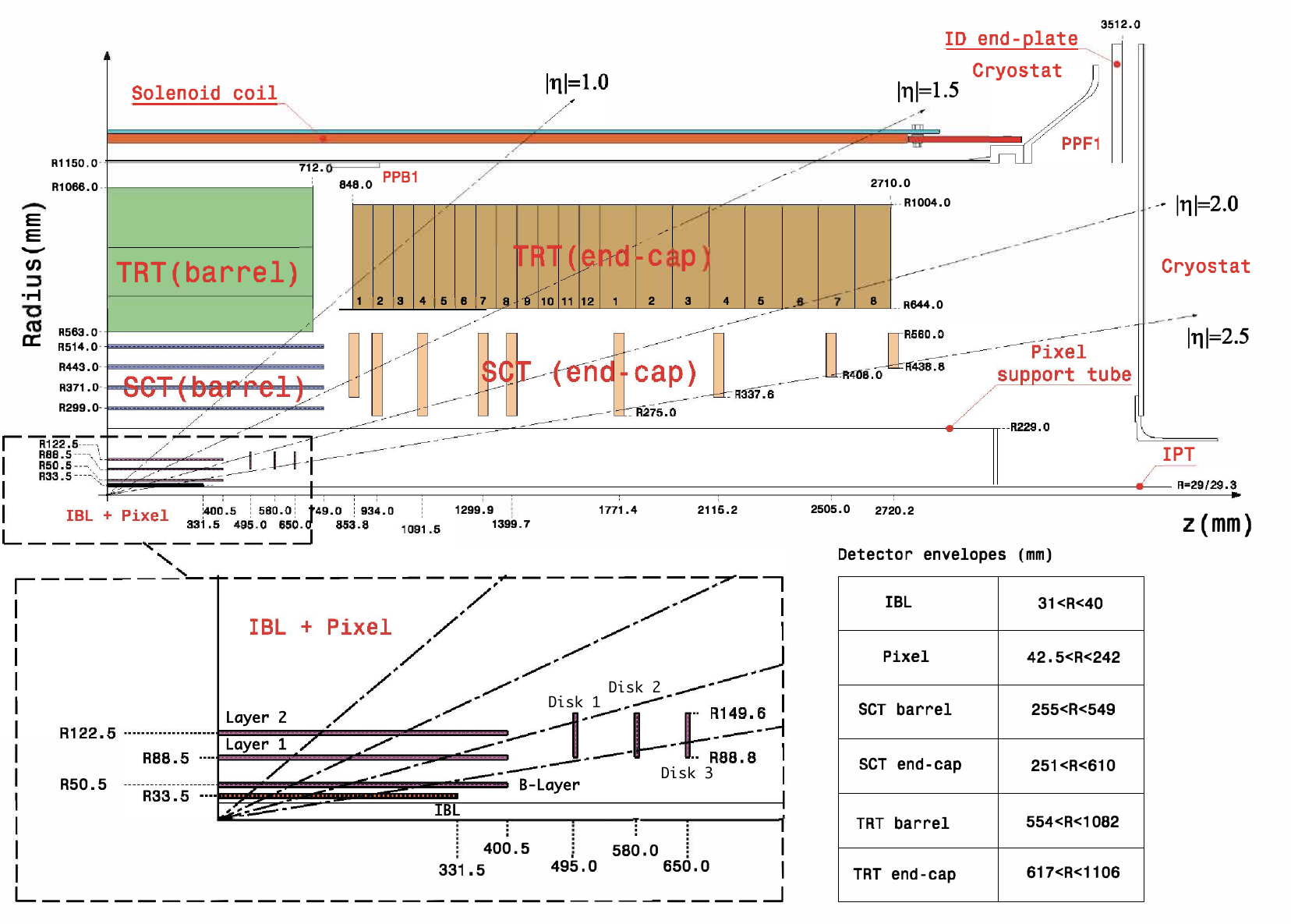}
\begin{center}
(b)
\end{center}
\caption{\label{fig:detector:idxsec} Cross sections through the ATLAS Inner Detector comprising the Pixel, SCT and TRT subsystems. (a) shows a 3D view of the detector~\cite{GENR-2019-02}, while (b) presents a quadrant in the $r-z$ plane~\cite{PIX-2018-001}.}
\end{figure}

The Inner Detector system provides charged-particle tracking capability
with high efficiency over the pseudorapidity range $\abseta < 2.5$. It is
located in the region of ATLAS closest to the interaction point and it is
contained within a cylindrical envelope covering the region up to $|z|=\qty{3512}{\mm}$ and of radius
\qty{1150}{\mm}. Two cross section views of the ID barrel are shown in
Figure~\ref{fig:detector:idxsec}. The entire ID is immersed in a \qty{2}{\tesla} magnetic
field generated by the superconducting central solenoid magnet.  The ID
comprises three subsystems based on different detector technologies.

The innermost region uses high-granularity silicon pixel detectors, whose
sensitive elements extend between a radius of about \qty{33}{\mm} and \qty{123}{\mm}. It
consists of sensor modules arranged in four concentric barrel layers and three endcap disks on each side,
for a total of about 86 million channels. The innermost barrel layer, known as
the Insertable B-layer (IBL)~\cite{ATLAS-TDR-19,PIX-2018-001}, was inserted at a radius of \qty{33.5}{\mm} prior to the
beginning of the LHC Run~2. The previous innermost pixel barrel layer (at a radius of \qty{50.5}{\mm}) is referred to as the B-layer, as in the zoomed view of the pixel region in Figure~\ref{fig:detector:idxsec}. The majority of the pixel detector uses planar sensors with a pitch (in the $\phi$ and $\eta$ directions respectively) of \qtyproduct[product-units = power]{50 x 400}{\um}; the IBL uses planar sensors
with \qtyproduct[product-units = power]{50 x 250}{\um} pixel pitch, with the exception of the highest $\abseta$ region
(largely outside the nominal tracker acceptance region $\abseta < 2.5$) which
employs 3D silicon technology.

The pixel detector is surrounded by a silicon microstrip tracker (SemiConductor Tracker, SCT), which
uses pairs of sensors with a typical strip pitch of \qty{80}{\um} mounted back-to-back on modules, with a \qty{40}{\milli\radian} stereo rotation between the sides to provide additional constraint on the position along the direction of the strips. The SCT detector is arranged in four concentric barrel layers and
nine endcap disks per side; the active elements span an average radius of about
\qty{299}{\mm} to \qty{514}{\mm}.

The outermost subsystem of the ID, the Transition Radiation Tracker (TRT),
contains up to 73 layers of gas-filled straws interleaved with fibres (barrel)
and 160 straw planes interleaved with foils (end-cap), which provide transition
radiation for electron identification~\cite{IDET-2015-01}. Most of the TRT system uses a Xenon-based
gas mixture, which provides optimal particle identification capabilities. However, such a mixture has been gradually substituted by an Argon-based gas mixture since the beginning of Run 2 (2015), due to cost considerations. The differences between the gas-mixtures have no significant impact on the tracking performance. The TRT provides typically 36 one-dimensional measurements
per track within the $\abseta < 2.0$ region.

Surrounding the ID, the calorimeter system covers the pseudorapidity range \(|\eta| < 4.9\).
Within the region \(|\eta|< 3.2\), electromagnetic (EM) calorimetry is provided by barrel and
endcap high-granularity lead/liquid-argon (LAr) calorimeters,
with an additional thin LAr presampler covering \(|\eta| < 1.8\)
to correct for energy loss in material upstream of the calorimeters.
Hadron calorimetry is provided by the steel/scintillator-tile calorimeter,
segmented into three barrel structures within \(|\eta| < 1.7\), and two copper/LAr hadron endcap calorimeters.
The solid angle coverage is completed with forward copper/LAr and tungsten/LAr calorimeter modules
optimised for electromagnetic and hadronic energy measurements respectively. The EM calorimeter is used in ID track reconstruction to identify regions where specialized settings for electron reconstruction can be applied, as discussed in Section~\ref{pattern:brem}.

The muon spectrometer (MS) comprises tracking chambers measuring the deflection of muons in a magnetic field generated by the superconducting air-core toroidal magnets. The field integral of the toroids ranges between \num{2.0} and \qty{6.0}{\tesla\metre} across most of the detector.
Three stations of precision chambers, each consisting of 3 or 4 layers of monitored drift tubes, cover the region \(|\eta| < 2.7\).
These were complemented in the innermost layer of the endcap region by cathode-strip chambers in Run~2,
which were replaced in Run~3 by layers of small-strip thin-gap chambers and Micromegas chambers, both providing precision tracking in the region \(1.3 < |\eta| < 2.7\).

An extensive software suite~\cite{ATL-SOFT-PUB-2021-001,SOFT-2022-02} is used in data simulation,
in the reconstruction and analysis of real and simulated data, in detector operations, and in the trigger and data acquisition systems of the experiment.

\begin{figure}[htbp]
\centering
\subfloat[][]{
\label{fig:detector:radlength}
\includegraphics[width=0.47\textwidth]{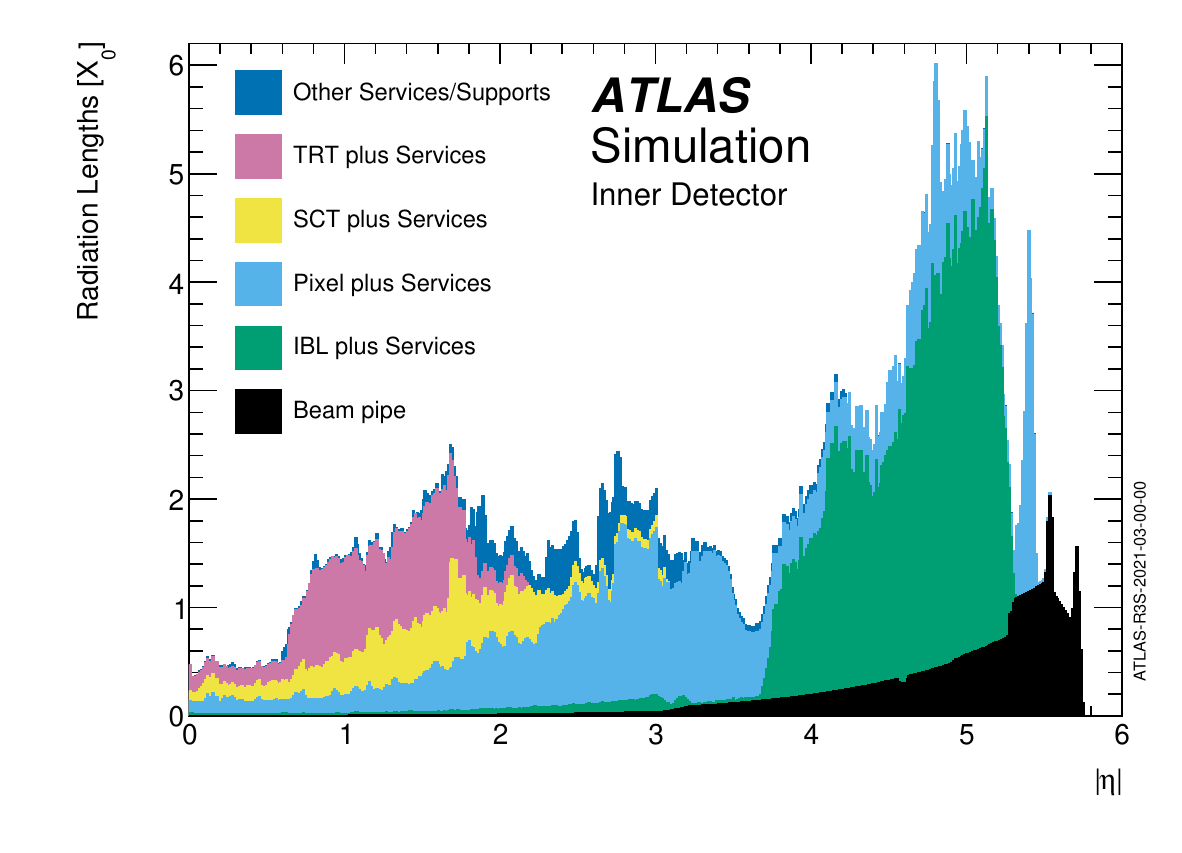}
}
\subfloat[][]{
\label{fig:detector:hadintlength}
\includegraphics[width=0.47\textwidth]{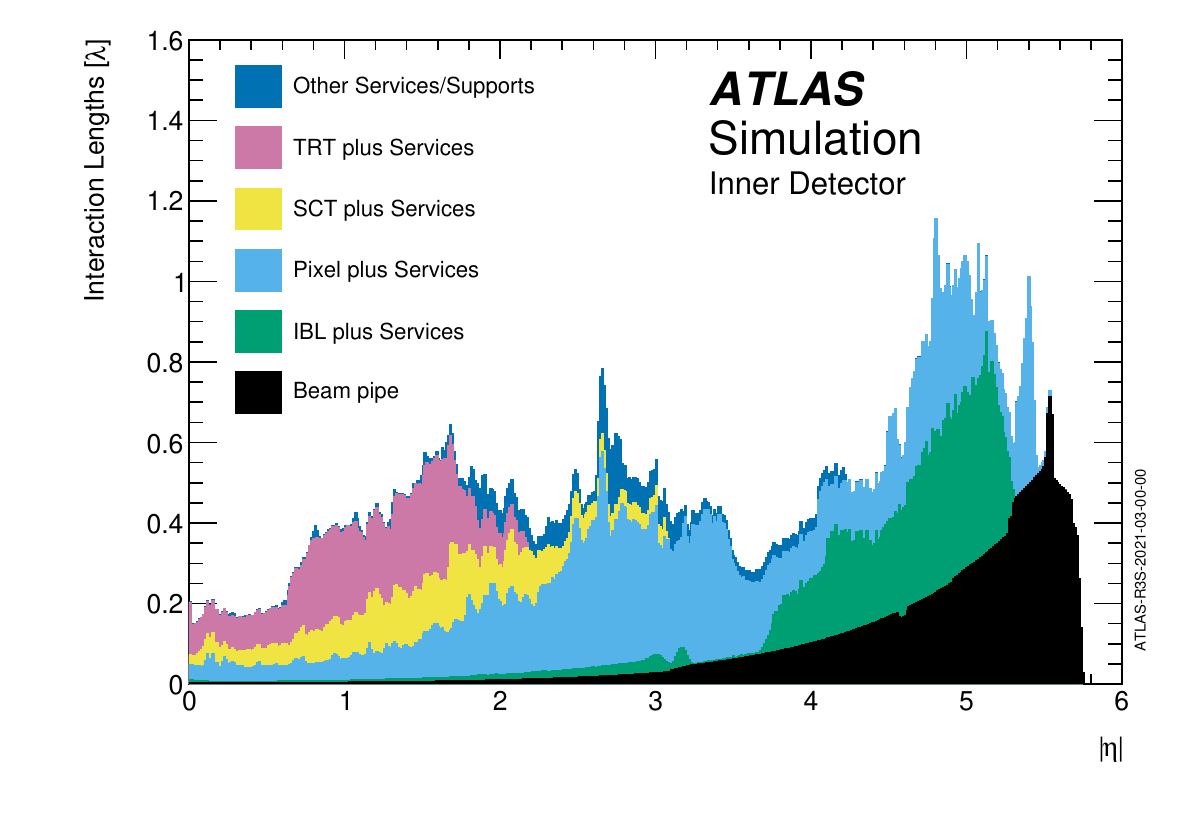}
}
\caption{\label{fig:detector:materialbudget} (a) Amount of radiation lengths ($X_0$)~\cite{GENR-2019-02} and (b) nuclear interaction lengths ($\lambda$) a particle intersects when originating from the nominal interaction point and traversing the entirety of the ID, as a function of $\abseta$ (averaged over $\phi$).}
\end{figure}

\begin{figure}[htbp]
\centering
\subfloat[][]{
\includegraphics[width=0.5\textwidth]{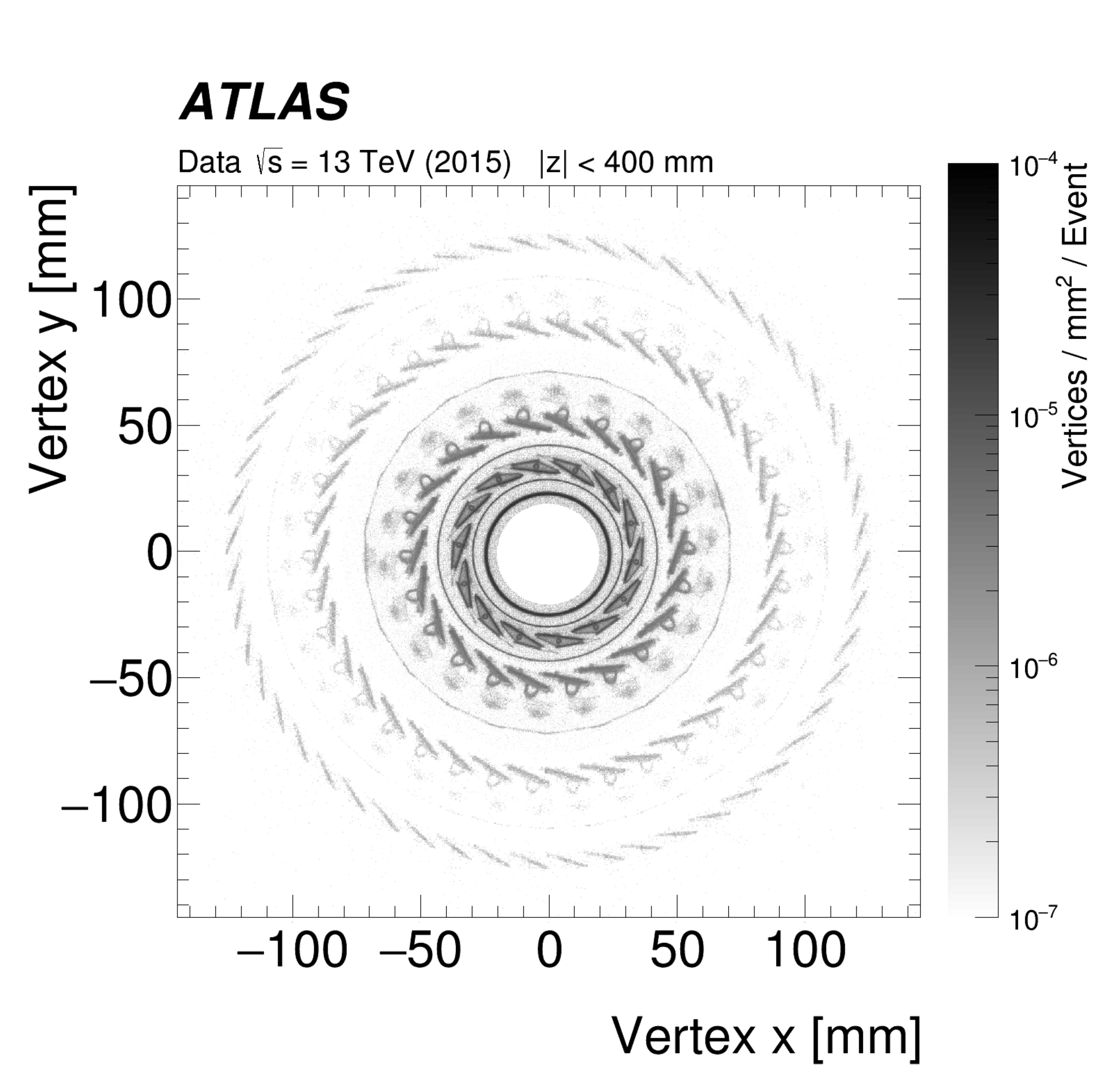}
}
\subfloat[][]{
\includegraphics[width=0.5\textwidth]{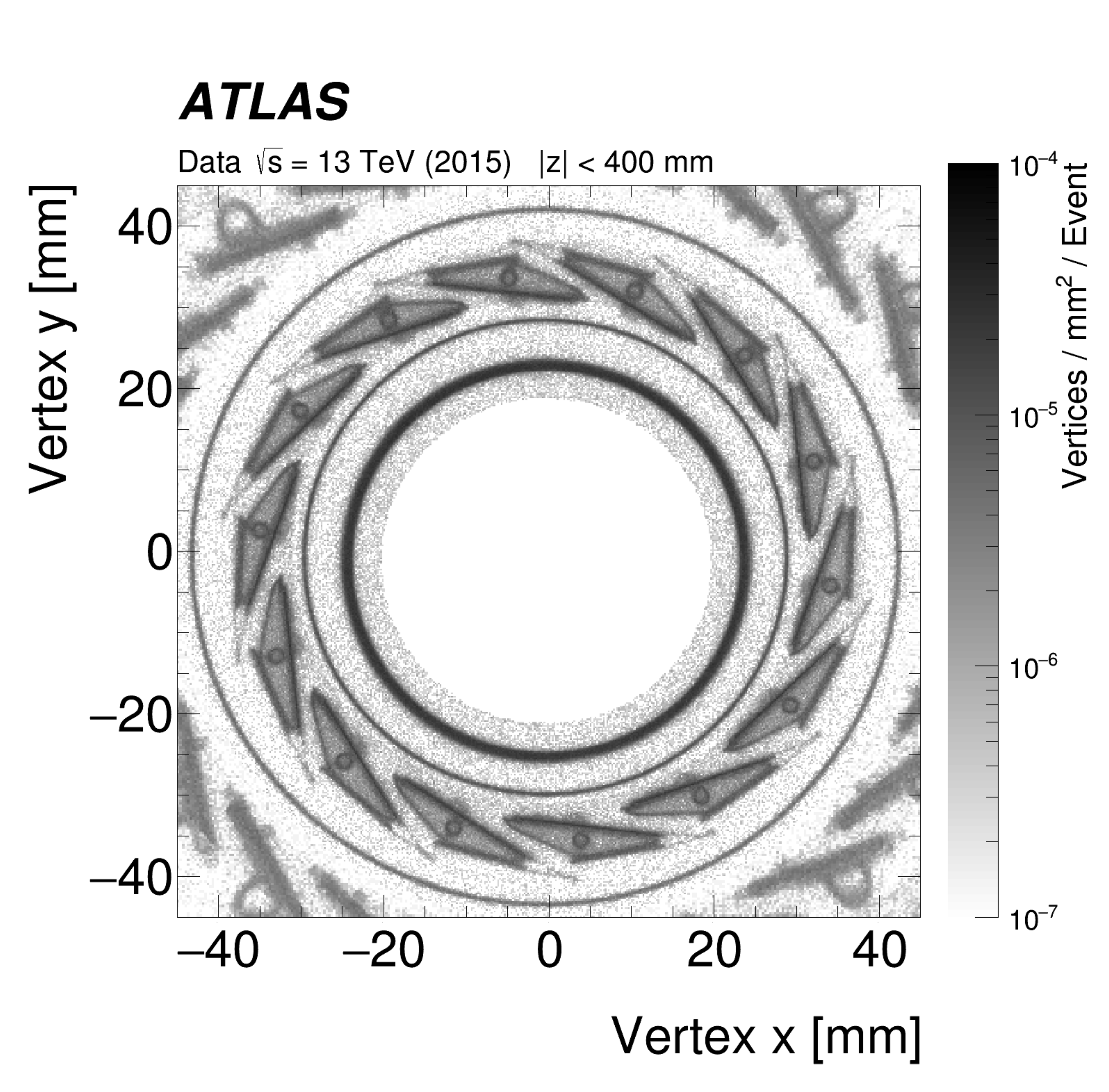}
}
\caption{Positions of hadronic interaction vertex candidates within  (a) the ATLAS Inner Detector barrel region ($|\eta|<2.4$ and $|z|<$\qty{400}{\mm}) in the $xy$-plane with $r>$\qty{20}{\mm} and (b) a zoomed-in view showing the beam pipe, Inner Positioning Tube, IBL staves, and Inner Support Tube~\cite{PERF-2015-07}.}
\label{fig:detector:hadronicinteractions}
\end{figure}

The reconstruction efficiency for charged hadrons is directly related to the amount of sensitive and passive material in the detector, and the precision to which it is known.
Figure~\ref{fig:detector:materialbudget} shows the material distribution in the ID, as a function of the pseudorapidity, as estimated using simulation. These estimates are validated with in-situ measurements, as detailed in Ref.~\cite{PERF-2015-07}.
Figure~\ref{fig:detector:hadronicinteractions} shows the reconstructed position of
hadronic interactions that reveal the as-built geometry and structures of passive and active elements of the innermost ID layers.
Accurate reconstruction of charged particle trajectories also requires precise knowledge of the detector geometry.
In-situ techniques are used to determine and monitor the alignment of the ID's sensitive elements throughout data-taking, as
described in detail in Ref.~\cite{IDTR-2019-05}.

\subsection{Measurement formation}

Charged particles passing through the silicon detectors deposit energy via ionisation and radiative processes. If the energy in a detector channel (i.e.\ the smallest possible unit of the sensor which can be individually read out, corresponding to an individual pixel or strip) exceeds a predefined threshold, the signal is recorded as a hit. Threshold values are chosen to provide high particle detection efficiency while suppressing electronic noise and correlated instrumental effects such as electronic cross-talk. These values vary across detector regions and have evolved over time due to accumulated radiation doses and increasing instantaneous luminosity. Neighbouring channels with hits are grouped together into clusters~\cite{PERF-2012-05,PERF-2015-08} using a connected component analysis (CCA) approach. The CCA method combines channels based on their connectivity, defined by shared edges, including diagonal connections in the implementation used here. Clusters form the fundamental measurements used for track reconstruction. In this paper, the term \textit{hits} refers to individual activated channels, while \textit{clusters} or \textit{measurements} refer to groups of hits that represent an estimate of the intersection of a charged particle with a sensor. Sensors with known performance issues can be excluded from the readout chain through being marked as \emph{inactive}. Such sensors cannot provide measurements, and so are explicitly accounted for in the reconstruction as discussed in Sections~\ref{sec:trk:ambi} and~\ref{sec:perf:quality}. Lists of inactive sensors are periodically updated and are also used in the production of Monte Carlo simulation samples to allow them to resemble data more closely.

The Pixel detector achieves position resolution on the order of \qty{10}{\um} in the most sensitive direction (the direction along $\phi$) and about \qty{60}{\um} in the orthogonal direction. Each pixel hit stores a Time-over-Threshold (ToT) value, defined as the time the amplifier output signal remains above threshold, measured in units of the LHC clock (\SI{25}{ns})~\cite{PIX-2018-001}.
This value is proportional to the deposited charge and is used in the computation of the cluster charge barycentre, which provides an estimate of the particle’s crossing position on the sensor and enables advanced cluster treatment as described in Section~\ref{sec:trk:ambi}.
In contrast, the SCT hits contain only binary information indicating whether a channel was activated. The typical SCT position resolution is of the order of \qty{20}{\um} in the sensitive direction, and \qty{580}{\um} in the orthogonal direction when combining measurements on both sides of a module, as done for space point formation discussed in Sec~\ref{sec:seeding}.

The sustained high particle flux integrated over the data-taking periods leads to significant radiation damage over the detector lifetime~\cite{IDET-2017-10,IDET-2020-01}. This damage manifests as increased leakage currents in the silicon sensors and a gradual reduction of the collected charge, which can affect measurement reconstruction and, at a modest level, the spatial resolution and tracking efficiency. To date, the innermost detector layer (closest to the interaction point) has received a fluence of approximately $2.0 \times 10^{16}$~$1\,\MeV$ neutron-equivalent particles per $cm^{2}$, resulting in a reduction of the charge-collection efficiency of nearly 60\%; nevertheless, the impact on measurement reconstruction efficiency remains minimal~\cite{IDET-2017-10,IDET-2020-01}.

Charged particle interactions in the TRT primarily produce ionisation in the Xe or Ar gas mixture, generating one-dimensional drift-circle measurements. The position is inferred from the drift time within each straw, with a resolution of approximately \qty{130}{\um} perpendicular to the straws. Both the drift time, and the relationship between the drift time and hit radius are calibrated using an in-situ track-based procedure~\cite{ATLAS-CONF-2011-006}.
Due to the symmetric structure of the straws and the technology used, a \textit{left–right} ambiguity remains, corresponding to the hit position being located on either side of the straw centre, and is resolved during track fitting~\cite{IDET-2015-01}.


\section{Data and simulation samples}
\label{sec:samples}

The primary dataset used in this paper was collected with the ATLAS detector in $pp$ collisions with a centre-of-mass energy $\sqrt{s}$ of \qty{13}{\TeV} during Run~2 (2015--2018) of the LHC.
Results particularly sensitive to radiation damage in Section~\ref{sec:perf:tide} also include a subset of Run~3 data (2022), collected with $\sqrt{s}=13.6$~\TeV.
Only events passing stringent data quality requirements~\cite{DAPR-2018-01} are used.
No significant updates were made to the ID during the data-taking period. Operational parameters such as the pixel sensor depletion voltage, readout thresholds, and others were adjusted to compensate for radiation damage effects, ensuring high operational efficiency~\cite{Battaglia:2023vik}. As a result, the track reconstruction performance shows little variation as a function of time.
The increasing instantaneous luminosity delivered by the LHC to the ATLAS experiment throughout this data-taking period results in $10 < \mu < 65$, with an average of $\langle \mu \rangle = 34$.
An unbiased sample of \pp\ collisions is collected using \emph{zero bias} data, where events are recorded one full LHC orbit after a reference trigger is fired~\cite{TRIG-2016-01}.

Monte Carlo (MC) simulated event samples are used to assess the performance of track and primary vertex reconstruction across a range of final states, pile-up conditions, and track topologies.
The production of top quark pairs $t\bar{t}$ was modelled using the \POWHEGBOX[v2]~\cite{Frixione:2007nw,Nason:2004rx,Frixione:2007vw,Alioli:2010xd} generator at next-to-leading order (NLO) in QCD~\cite{ATL-PHYS-PUB-2016-020,ATL-PHYS-PUB-2016-004, ATL-PHYS-PUB-2017-007,ATL-PHYS-PUB-2018-009} and \PYTHIA[8.186] for showering and hadronisation. The final state is selected such that at least one of the $W$ bosons from the top quark decays leptonically. This sample is used to assess basic performance of track and primary vertex reconstruction in a final state containing both leptonic and hadronic particles. The same generator setup was also used to produce samples of $Z \to \mu \mu$ and $Z^*/\gamma^* \to \mu \mu$ events for use in efficiency studies of ID tracks associated to muons reconstructed in the MS.
Dijet events were generated using \PYTHIA[8.230]~\cite{Sjostrand:2014zea} with leading-order matrix elements for dijet production and interfaced to a \pt-ordered parton shower~\cite{ATL-PHYS-PUB-2019-017}. This sample is primarily used to assess performance within high-\pt jets.
A sample of SM Higgs boson events produced via vector-boson fusion and decaying to four neutrinos (VBF $H\to 4\nu$) was simulated using the \POWHEGBOX[v2] generator at NLO in QCD and \PYTHIA[8.186] for showering and hadronisation. This sample is primarily used to assess the performance of primary vertex reconstruction for processes where a substantial fraction of the collision energy is carried away by particles that are invisible to the tracker.
To assess performance for non-prompt particle reconstruction, a simplified model is used in which a Higgs boson produced in association with a $Z$ boson decays into a pair of BSM electrically neutral pseudoscalar and long-lived $a$ bosons, each subsequently decaying exclusively into a pair of $b$-quarks ($H \to aa \to bbbb$). The non-prompt decay results in a displaced vertex signature. The $a$ boson mass is \qty{55}{\GeV} and the proper lifetime $c\tau$ is \qty{100}{\mm} for this sample. This sample was generated using \POWHEGBOX[v2] at NLO with up to one additional jet, with the decays of particles simulated with \PYTHIA[8.212].

All simulated particles from collisions were processed with the ATLAS detector simulation~\cite{SOFT-2010-01} based on \GEANT~\cite{Agostinelli:2002hh}. In addition to the detailed model of the ID used in event simulation, a simplified geometry representation, referred to as the \emph{Tracking Geometry}~\cite{ATL-SOFT-PUB-2007-004}, is created for use during reconstruction. The tracking geometry consists of simple layers and volumes, onto which a projection of the simulated material traversed prior to that layer (or contained within the volume) is mapped,
with an appropriate binning. This projection is obtained in a dedicated simulation step by recording, as a function of direction, the amount of material traversed between successive tracking geometry layers, and storing it as a look-up table used during reconstruction. Such layers are present for all measurement surfaces (i.e.\ the representations of detector planes), as well as for other relevant material structures. The tracking geometry provides sufficient information about the material distribution to allow track reconstruction to be performed without the high computational cost of loading and navigating through the fully detailed simulation geometry.

The effect of multiple interactions in the same and neighbouring bunch crossings (\pileup) was modelled by overlaying~\cite{SIMU-2020-01} the simulated hard-scattering event with inelastic \(pp\) events generated from a mix of \EPOS[2.0.1.4]~\cite{Werner:2005jf} and \PYTHIA[8.308]~\cite{Bierlich:2022pfr}. The \EPOS\ events were generated with the \EPOS\ LHC tune~\cite{Pierog:2013ria} and the \PYTHIA\ events with the A3 tune~\cite{ATL-PHYS-PUB-2016-017} and the \NNPDF[2.3lo]~\cite{Ball:2012cx} set of parton distribution functions (PDF).\PYTHIA\ pileup events include either a high transverse momentum (\pt) jet, a prompt photon, or a lepton from a $b$-hadron decay, while \EPOS\ was filtered to simulate all remaining pileup events in the overlay sample. The individual simulations were first reweighted to ensure a smooth connection across jet \pt then the combination reweighted to match the distribution of the actual number of interactions per bunch crossing (\(\mu\)) measured in data. However, for several of the studies in this paper, a sample was used comprising events with a wide range of~$\mu$ following a roughly uniform distribution (from~$\mu=0$ up to~$\mu=80$), in order to assess the performance as a function of \pileup. Where such a sample is used it is explicitly labelled as such, except in cases where the observable is displayed as a function of $\mu$ thus making the dependence explicit. While the time response of the ID means that it is primarily sensitive to additional interactions within the same bunch crossing as the event of interest (referred to as \emph{in-time} \pileup), other subdetectors with different time responses can be sensitive to the presence of additional interactions in earlier or later bunch crossings (referred to as \emph{out-of-time} \pileup). Both the in-time and out-of-time contributions are included in the simulated events.

\subsection{Truth-matching in simulated samples}
\label{sec:perf:truthmatching}
In order to establish the performance of the track reconstruction, a requirement is often applied that a reconstructed track is matched to a particle produced during MC generation, referred to as \emph{truth particles}. This ensures correspondence between the simulated input particles and reconstructed output tracks, allowing meaningful comparisons to be made.

In order to perform this matching, information is stored about the truth particles contributing to the simulated energy deposits, thereby mapping the true trajectory of a given simulated particle through the detector. From this information, the contribution of a given truth particle to the measurements on a reconstructed track can be quantified, allowing a hit-purity-based truth association to be defined. A reconstructed track is matched to a truth particle by comparing the number of measurements common between the truth particle trajectory and the reconstructed track. This comparison is based on the quantity $R_{\mathrm{match}}$, defined as

\begin{linenomath}
\begin{equation}
R_{\mathrm{match}} = \frac{10 \times N^{\mathrm{pix}}_{\mathrm{common}} + 5 \times N^{\mathrm{SCT}}_{\mathrm{common}} +  N^{\mathrm{TRT}}_{\mathrm{common}}}{10 \times N^{\mathrm{pix}}_{\mathrm{reco}} + 5 \times N^{\mathrm{SCT}}_{\mathrm{reco}} + N^{\mathrm{TRT}}_{\mathrm{reco}}},
\end{equation}
\end{linenomath}

where $N^{\mathrm{pix/SCT/TRT}}_{\mathrm{reco}}$ is the number of measurements in the different subsystems which are present on the reconstructed particle trajectory. The numbers of measurements in the different subsystems which are present on both the truth particle trajectory and the reconstructed track are given by $N^{\mathrm{pix/SCT/TRT}}_{\mathrm{common}}$. The different weights applied to the different subsystems reflect the different numbers of possible measurements and the precision of the measurements which can be provided by each subsystem. These considerations result in a relative downweighting of TRT measurements with respect to those from the Pixel and SCT.

Reconstructed tracks with high values of $R_{\mathrm{match}}$ are considered well-matched to a truth particle, and can therefore be used to evaluate reconstruction performance. Specifically, they contribute to the numerator in efficiency calculations, and their parameters can be compared to truth for resolution measurements. Unless otherwise stated, tracks with $R_{\mathrm{match}} \ge 0.5$ are considered successfully matched to a truth particle. This threshold is chosen based on an observed correlation between reconstructed and true track parameters. Conversely, tracks with $R_{\mathrm{match}} < 0.5$ are considered poorly matched and are typically the result of incorrect combinations of clusters from multiple particles and interactions. These are referred to as \emph{fake tracks} or just \emph{fakes}. This definition does not exclude that a fake track can be the single best reconstructed match to a given particle. The distribution of $R_{\mathrm{match}}$ is strongly peaked at values close to $1.0$ but is otherwise relatively uniform for values $>0.5$.%

When computing quantities such as efficiencies using truth particles, a selection is applied to identify the set of truth particles against which performance is measured. Only particles produced during event generation are considered. This definition includes decay products of long-lived neutral hadrons such as $K^0_S$ and $\Lambda$, despite originating from displaced vertices. Particles produced in material interactions during detector simulation, such as from photon conversions, are excluded. In addition, selected truth particles must be electrically charged, have a transverse momentum $\pt > \qty{500}{\MeV}$, and lie within the pseudorapidity range $\vert\eta\vert < 2.5$, unless otherwise stated.
No such restriction is applied when classifying fake tracks: tracks matched to particles produced during simulation are excluded from the fake category. This ensures that the fake rate reflects only reconstructed tracks that are not associated with any truth particle originating from a \pp collision.


\section{Track reconstruction}
\label{sec:tracking}

\label{sec:trk:overview}

\begin{figure}[htbp]
\begin{center}
\subfloat[]{\includegraphics[width=0.49\textwidth]{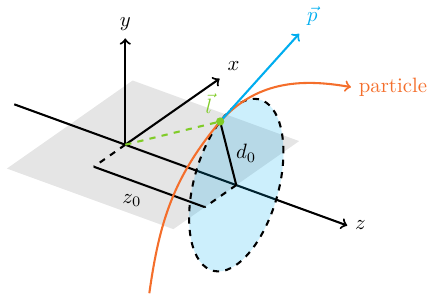}}
\subfloat[]{\includegraphics[width=0.49\textwidth]{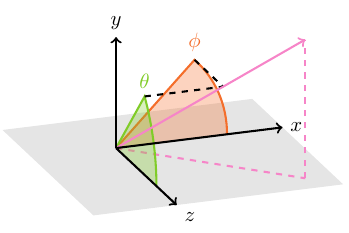}}
\caption{The parameters used to represent tracks showing the definitions of (a) the longitudinal and transverse impact parameters ($\zZero$ and $\dZero$ respectively), and (b) the azimuthal and polar angular parameters. The vector $\vec p$ represents the momentum. A vector from the reference point to the point of closest approach is denoted by $\vec l$. Applying the definitions in (b) to $\vec p$ yields the standard angular parameters $\phi_0$ and $\theta_0$, while applying them to $\vec l$ results in angles denoted as $\phi$ and $\theta$. Diagrams courtesy of the ACTS (A Common Tracking Software) project~\cite{ACTS}.}
\label{fig:overview:trkparam}
\end{center}
\end{figure}

Track reconstruction, or \textit{tracking}, refers to the reconstruction of charged-particle trajectories from clusters of detector measurements.
A charged particle propagating in a uniform magnetic field in vacuum follows a helical trajectory. Accordingly, the most common track representation for ID-based tracking is a local helicoidal (perigee) parameterisation~\cite{ATL-SOFT-PUB-2006-004}, expressed by five parameters relative to a chosen reference line and reference point: $\left(\dZero, \zZero, \phi_0, \theta_0, q/p \right)$. In this parameterisation, illustrated in Figure~\ref{fig:overview:trkparam}, $q/p$ is the ratio of charge ($q$) and momentum ($p$), $\phi_0$ and $\theta_0$  are the azimuthal and polar angles of the momentum vector at the point of closest approach, $\dZero$ is the transverse impact parameter, and $\zZero$ the longitudinal impact parameter. The point of closest approach of the trajectory to the reference line is commonly referred to as the \textit{perigee}. The transverse impact parameter $\dZero$ is defined as the signed distance in the transverse plane between the perigee and the chosen reference line. If $\phi$ denotes the azimuthal angle of the perigee and $\phi_0$ the azimuthal angle of the momentum vector at that point (with $\theta_0$ the corresponding polar angle), the sign of $\dZero$ is defined as $\operatorname{sign}(\dZero) = \operatorname{sign}\big(\sin(\phi - \phi_0)\big)$. This implies $\dZero > 0$ when $0 < \phi - \phi_0 < \pi \ (\mathrm{mod}\ 2\pi)$. This definition corresponds to the ATLAS software convention where $\phi - \phi_0 \approx \pi/2 \ (\mathrm{mod}\ 2\pi)$ for positive $\dZero$. Further details are provided in Appendix~8.2 of Ref.~\cite{ATL-SOFT-PUB-2006-004}.
The longitudinal impact parameter $\zZero$ represents the distance along the $z$-axis between the perigee and the reference point.

The default reference line is the beamline, defined as the line best representing the average position and direction of \pp collisions, commonly referred to as the \textit{luminous region}~\cite{DAPR-2021-01}. The reference point is taken as the centre of the luminous region. These quantities are determined using in-situ calibration techniques with a typical time granularity of approximately \SI{60}{s}~\cite{ATLAS-CONF-2010-027}.

The magnetic field also plays a critical role in track reconstruction.
The charge-to-momentum ratio is determined by the curvature of the particle’s trajectory, which is estimated using the \textit{sagitta}, defined as the perpendicular distance between the particle’s curved path and a straight line connecting two reference points along the trajectory. In a uniform magnetic field $B$, the sagitta $s$ of a charged particle’s trajectory is related to its transverse momentum \pt\ by:  $s \propto \frac{L^2 \, qB}{8\pt}$ where $L$ is the effective lever arm, determined by the radial separation of the measurements contributing to the track fit (typically of the order of the tracking volume radius) and $q$ is the particle charge.
Spatial variations in the magnetic field are accounted for using a detailed map measured during the detector installation. The magnetic field value at any point within the ID is interpolated with a relative precision of $7 \times 10^{-4}$, resulting in a negligible (0.02--0.12\%) contribution to the sagitta uncertainty~\cite{Aleksa:2008zza}.

The tracking algorithms must account for several physical effects, which increase their complexity and influence the resultant performance.
Hadronic interactions with detector material can prevent a particle from reaching outer layers, thus limiting the number of usable measurements.
The probability of hadronic interactions depends on the amount of nuclear interaction lengths traversed, shown in Figure~\ref{fig:detector:hadintlength}.
Multiple scattering alters the particle trajectory, increasing the direction uncertainty and making cluster-to-track assignment more challenging especially in the presence of multiple possible clusters.
This effect depends on the amount of radiation lengths traversed, shown in Figure~\ref{fig:detector:radlength}, and is approximately inversely proportional to the particle's momentum.
At low momentum (up to a few \GeV), the resolution on track parameters is dominated by multiple scattering, whereas at high momentum, it is limited by the intrinsic position resolution of the detector.
For electrons, and very high-momentum muons, radiative energy losses are a limiting effect on reconstruction efficiencies, as described in Section~\ref{pattern:brem}.

The ATLAS track reconstruction strategy starts from the silicon detectors.
The tracking strategy consists of multiple dedicated passes, each optimised for different topologies or particle origins.
The main track reconstruction sequence, referred to as the \textit{primary sequence} or \textit{inside-out sequence}, is designed to reconstruct primary charged particles with high efficiency, while keeping the number of \emph{mis-reconstructed} tracks extremely low. Mis-reconstructed tracks are reconstructed trajectories that include measurements originating from more than one charged particle. In particular, fake tracks are those that cannot be matched to any single generated particle with sufficient purity (see Section~\ref{sec:perf:truthmatching}). Tracks that are mostly correct but include a small number of measurements from other particles are not classified as fakes, but may still be considered mis-reconstructed depending on the impact of the incorrect measurements on reconstructed parameters.

\begin{figure}[htbp]
\begin{center}
\includegraphics[width=0.9\textwidth]{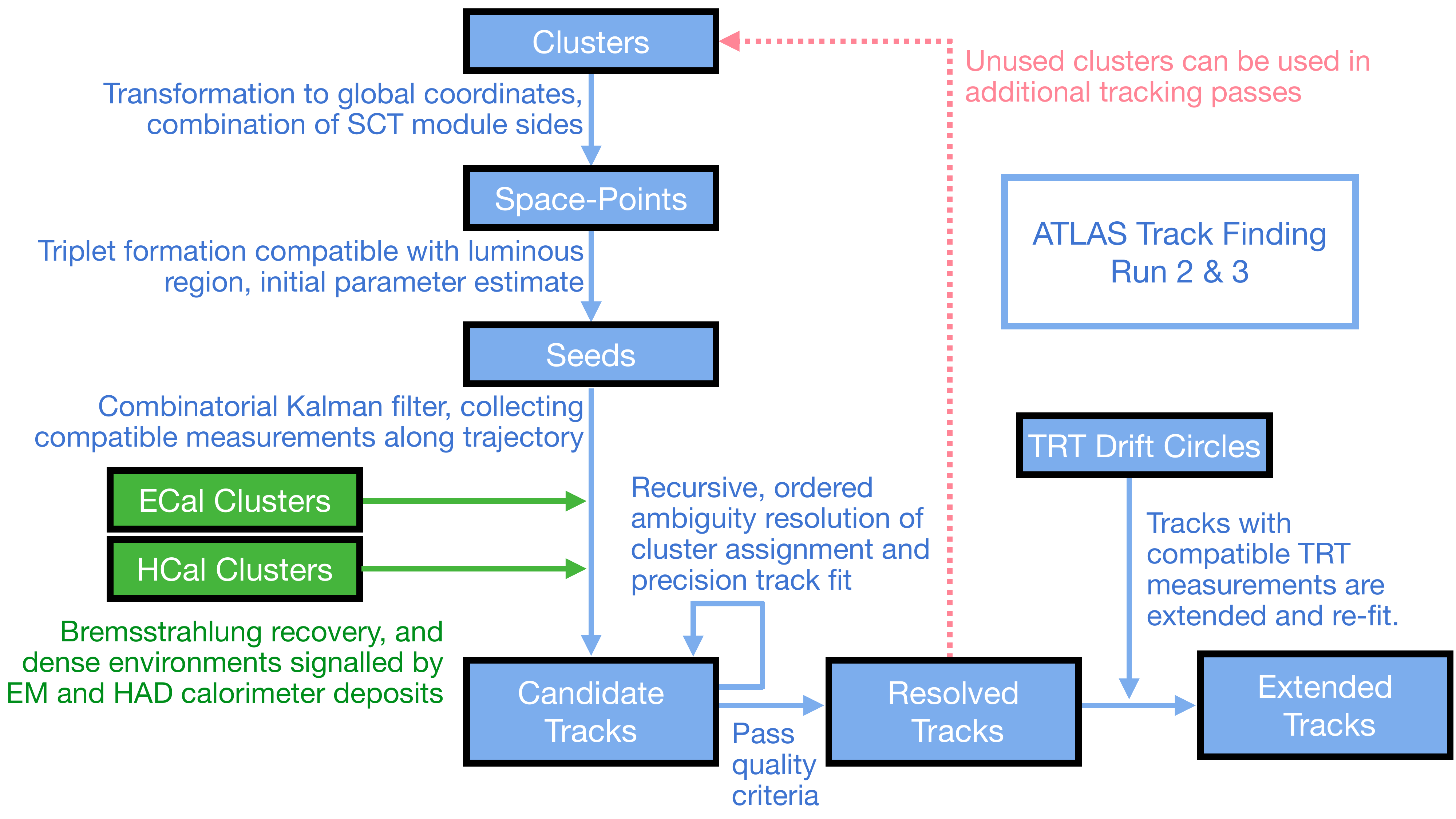}
\caption{Flow diagram depicting the interaction of the different stages of the track reconstruction process. Blue boxes represent various data collections constructed from Inner Detector information which are processed as described in the order indicated by the arrows. The green boxes represent data collections created from information in the calorimeter systems.}
\label{fig:overview:tracking_flow}
\end{center}
\end{figure}

\begin{figure}[htbp]
\begin{center}
\includegraphics[width=0.6\textwidth]{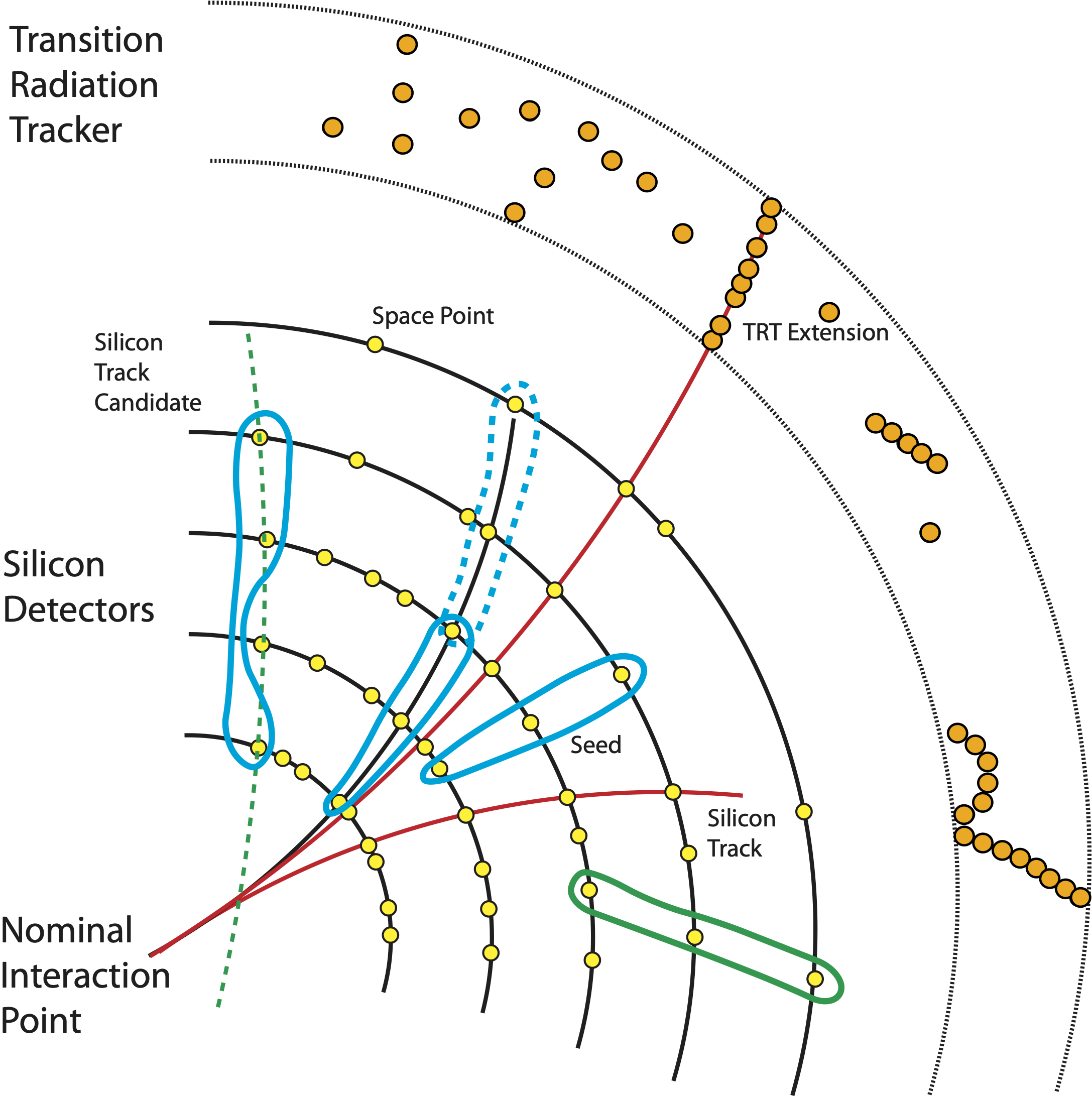}
\caption{Illustration of the stages of inside-out track pattern recognition in a simplified representation of the Inner Detector transverse plane. The silicon space points are shown in yellow, and TRT drift circles in orange. The seeds reconstructed from combinations of space points are indicated in blue. The dashed blue seed illustrates a case where two seeds correspond to the trajectory of the same charged particle. The green seed is rejected by the requirement that the seed be consistent with a particle from the nominal interaction point. The track candidates are shown with lines. The green dashed track candidate is rejected because it is inconsistent with the nominal interaction point. The black track candidate is reconstructed up to the CKF stage. The red track candidates are fully reconstructed silicon tracks fit with the global $\chi^2$ where an extension into the TRT is possible.}
\label{fig:overview:tracking}
\end{center}
\end{figure}

The first step involves forming three-dimensional \emph{space points} from detector measurements. Groups of three space points form \emph{track seeds}, or simply \emph{seeds}, which provide an initial trajectory estimate. Seeds meeting quality requirements are passed on to the \emph{track finding} stage, where additional clusters are added using a Kalman filter~\cite{Kalman}. These steps constitute the \emph{pattern recognition} sequence. The resulting \emph{track candidates} are passed through an \emph{ambiguity resolution} algorithm that selects the most promising track candidates and resolves cases where clusters are assigned to multiple track candidates. A global $\chi^2$ track fitting algorithm is run on the \emph{resolved} track candidates to determine the parameters of the tracks. Tracks may be further extended into the TRT detector by adding compatible TRT measurements and refitting them together with the silicon measurements for the final trajectory determination.
These steps are illustrated in terms of data flow in Figure~\ref{fig:overview:tracking_flow} and diagrammatically within a simplified Inner Detector cross-section in Figure~\ref{fig:overview:tracking}.
The general steps are described in Sections~\ref{sec:trk:pattern}--\ref{sec:trk:trt}; however, the exact implementation and behaviour depend on the specific tracking pass and the properties of the track candidates.

The primary sequence is complemented by a TRT-seeded pass, where TRT drift circles are first grouped into track segments and extrapolated inward to search for compatible silicon clusters which were not used previously. This sequence is discussed in Section~\ref{sec:trk:trkpasses}.
Specialised tracking passes are also employed to target signatures that evade common assumptions. A specific example is reconstructing particles coming from the decay of particles that travel macroscopic distances in the detector; this sequence is run after the primary sequence and it is described in Section~\ref{sec:trk:trk_LRTpass}.

The output of the track reconstruction algorithms serves a large variety of higher-level object reconstructions and is also used in physics analyses directly. Depending on the application, additional quality selections are needed to restrict the track sample.
Section~\ref{sec:perf:quality} defines additional track quality selections provided to allow different compromises between high efficiency and low rate of mis-reconstructed tracks.

\subsection{Pattern recognition}
\label{sec:trk:pattern}

\subsubsection{Space point formation and seed finding}
\label{sec:seeding}
Space points provide three-dimensional positional information formed from clusters in the silicon detectors.
Pixel measurements correspond to two-dimensional positions on sensor surfaces, which are transformed into three-dimensional information in the global frame using the known sensor alignment and orientation.
SCT sensors provide effectively only one-dimensional measurements due to the very different precision in the two sensor directions. Clusters on both sides of the SCT modules, with each side being considered an independent \textit{measurement surface}, are combined into a single space point. This combination exploits the \qty{40}{mrad} stereo angle between the sides to resolve the second coordinate\footnote{The \qty{40}{mrad} stereo angle balances improved spatial precision against the increased probability of \emph{ghost clusters} when multiple particles leave overlapping clusters.}. This explicit combination of measurements on the two sides of the module is only done for space points. Elsewhere, the information from the stereo rotation is incorporated implicitly via the relative rotation of measurement surfaces and thus included in the expression of track parameters and covariances.

Seeds are formed by identifying triplets of compatible space points across three separate layers of the silicon detectors, subject to geometrical constraints reflecting the detector layout and acceptance. The seeds are either formed out of three pixel (PPP) or three SCT space points (SSS). Seeds formed by a combination of pixel and strip space points are not used in the primary sequence due to their lower purity under nominal \pileup conditions, a result of the relatively large distance and amount of material between the outermost pixel and innermost SCT layers.

SSS seeds are considered first due to their higher purity. They are required to point back to a $\qty{250}{\mm}$ window along the beamline ($|z(r=0)|<\qty{250}{\mm}$), ensuring consistency with an origin the luminous region. The SSS seeds are processed through \emph{track finding}, discussed in Section~\ref{sec:tracking:finding}, and the resulting track candidates are extrapolated to the beam line to estimate, on an event-by-event basis, the longitudinal extent of the interaction region. This estimate is then used as a constraint in the subsequent processing of PPP seeds, rejecting candidates pointing outside the observed $z(r=0)$ range before PPP track finding is performed.

A quality score is assigned to seeds in order to prioritise those with small transverse impact parameter and large transverse momentum for subsequent processing.
Compatibility of a fourth space point in a different layer further increases priority as such \emph{confirmed} seeds are promising candidates.
Seeds are further required to pass $|\dZero| < \qty{5}{\mm}$, $|\zZero| < \qty{200}{\mm}$, and $\pT > \qty{500}{\MeV}$. For further detail on the seeding algorithm, see Ref.~\cite{ATL-PHYS-PUB-2023-034}.

\subsubsection{Track finding}\label{sec:tracking:finding}
Track finding is a combined measurement association and fitting stage. Seeds define the initial parameters and form a basis for generating \textit{track candidates}. Seeds are skipped if all their space-points are already used in reconstructed track candidates. Additionally, a PPP seed is not considered if all its space points are on other higher-quality seeds.

Remaining seeds are extrapolated along the initial trajectory estimate through the tracking volume to build a region of compatible detector elements, or a \textit{road}. Any detector elements within a \textit{road width}, \qty{12}{mm} for the primary sequence, of the extrapolated trajectory at each layer is collected.

Space-points are used exclusively at the seeding stage. Once track finding proceeds beyond seed formation and road building, the reconstruction operates on individual detector measurements (clusters), with SCT clusters treated independently on each sensor side. Track candidates are then created using a combinatorial Kalman filter (CKF)~\cite{Kalman,FRUHWIRTH1987444}, which incrementally associates individual cluster measurements to the evolving trajectory. Track candidates are represented as a list of \emph{track states} of the following types:

\begin{itemize}
\item \emph{Measurement} states correspond to individual detector clusters and the associated track-parameter estimate at the corresponding measurement surface,
\item \emph{Hole} states represent missing expected clusters on measurement surfaces between the innermost and outermost radius elements,
\item \emph{Scatterer} or \emph{bremsstrahlung} states represent significant changes in direction or momentum and are treated during fitting (Section~\ref{sec:trk:fitting}),
\item \emph{Outlier} states represent measurements with poor compatibility with the trajectory due to a large contribution to the total track $\chi^{2}$. They are only present after the precision track fitting discussed in Section~\ref{sec:trk:fitting}. They are excluded from the fit but retained for book-keeping and possible future refits under different assumptions, e.g.\ a different particle hypothesis.

\end{itemize}

The CKF extends the seeds both towards and away from the interaction point, performing smoothing steps incorporating measurements and subsequently refining the trajectory parameters.
Track propagation uses the Runge-Kutta-Nystr\"{o}m method~\cite{nystrom,Lund_2009} to transport the track parameters and covariances layer-by-layer according to the equation of motion, using magnetic field maps and accounting for detector material via contributions to the covariance matrix.

The propagation direction depends on seed type: SSS seeds are initially propagated inwards, PPP seeds outwards.
If a candidate shares most of its measurements with a better-quality candidate, or if empirical metrics such as measurement multiplicity, hole count, or $\chi^2$ estimates deteriorate beyond tunable thresholds, the extension is terminated early and the algorithm moves to the next available seed.

The remaining track candidates are required to pass quality criteria, the most important being a transverse momentum of \pT $> \qty{500}{\MeV}$, at least eight measurements, and no more than two holes on the track. In addition, the overall track-fit quality is required to satisfy $\chi^2/\mathrm{DoF} < 9$, where the $\chi^2$ is accumulated from the cluster-to-track residuals during the Kalman filter procedure and the number of degrees of freedom (DoF) is defined by the number of measurements contributing to the candidate.

\subsubsection{Bremsstrahlung recovery}\label{pattern:brem}
\label{sec:brem}

Due to the typical particle composition of events, and in the absence of information from other detectors, all particles are assumed to be charged pions. However, electrons have different energy loss properties in material, in particular the production of bremsstrahlung photons, resulting in losses in reconstruction efficiency when treated as pions. Electron inefficiencies could be recovered by allowing for larger energy losses during track finding, but this would degrade reconstruction performance for other particle types.
Instead, \emph{bremsstrahlung recovery} is performed only when electromagnetic calorimeter energy deposits consistent with bremsstrahlung are detected. These deposits define a \emph{Region of Interest} (RoI) in an $\eta-\phi$ cone, within which the ambiguity resolution described in the next section is relaxed, and the track search road is expanded to accommodate electron-like energy losses (discussed in more detail in Section~\ref{sec:GSF}).

%


\subsection{Ambiguity solving}
\label{sec:trk:ambi}

The fine granularity of the ID means that the probability of multiple, independent charged particles producing hits in the same sensor element is small across most kinematic regimes.
This informs the reconstruction strategy: a high-efficiency pattern recognition stage is designed to identify all plausible combinations of measurements consistent with a charged-particle trajectory, ensuring genuine tracks are not lost.
As a result, ambiguities, such as individual measurements being shared across multiple candidates, can arise.
The next stage in track reconstruction is designed to resolve such conflicts and remove track candidates inconsistent with sensor occupancies or kinematic expectations to suppress mis-reconstructed tracks without losing efficiency.

\begin{figure}[htbp]
\begin{center}
\includegraphics[width=0.9\textwidth]{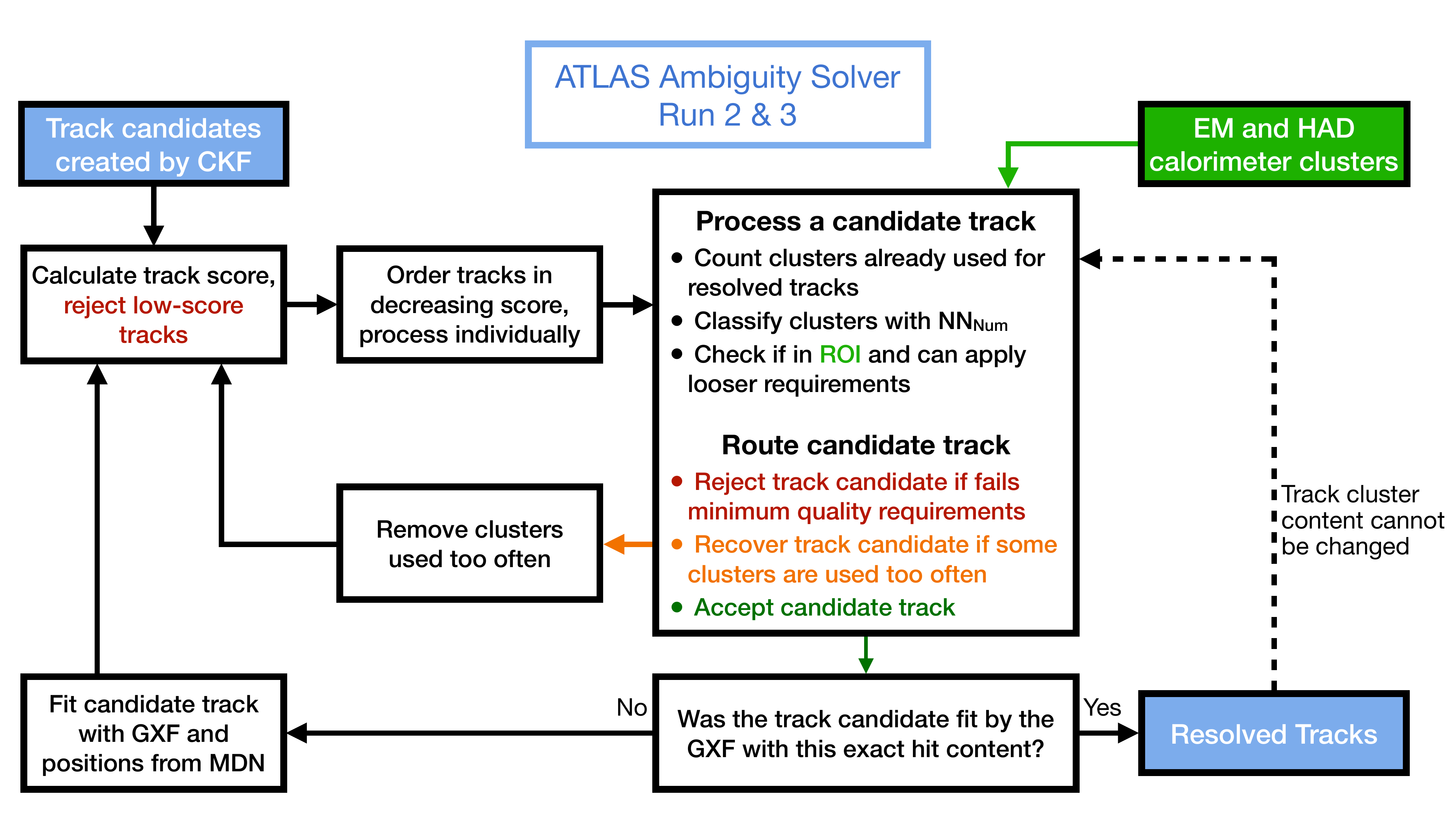}
\caption{Flow diagram depicting the ambiguity solver. Blue boxes represent various data collections constructed from Inner Detector information. The green boxes represent data collections created from information in the calorimeter systems. The GXF is the Global $\chi^2$ Fitter, and MDN is a Mixture Density Network, both described in Section~\ref{sec:trk:fitting}.}
\label{fig:ambi:flow}
\end{center}
\end{figure}

The \emph{ambiguity solver} is designed to determine the optimal set of tracks for downstream algorithms.
It processes track candidates ordered by a quality score, resolving measurement competition with previously accepted tracks, eliminating low-quality candidates, and initiating the global $\chi^2$ fit described in Section~\ref{sec:trk:fitting}.
The flow of tracks through the solver is described below and summarised in Figure~\ref{fig:ambi:flow}.

To be accepted, a track candidate must satisfy requirements on the number of total and \emph{unique clusters} and \emph{shared modules}. For example, in the primary sequence, a candidate must have at least six unique silicon clusters and no more than one shared module. A cluster is considered \emph{unique} if it is used on a number of tracks consistent with the cluster classification scheme described below; otherwise, it is marked as \emph{shared}. For Pixel clusters, a shared cluster corresponds to one shared module. In the SCT, shared clusters are combined into an effective module-level quantity, such that two shared (single-sided) SCT clusters correspond to one shared module.
Inactive sensors along the trajectory are counted as clusters to ensure robust performance under variable detector conditions.

Clusters appearing on multiple tracks are often the result of incorrect assignment. However, in dense environments, such as the core of high-\pt jets ($\pT \gtrapprox \qty{300}{\GeV}$), the charged particle separation can be comparable to the sensor granularity, and multiple charged particles may contribute to a single \emph{merged} cluster. This effect is more pronounced closer to the beamline, where the charged particle density is highest. An artificial neural network (\nnNum) is trained to identify merged Pixel clusters and classify them as compatible with one (P1), two (P2), or three or more (P3) particles, based on the charge distribution pattern, candidate incident angle, and sensor location~\cite{PERF-2015-08}. A configurable threshold is applied on the class probabilities to define the cluster type. This approach exploits the analogue charge information available in the Pixel detector and is therefore not applicable to the SCT, where the hit information is binary. The network is re-trained for each data-taking and simulation campaign to account for variations in charge collection due to radiation damage and other time-dependent detector effects~\cite{IDET-2017-10,IDET-2020-01}. As the classification relies on the total collected charge and the cluster shape, the progressive degradation of charge collection in the Pixel sensors due to radiation damage leads to a corresponding reduction in the efficiency of the re-trained model to identify merged clusters. Considering the fluence at the end of 2025 data taking relative to the pre-irradiation performance, the \nnNum~performance degradation is on the order of 10\%.

Pixel cluster assignment in track reconstruction is guided by the classification provided by the \nnNum\ neural network. P1 clusters should belong to only one track but can be shared with one additional track (i.e., used on two tracks at most).
P2 and P3 clusters can be used on up to two and four tracks, respectively, without being considered shared. P2 and P3 clusters cannot be shared beyond these limits (i.e.\ used on a third or fifth track, respectively).
In cases where two tracks compete for clusters on consecutive Pixel layers, the \nnNum\ often predicts the clusters on both layers were created by multiple particles (P2 or P3). However, to avoid unnecessarily rejecting a valid track, an inner layer cluster originally labelled P1 is re-classified to P2 if the outer layer cluster is labelled P2 or P3.
SCT clusters have the same sharing limits as P1 pixel clusters, as without analogue charge information merged SCT clusters are difficult to identify~\cite{ATLAS-CONF-2013-005}.

Track candidates are prioritised by a score~\cite{ATLAS-CONF-2010-072} constructed from basic track-quality and reconstruction observables. The score increases with the number of clusters, with weights of 20, 10, and 1 for Pixel, SCT, and TRT measurements. The score decreases with the number of holes, using a weight of 10 per Pixel module, and weights of 5 or 15 for holes on one side or both sides of an SCT module, respectively. Poor fits are penalised based on the $\chi^2$ of the track once a fit has been performed. Additionally, a term increases the score based on the logarithm of the candidate \pT\ since high \pT\ tracks are typically prioritised as physics objects and are less susceptible to pattern recognition issues.

Candidates are processed in descending score order until all are accepted or removed. Once a track candidate is accepted, its cluster content is fixed and is not modified by any subsequently processed candidates. Only clusters associated with already accepted tracks are considered shared when evaluating a new candidate. Shared clusters are then subject to usage limits.
If a candidate track has $\pT>\qty{1}{\GeV}$, at least nine clusters, and no more than one Pixel hole, the \nnNum\ is evaluated on shared clusters. Clusters are removed from a candidate if they exceed the usage limit. A candidate must also have a minimum of four unique SCT clusters to retain any shared SCT clusters. Shared clusters are removed from a candidate if the fit yields a $\chi^2/\mathrm{DoF} > 3$.
Clusters are also removed from a candidate if accepting it would cause a cluster that is currently unique on an accepted track to become shared and cause that accepted track to fall below its minimum unique cluster requirement or exceed the shared module limit.
This last constraint is relaxed if the candidate has at least two Pixel clusters and the innermost cluster is not shared.

As part of the bremsstrahlung recovery procedure described in Section~\ref{sec:brem}, dedicated ambiguity resolution rules are applied to ensure efficient reconstruction of electrons and photon conversions. By default, shared clusters are penalised throughout the detector; however, to maintain efficient reconstruction of electrons and photon conversions, tracks are not penalised for the first shared Pixel cluster \emph{after} the IBL or for shared SCT clusters if they are compatible with an EM calorimeter cluster RoI and pass additional selection criteria.
Shared Pixel clusters on the IBL are not included in this relaxation of requirements, as sharing at the innermost layer leads to significantly increased combinatorial ambiguity.
Also, to target the core of high-\pT\ hadronic objects, the same relaxed requirements are applied to candidates within RoIs defined by high-energy calorimeter clusters.
Within either type of RoI, a candidate must satisfy the following additional criteria: it must have four unique SCT clusters to allow one shared cluster, no more than one hole across the Pixel and SCT detectors, $\pT > \qty{1}{\GeV}$, and must overlap significantly with only one accepted track. Specifically, all shared clusters must be with the same accepted track, and there must be at least two such shared clusters.

Clusters on a candidate track that are used too often according to the criteria described below are removed. Otherwise, on satisfying all the requirements of the cluster-based ambiguity resolution, the track candidate enters the precision track fit as described in Section~\ref{sec:trk:fitting}. This fit refines the track parameters, may identify outlier measurements that modify the cluster content, and produces a track $\chi^2/\mathrm{DoF}$. The fitted candidate is then re-scored, including the $\chi^2/\mathrm{DoF}$, and returned to the ambiguity solver input list.
On its next pass through the solver, cluster counting and sharing limits are re-calculated following the removal of shared outliers. If the candidate still satisfies all ambiguity resolution criteria without further modification, it is added to the set of output tracks available for downstream algorithms.

\subsection{Track fitting}
\label{sec:trk:fitting}

Once a suitable set of track candidates has been selected, the next step in reconstruction is \emph{track fitting}. The goal is to provide the best estimate of the track parameters using the full set of measurements assigned to a single charged-particle trajectory.

Each measurement on a candidate track can be refined using trajectory-dependent information such as the incident angle on the sensor, allowing improved estimates of the measurement uncertainties and cluster positions for the precision fit.
In particular, pixel cluster positions and uncertainties are refined using Mixture Density Networks (MDNs)~\cite{Khoda_MDN} which provide more accurate measurement position estimates than the centre-of-gravity method used during pattern recognition.
The networks use the same input features and are trained following the same schedule as the \nnNum\ networks (Section~\ref{sec:trk:ambi}), i.e. whenever a set of simulated samples is produced that include changes affecting cluster properties. Training the MDN on simulated samples that incorporate radiation-damage effects partially mitigates the degradation of the spatial resolution, limiting the impact on the average local measurement resolution to 25\% ($\Delta \sigma_{x/y}/\sigma_{x/y}$) compared to a degradation of about 45\% without corrections, when evaluated at the fluence corresponding to the end of 2025 data taking relative to the pre-irradiation performance.
For each cluster on a track, the MDNs estimate one, two, or three possible crossing points, with associated uncertainties, for clusters classified as P1, P2, or P3, respectively, as discussed in Section~\ref{sec:trk:ambi}. These refined positions and uncertainties are then used in the track fit.

While Kalman filter techniques (Section~\ref{sec:trk:pattern}) are suitable for trajectory propagation and seed-based fitting, the precision fit in ATLAS uses a least-squares regression method referred to as the \emph{Global $\chi^2$ Fitter} (GXF)~\cite{Cornelissen:2008zza, BILLOIR1984352}. This choice offers two main advantages:

\begin{itemize}
\item The GXF provides a straightforward treatment of left–right ambiguities in straw tube detectors such as the TRT. A Kalman-based approach would otherwise require multiple fit hypotheses or inflated uncertainties.

\item The GXF accounts for large material effects by incorporating explicit scattering angles into the fit rather than relying on uncertainty inflation. Degrees of freedom are introduced for both measurement states and dedicated \emph{scatterer} track states at known material structures.
\end{itemize}

These features are also critical for \emph{combined} track fits using measurements from both the ID and MS, where large amounts of material and gaseous detectors must be handled consistently.

\subsubsection{The Global $\chi ^ 2$ Fitter}

The GXF fit is based on minimising the residuals between predicted trajectories and measured positions. The residuals represent the differences between the intersection of the predicted trajectory for a given set of track parameters with the measurement surfaces and the measured positions on those surfaces. The trajectory prediction used in the residual calculation is performed accounting for the full magnetic field information and assumes Gaussian-distributed energy losses at each material surface.

Energy loss is estimated under a \emph{particle hypothesis}. By default, tracks are assumed to be pions, but alternative hypotheses can be utilised. Electron tracks can also be refit using a dedicated fitting algorithm (Section~\ref{sec:GSF}) for improved performance. The $\chi^2$ expression to be minimised is calculated using the residuals, the scattering angles on any material surfaces, and their respective covariance matrices. An iterative solution using the Newton--Raphson method is applied to determine the best-fit track parameters $\per_{\mathrm{fit}}$. Further details can be found in Ref.~\cite{Cornelissen:2008zza}; its use within the Inner Detector alignment procedure is explored in Ref.~\cite{IDTR-2019-05}.

\subsubsection{The Gaussian Sum Filter}\label{sec:GSF}

Electrons frequently undergo bremsstrahlung in material, resulting in large and non-Gaussian energy losses. The Bethe–Heitler formula~\cite{doi:10.1098/rspa.1934.0140} describes this process, with additional corrections such as the Landau–Pomeranchuk–Migdal effect becoming relevant for energies above a few \GeV~\cite{PhysRev.103.1811}. A single-Gaussian model cannot adequately describe significant energy loss from these non-linear bremsstrahlung effects.

To address this, identified electron tracks~\cite{PERF-2017-01} within the silicon detectors are refit using the \emph{Gaussian Sum Filter} (GSF)~\cite{FRUHWIRTH2003131, ATLAS-CONF-2012-047}. The GSF is an extension of the Kalman filter that models the bremsstrahlung energy loss probability density function as a weighted sum of Gaussian components. Each component is treated as an independent Kalman filter process, and the weighted sum of these components provide an effective description of the non-Gaussian energy-loss distribution.

The GSF operates on track states processed by the GXF which may include outlier states. During GSF fitting, outliers that become compatible can be reintegrated, while previously compatible measurements may be reclassified as outliers.

The use of the GSF leads to a significant improvement in the reconstruction of electron track parameters in the bending plane. Compared to standard GXF fits, the GSF reduces biases in the reconstructed $q/p$ by up to a factor of four and improves the resolution of the track $\phi$ and $\dZero$ by approximately 10--70\% for electrons undergoing substantial bremsstrahlung, with the largest gains observed at high pseudorapidity where the material budget is greatest. These improvements result in better track--calorimeter matching, a narrower and more symmetric $E/p$ distribution, and increases in electron reconstruction and identification efficiencies of a few percent, particularly at low transverse momentum and in the endcap regions.
A detailed comparison of the GSF and GXF approaches is presented in Ref.~\cite{ATLAS-CONF-2012-047}.
%
%
%


%
%
%

%
%

\subsection{TRT extensions}
\label{sec:trk:trt}

Following the silicon-only track finding and fitting, an attempt is made to extend tracks into the TRT.
Adding measurements from the outermost tracking detector improves momentum resolution by increasing the lever arm.
The track parameters at the entrance to the TRT detector are used to build an extension road, within which drift tube measurements can be collected.
If the road contains at least nine TRT measurements, the track is refit using the Global $\chi^2$ Fitter. %
The extension is rejected if the fit quality degrades after including the TRT measurements.
To be accepted, the extended track must also satisfy: $\pT > \qty{1}{\GeV}$ (in order to reject tracks that could loop within the TRT), and if more than 15 TRT measurements are used, the fraction of \emph{precision} measurements must be greater than $0.3$.
A precision measurement is defined as a drift-circle radius measurement that is consistent with the track hypothesis, requiring the measured radius to be within $1.75\sigma$ of the distance of closest approach of the track to the corresponding wire\footnote{Since the trajectory is updated during the fit, the estimated distance to the wire can change during the fit procedure, and so the precision classification may need to be updated. However, this is only done after three minimisation iterations within the fitter.}, where $\sigma$ represents the width of the unbiased residual between the drift circle and the track trajectory. The values for $\sigma$ are calibrated depending on the drift time and the detector layer. Non-precision measurements receive a lower weight in the fit. Over three quarters of tracks from the primary sequence within TRT acceptance are successfully extended into the TRT.

\subsection{TRT-seeded track finding}
\label{sec:trk:trkpasses}
In addition to the silicon-based tracking sequence, a TRT-seeded tracking pass is run to recover secondary tracks.
It is tuned primarily to recover tracks originating from $\gamma$-conversions, which may otherwise have a low reconstruction efficiency due to originating within the ID volume and traversing few or no silicon detector layers.
This sequence uses only measurements not previously used on any other successfully reconstructed track. It proceeds from the outer radius of the tracking detectors inward and
is referred to as \emph{TRT-seeded track finding}, or \emph{back-tracking}.

Calorimeter clusters satisfying a transverse energy requirement of $\ET> \qty{6}{\GeV}$ are used to define RoIs for back-tracking. TRT segments are formed within each RoI using a Hough transformation of TRT hits to identify compatible measurements.
Since the TRT straw measurements lack $z$ information in the barrel and $r$ information in the endcaps, the same measurements can be used in multiple RoIs.
TRT segments with at least 15 measurements and $\pT > \qty{1}{\GeV}$ are extrapolated inward into the silicon detectors.
If a suitable TRT segment is found, a silicon seeding procedure is initiated using only SCT modules in the same RoI, initiating the procedure by defining a region for an additional tracking pass.
Seeds with two space points are formed and used to search for compatible measurements on a road inwards toward the interaction point. If no silicon seed is found, the track is retained as a \emph{TRT-only track}. If the track candidate formed during the inwards search contains at least four silicon clusters and at most one hole, it is extrapolated back into the TRT and a TRT extension procedure as described in Section~\ref{sec:trk:trt} is performed. In this case, the resulting TRT extension can differ from the original TRT segment that seeded the inwards search. In all other cases where compatible silicon clusters are found, a GXF fit is performed using the silicon clusters and the original TRT segment.
%
%


\subsection{Tracking for long-lived particles}
\label{sec:trk:trk_LRTpass}

The \textit{Large-Radius Tracking} (LRT) sequence targets the reconstruction of charged particles produced in the decays of long-lived particles (LLPs), which can have macroscopic displacements from the primary interaction point. This reconstruction strategy is particularly important for searches involving unstable LLPs, as predicted in various extensions of the Standard Model.
The standard inside-out sequence is not efficient in reconstructing charged particles from decays occurring beyond about \qty{5}{\mm} in the transverse direction from the IP. For SM processes such as $b$-hadron decays, displacements are typically associated with a corresponding boost which mitigates this problem, but for BSM LLP scenarios displaced particles could be produced without significant boost. LRT addresses this as an additional tracking sequence using only the detector clusters not already assigned to primary tracks, ensuring an orthogonal track collection. It uses the same logic and algorithms as the primary sequence but is specifically optimised for displaced topologies.
The large LRT tracking phase space dramatically increases the track combinatorics. Therefore strict selection criteria are applied to maintain manageable event processing times.

LRT tracks are seeded from SCT space points only. Seeds are ranked using a quality score assignment, and a maximum of one seed per middle space point (i.e.\ the central space point in the triplet) is allowed (this requirement is not applied to seeds confirmed with a fourth space point, as described in Section~\ref{sec:seeding}). To reflect the extended phase space of LRT tracks, the impact parameter selections are significantly relaxed to $|\dZero| < \qty{300}{\mm}$ and $|\zZero|<\qty{500}{\mm}$.
The same relaxed impact parameter selections are maintained through the track finding and ambiguity resolution steps. Other criteria, however, are significantly stricter to control the mis-reconstructed track rate.
The minimum \pt threshold is increased to \qty{1}{\GeV} to suppress the large number of soft charged pion tracks created in material interactions. The seed ordering is adapted to reflect the correlated direction of the LLPs and their decay products.
LRT tracks are required to satisfy the same strict silicon cluster multiplicity and track $\chi^2$ requirements as primary tracks\footnote{Typically, the number of available measurements for primary particles exceeds the required number of silicon clusters, allowing the same criteria to be applied, even for particles that decay within the detector volume, at the cost of some efficiency.}. In addition, the maximum number of allowed silicon holes is reduced to one, and the search road width during track finding is narrowed. The resulting tracks (both with and without TRT extensions) form the LRT track collection. More details can be found in Ref.~\cite{IDTR-2021-03}.


%
%
%
%

\subsection{Quality requirements}
\label{sec:perf:quality}

Track quality Working Points (WPs) are defined to further refine the selection of reconstructed tracks, targeting different trade-offs between efficiency and mis-reconstruction rate. Two WPs, called \emph{Loose} and \emph{Tight Primary}, are defined and summarised in Table~\ref{tab:selections}.

\begin{table}
\centering
\caption{Loose and Tight Primary selection definitions, where $N_{\mathrm{Si}}$ is the number of silicon measurements (i.e.\ Pixel + SCT) including inactive sensors, $N^{\mathrm{shared}}_{\mathrm{mod}}$ is the number of shared modules ($N^{\mathrm{shared}}_{\mathrm{Pix}} + N^{\mathrm{shared}}_{\mathrm{SCT}}/2$), $N^{\mathrm{hole}}_{\mathrm{Pix}}$ is the number of Pixel holes, $N^{\mathrm{hole}}_{\mathrm{Si}}$ is the number of silicon (Pixel + SCT) holes, and $N_{\mathrm{L0}}$ and $N_{\mathrm{L1}}$ are the number of innermost and next-to-innermost layer measurements, respectively.}
\renewcommand*{\arraystretch}{1.25}
\begin{tabular}{c | c | c}
\toprule
Selection Criteria & Loose & Tight Primary \\
\midrule
$\pt$ & $> \qty{500}{\MeV}$ & $> \qty{500}{\MeV}$ \\
$|\eta|$ & $< 2.5$ & $< 2.5$ \\
$N_{\mathrm{Si}}$ & $\geq 8$ & $\geq 9$ if $|\eta| \leq 1.65$, $\geq 11$ elsewhere \\
$N^{\mathrm{shared}}_{\mathrm{mod}}$ & $\leq 1$ & $\leq 1$ \\
$N^{\mathrm{hole}}_{\mathrm{Pix}} $ & $ \leq 1$ & $=0$ \\
$N^{\mathrm{hole}}_{\mathrm{Si}}$ & $\leq 2$ & $\leq 2$ \\
$N_{\mathrm{L0}} + N_{\mathrm{L1}}$ & - & $> 0$ (if clusters are expected on both layers) \\
\bottomrule
\end{tabular}
\label{tab:selections}
\end{table}

The Loose WP closely matches the track quality requirements applied during the primary
reconstruction sequence and prioritises high efficiency, accepting a moderate rate of mis-reconstructed tracks. It was optimised based on studies of physics performance and computational cost under Run~3 \pileup conditions~\cite{IDTR-2022-04}. The Tight Primary WP imposes stricter requirements aimed at reducing the mis-reconstructed track rate with a corresponding reduction in efficiency. In particular, this working point introduces a criterion on the total number of measurements on the two innermost Pixel layers, to target a purer selection of tracks from particles produced in the primary interaction.


\FloatBarrier

\section{Primary vertex reconstruction}
\label{sec:vertex}


%
Primary vertices represent the reconstructed position of \pp interactions in the ATLAS detector. Accurate and efficient primary vertex reconstruction is essential for determining the event kinematics. Usually only a single \pp interaction per recorded bunch crossing produces physics of interest. Typically, this interaction is the one with the largest sum of transverse momenta of its final state objects and is referred to as the \emph{hard-scatter} vertex.

The primary vertex reconstruction algorithm proceeds through four stages: seed finding, association of tracks with seeds, vertex fitting, and vertex acceptance or rejection.
The first stage begins by selecting a subset of tracks that pass quality criteria designed to remove poorly measured and mis-reconstructed tracks. For the standard primary vertex configuration, the Tight Primary working point discussed in Section~\ref{sec:perf:quality} is used.
The reconstruction algorithm follows a \emph{finding-through-fitting} approach, in which new vertices are seeded from tracks not yet compatible with an existing vertex. This strategy allows maximal exploitation of the available tracks and is implemented via an iterative procedure known as the \emph{Adaptive Multi-Vertex Finder} (AMVF).
Reflecting the small transverse extent of the luminous region, which has a width of order $7$--$\qty{15}{\mu m}$, the transverse coordinates of reconstructed vertices are constrained using the beam-spot position and width.

\begin{figure}[htbp]
\centering
\includegraphics[width=0.75\textwidth]{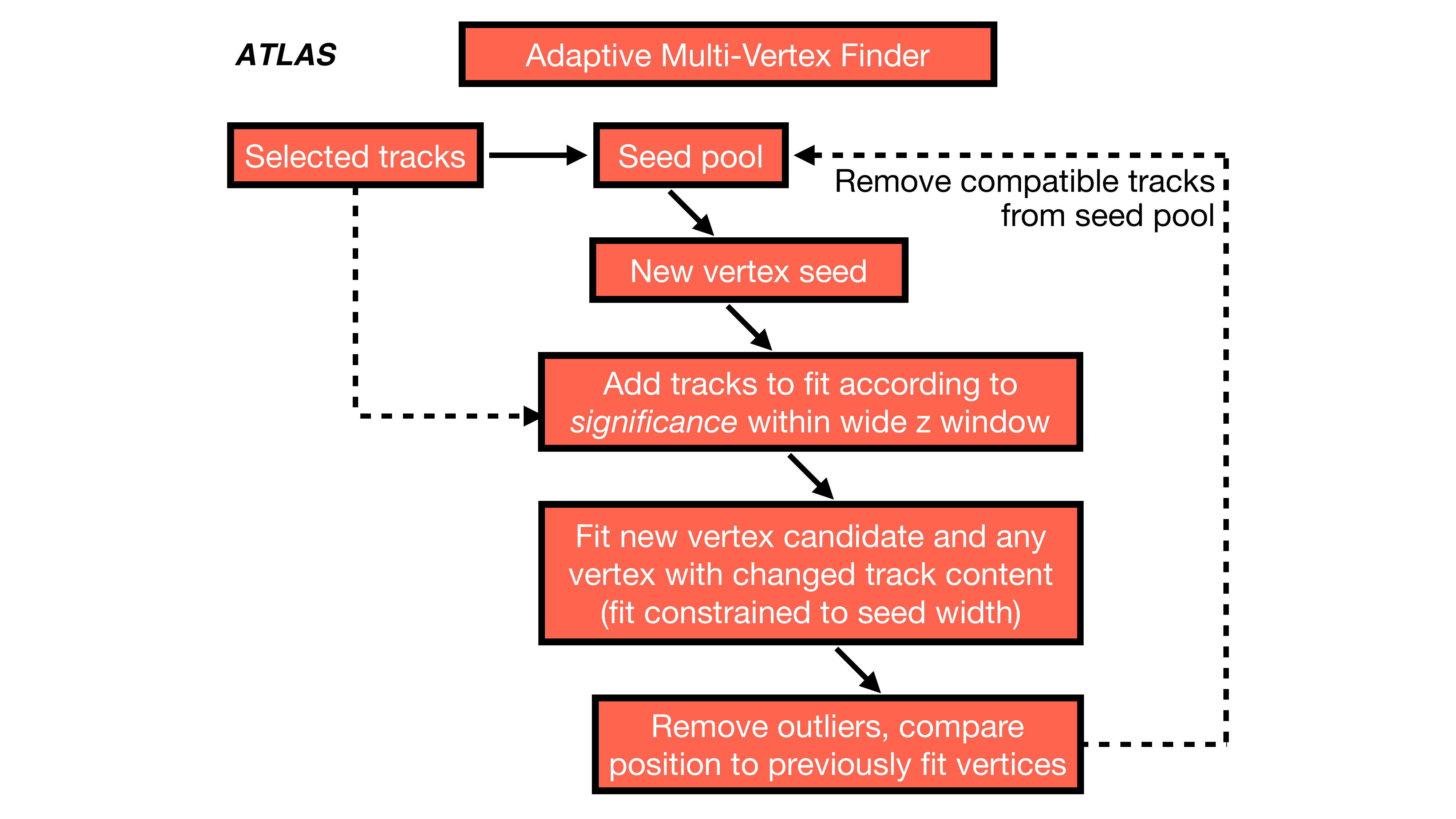}
\caption{\label{fig:vertex:amvf} Schematic view of the steps of the ATLAS algorithm used for primary vertex finding and fitting as it is configured within the ACTS toolkit.}
\end{figure}

These steps are repeated until no further vertex seed can be found or until a maximum number of iterations is reached. The latter condition, currently set to 100, acts as a fail-safe in case of unexpected event conditions.
The primary vertex reconstruction procedure in ATLAS is illustrated in Figure~\ref{fig:vertex:amvf} and implemented using the open-source, experiment-independent ACTS (A Common Tracking Software) toolkit~\cite{ACTS}, with ATLAS-specific configuration and input handling. Further technical details are provided in Ref.~\cite{ATL-PHYS-PUB-2019-015}.

Once vertex finding is finished, the hard-scatter vertex is selected as the one with the largest sum of the squared transverse momenta, $\sum \pt^2$, of the tracks. Some analyses may employ a different choice later due to the specific decay kinematics of interest.

%

\subsection{Seed finding}
A vertex seed is the initial estimate of a vertex position.
ATLAS employs a Gaussian track density seed finder~\cite{ATL-PHYS-PUB-2019-015} for this purpose. Each track is represented by a two-dimensional correlated Gaussian probability distribution $P(r,z)$ centred at its point of closest approach to the beamline $(d_0,z_0)$. The Gaussian widths are set by the track impact parameter uncertainties, derived from the covariance matrix of the track fit. The local maxima of the sum of the track densities along the $z$-axis are then identified using the Gauss–Newton method~\cite{bjorck1996}. The identified maxima are used as the initial vertex seed positions.

A key feature of the vertex finding algorithm is that only tracks incompatible (according to the criteria discussed in the following sections) with previously fitted vertices are used in the density calculation for new seeds. In this way, high-multiplicity vertices are removed from the track density distribution which allows lower-multiplicity vertices to be found in later iterations.


\subsection{Track association}
Once an initial estimate of the vertex position has been found, nearby tracks are associated with the seed for fitting.
The beamspot position and width are used to constrain the transverse coordinates of the vertex, to avoid attempting to reconstruct vertices far from the nominal interaction region.
This strong geometric constraint significantly stabilises vertex finding and fitting in high pile-up environments.
The longitudinal coordinate is not constrained.

A weighted least-squares approach is used to fit the vertex positions based on a set of tracks. To improve the quality of the fit, a pre-selection is applied to limit the number of incompatible tracks given to the vertex fitter.
While the fitting procedure down-weights mis-measured or unrelated tracks (as discussed in the following section), its performance improves when the input tracks are already compatible with the seed.

Tracks are associated to a vertex seed for fitting if they satisfy  $\sqrt{(z/\sigma_z)^2 + (d/\sigma_d)^2} < 5 $,
where $z$ and $d$ are the longitudinal and radial distances from the vertex seed, and $\sigma_z$ and $\sigma_d$ are their respective uncertainties.
A loose longitudinal cut of $z < \qty{3}{\mm}$ is applied to reject tracks that have both large distances and large uncertainties. The vertex seed and associated tracks form a \emph{vertex candidate} which is passed to the fitting stage.

Unlike during vertex seed finding, tracks may be associated with previously fitted vertices. If a vertex candidate shares tracks with other vertices, then these vertices are fit simultaneously to resolve all associated positions consistently. In effect, the vertex fit acts as a global fit in vertex-dense regions. If a vertex candidate does not share any tracks, then it is fit independently and other vertices are not refitted.


\subsection{Adaptive multi-vertex fitting}
The goal of vertex fitting is to determine the vertex position and its associated covariance matrix by fitting a set of input tracks. Primary vertex fitting is performed using the AMVF~\cite{Waltenberger:2005zz, Piacquadio:2008zzb, ATL-PHYS-PUB-2019-015}.

The AMVF uses a weighted least-squares minimisation, where the objective function is the weighted sum of squared standardised Euclidean distances in three-dimensional space ($\chi = d/\sigma_d$) between the fitted vertex position and the point of closest approach of each input track.
The weight of a given track is calculated as:

\begin{equation}
\omega_i(\chi_i^2) = \frac{e^{-\frac{1}{2}\chi_i^2}}{\Sigma_je^{-\frac{1}{2}\chi_j^2}+e^{-\frac{1}{2}\chi_0^2}},
\label{eq:vertex:omega}
\end{equation}

where $\chi_i$ is the normalised Euclidean distance between the vertex position and the track’s closest approach, $\chi_j$ are the distances of that track to all associated vertices (including $\chi_i$), and $\chi_0$ is a configurable cutoff parameter, nominally set to a value corresponding to three standard deviations ($3\sigma$).
This formulation allows the weight of a track to a given vertex to depend not only on its compatibility with that vertex but also on its compatibility with all other nearby vertices. The vertices are, in effect, competing for tracks.

Once a vertex has been fit, it is accepted if it satisfies the following criteria:
\begin{itemize}
\item The vertex must include at least two distinct tracks that are not assigned to any other vertex and are mutually compatible with the fitted vertex position. This requirement ensures that at least two of the original seed tracks remain consistent after the fit, confirming that the resulting vertex corresponds to the seeded vertex. It also prevents the vertex from being composed solely of tracks that are already associated with previously reconstructed vertices.
\item The fitted vertex position cannot be within $3\sigma$ of any previously accepted vertex, where $\sigma$ is calculated from the fit errors of both vertices.
\item The weighted average of the track weights in the fit, defined as $\Sigma\omega^2/\Sigma\omega$, must be greater than two-thirds.
\end{itemize}
If a fitted vertex is accepted, all of its compatible tracks are removed from the seed pool. If the vertex is rejected, the track most compatible with the rejected seed position is removed from the seed pool. This guarantees that the number of available seed tracks is reduced on each iteration, ensuring that the algorithm will reliably terminate.


\subsection{Track-To-Vertex Association}

The principal task of vertex reconstruction is to accurately locate vertex positions. While the primary vertex finding and fitting algorithm assigns tracks to vertices during reconstruction, this association may not always yield the optimal balance between efficiency and purity for downstream reconstruction tasks and physics analyses. For this reason, a separate track-to-vertex association (TTVA) procedure can be applied after vertex reconstruction, with the choice of working point optimised for the requirements of a given use case.

Track association requirements rely primarily on the impact parameters of the tracks ($\dZero$ and $\dzSinTheta$) and on the track weight ($\omega$) obtained from a given vertex fit, defined in~\cref{eq:vertex:omega}. These quantities are used to assess the compatibility of each track with candidate vertices under different selection strategies. Where an explicit cut is applied on $\omega$, a threshold of $\omega > 0.03$ is employed, corresponding to a $4\sigma$ deviation (i.e.\ $\chi^2 = 16$ for one degree of freedom) from the null hypothesis that the track does not originate from the \pp interaction position represented by the reconstructed vertex in question.

Two families of TTVA WPs are defined to reflect different approaches to track–vertex association:
\begin{itemize}
\item The \emph{d0Sig} WPs (Table~\ref{tab:perf:ttva:defs:d0Sig}) use criteria based primarily on the transverse impact-parameter significance $\dZeroSig$, supplemented by requirements on the absolute impact parameters.
\item The \emph{MaxWeight} WPs (Table~\ref{tab:perf:ttva:defs:MaxWeight}) associate tracks to the vertex for which the track has the largest fit weight $\omega$ among all reconstructed vertices.
\end{itemize}
For the ``d0Sig'' WPs, a minimum requirement on the track weight is applied to ensure basic compatibility with the vertex fit, while the association itself is driven primarily by impact-parameter-based criteria. In the ``MaxWeight'' scheme, tracks that participate in one or more vertex fits are associated to the vertex for which they have the highest fit weight. Tracks not used in any vertex fit are instead associated using impact-parameter-based fallback criteria consistent with the corresponding ``d0Sig'' WP.
Additional WPs optimised for specific downstream algorithms are also used.

In both association schemes, an identical set of WP labels is defined for prompt and non-prompt track selections.
Within the \emph{Non-prompt} WPs, the \emph{Non-prompt Hard}, \emph{Non-prompt Medium}, and \emph{Non-prompt All} WPs form a hierarchy with progressively looser impact-parameter requirements, corresponding to different efficiency–purity trade-offs for tracks originating from displaced decays\footnote{No attempted is made here to associate tracks to the secondary vertex occurring in such decays. The aim of the association is to match tracks to the PV from which the decaying particle originated.}. Tighter WPs are optimised for high-purity selections of hard non-prompt signatures, while looser WPs maximise acceptance in analyses requiring inclusive track selections.
Examples of TTVA use cases for which the Non-prompt All WP is intended include jet reconstruction~\cite{PERF-2015-09,JETM-2018-06}, pile-up jet identification~\cite{JETM-2023-08,PERF-2014-03}, and missing transverse momentum reconstruction~\cite{JETM-2020-03}. Hard non-prompt physics, targeted by the Non-prompt Hard WP, includes jet flavour tagging~\cite{FTAG-2019-07} and hadronically decaying $\tau$ leptons~\cite{ATL-PHYS-PUB-2022-044,ATLAS-CONF-2017-029}.

In contrast, the \emph{Prompt} WP targets tracks that are compatible with, and used in, the primary vertex fit. Since only high-quality Tight Primary tracks contribute to vertex reconstruction, this WP provides a selective choice optimised for prompt particles originating from a primary interaction. As a result, the Prompt WP should not be interpreted as looser or tighter than the Non-prompt WPs, but rather as optimised for a different class of physics signatures. Prompt physics applications, such as track-based lepton isolation, typically employ these WPs~\cite{EGAM-2021-01,EGAM-2019-01,MUON-2018-03}.

\begin{table}
\centering
\caption{Track-To-Vertex Association working points using the ``d0Sig'' scheme, based primarily on the significance of the track’s transverse impact parameter with respect to the vertex in question. In this scheme, the transition from Non-prompt Hard to Non-prompt All corresponds to progressively relaxed absolute impact-parameter requirements, while maintaining a minimum level of vertex compatibility.}
\begin{tabular}{c | c}
\toprule
Working Point     & Definition \\
\midrule
Prompt            & $\omega > 0.03$ \\ \noalign{\vskip 2pt}
Non-prompt Hard   & $|\dZero| < \qty{1}{\mm}$, and
$\begin{cases}
\omega > 0.03, & \text{if } |\dZeroSig| < 3 \\
|\dzSinTheta| < \qty{1}{\mm}, & \text{otherwise}
\end{cases}$ \\ \noalign{\vskip 2pt}
Non-prompt Medium & $|\dZero| < \qty{2}{\mm}$, and
$\begin{cases}
\omega > 0.03, & \text{if } |\dZeroSig| < 3 \\
|\dzSinTheta| < \qty{2}{\mm}, & \text{otherwise}
\end{cases}$ \\ \noalign{\vskip 2pt}
Non-prompt All    & $|\dZero| < \qty{5}{\mm}$, and
$\begin{cases}
\omega > 0.03, & \text{if } |\dZeroSig| < 3 \\
|\dzSinTheta| < \qty{5}{\mm}, & \text{otherwise}
\end{cases}$ \\ \noalign{\vskip 2pt}
\bottomrule
\end{tabular}
\label{tab:perf:ttva:defs:d0Sig}
\end{table}

\begin{table}
\centering
\caption{Track-To-Vertex Association working points using the ``MaxWeight'' scheme, based on comparing the weight of the track across different vertex fits ($\omega$). The parameter $N_\textrm{fits}$ represents the number of vertex fits in which the track was used. Tracks not used in any vertex fit are instead associated using impact-parameter-based fallback criteria.}
\begin{tabular}{c | c}
\toprule
Working Point     & Definition \\
\midrule
Prompt            & $\omega > 0.03$, and $\omega$ is the highest weight among all vertex fits. \\ \noalign{\vskip 2pt}
Non-prompt Hard   &
$\begin{cases}
\omega \text{ is the highest weight among all vertex fits}, & \text{if } N_\textrm{fits} > 0 \\
|\dZero| < \qty{1}{\mm} \text{ and } |\dzSinTheta| < \qty{1}{\mm}, & \text{if } N_\textrm{fits} = 0
\end{cases}$ \\ \noalign{\vskip 2pt}
Non-prompt Medium &
$\begin{cases}
\omega \text{ is the highest weight among all vertex fits}, & \text{if } N_\textrm{fits} > 0 \\
|\dZero| < \qty{2}{\mm} \text{ and } |\dzSinTheta| < \qty{2}{\mm}, & \text{if } N_\textrm{fits} = 0
\end{cases}$ \\ \noalign{\vskip 2pt}
Non-prompt All    &
$\begin{cases}
\omega \text{ is the highest weight among all vertex fits}, & \text{if } N_\textrm{fits} > 0 \\
|\dZero| < \qty{5}{\mm} \text{ and } |\dzSinTheta| < \qty{5}{\mm}, & \text{if } N_\textrm{fits} = 0
\end{cases}$ \\ \noalign{\vskip 2pt}
\bottomrule
\end{tabular}
\label{tab:perf:ttva:defs:MaxWeight}
\end{table}



\FloatBarrier

\section{Performance}
\label{sec:perf}

The performance of track and primary vertex reconstruction is assessed using a combination of Monte Carlo simulations and data-driven methods. This section presents key metrics such as reconstruction efficiency, resolution, and mis-reconstruction rates, covering both inclusive measurements and specific phase-space regions of interest. The tracking performance is examined at different stages of the reconstruction chain, from initial track seeds to final reconstructed tracks, highlighting effects such as material interactions, pile-up, and tracking in dense environments. The impact of reconstruction choices on timing and computational efficiency is also evaluated.

A fundamental quantity used throughout this section is the track reconstruction efficiency, which quantifies the fraction of generated charged particles that are successfully reconstructed as tracks. Specifically, track reconstruction efficiency is defined as the number of simulation-level particles which can be successfully matched to a reconstructed track, $N_{\text{rec}}^{\text{matched}}$, divided by the total number of charged particles generated in simulation, $N_{\text{gen}}$:
\begin{equation}
\epsilon_{\text{trk}}(\pT, \eta) = \frac{N_{\text{rec}}^{\text{matched}}(\pT, \eta)}{N_{\text{gen}}(\pT, \eta)}
\label{eq:eff}
\end{equation}
where \pT and $\eta$ are the truth particle parameters directly from the MC generator.
The criteria for truth matching are defined in \Sect{\ref{sec:perf:truthmatching}}.

The reconstruction efficiency is one of the most critical parameters to monitor at each stage of the track reconstruction sequence. While the subsequent stages in track reconstruction can tolerate a certain level of wrong or duplicate seeds and track candidates, any true particle that does not have any good seeds will not be reconstructed and is therefore permanently lost.


%
%
%

\subsection{Performance of track reconstruction stages}
\label{sec:perf:lowlevel}

\begin{figure}[htbp]
\begin{center}
\includegraphics[width=0.7\textwidth]{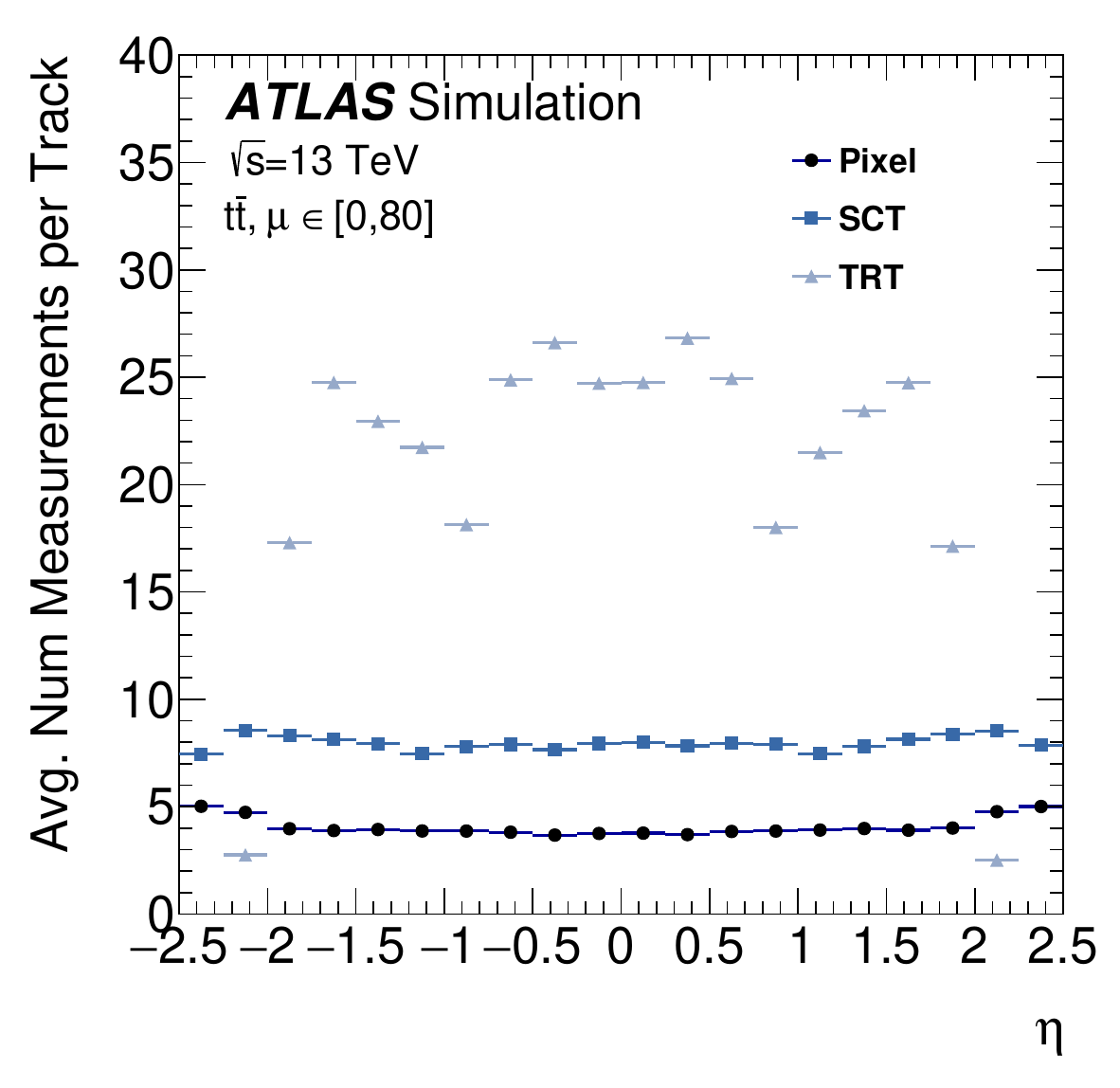}
\caption{Average number of measurements on track, separately for each sub-system.}
\label{fig:pattern:nhits}
\end{center}
\end{figure}

The following section examines tracks at various stages of the primary sequence in the track reconstruction chain, specifically track seeds from the seeding stage, track candidates from the track-finding stage, and the final (resolved) tracks selected in the ambiguity resolution stage (see~\cref{fig:overview:tracking_flow}).
On average, reconstructed tracks contain approximately four pixel clusters, eight SCT clusters, and between 18 and 27 TRT measurements, depending on $\eta$. Figure~\ref{fig:pattern:nhits} explicitly illustrates the $\eta$ dependence of the average number of measurements in each subdetector, in a simulated sample of $t\bar{t}$ events with a uniform distribution of $\mu$ as described in Section~\ref{sec:samples}.

\begin{figure}[htbp]
\begin{center}
\subfloat[]{\includegraphics[width=0.49\textwidth]{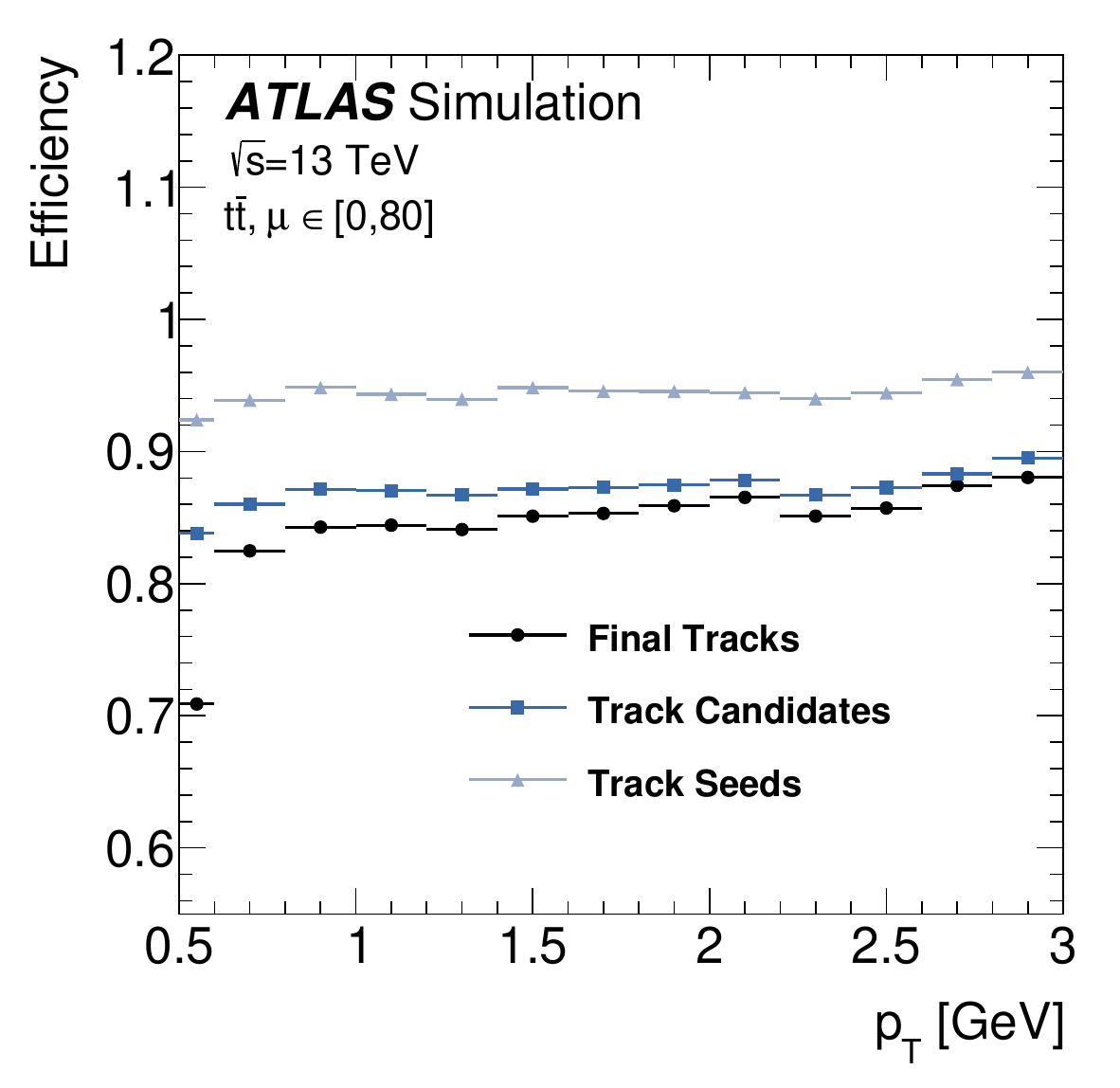}}
\subfloat[]{\includegraphics[width=0.49\textwidth]{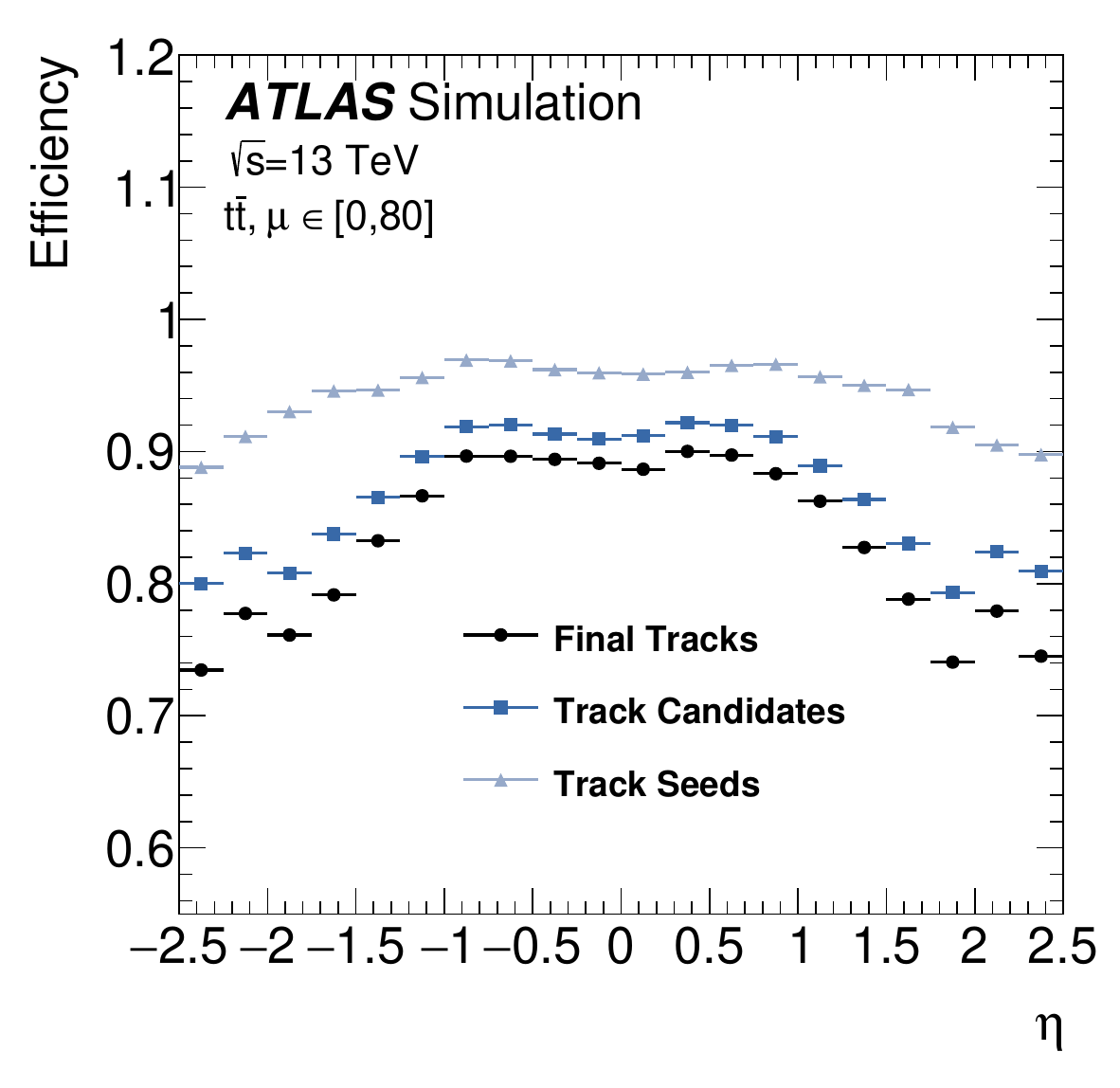}}
\caption{Efficiency to reconstruct a track seed, a track candidate and a final silicon track as a function of the true particle (a) \pT and (b) $\eta$.}
\label{fig:pattern:seedcandEff}
\end{center}
\end{figure}

\begin{figure}[htbp]
\begin{center}
\subfloat[]{\includegraphics[width=0.49\textwidth]{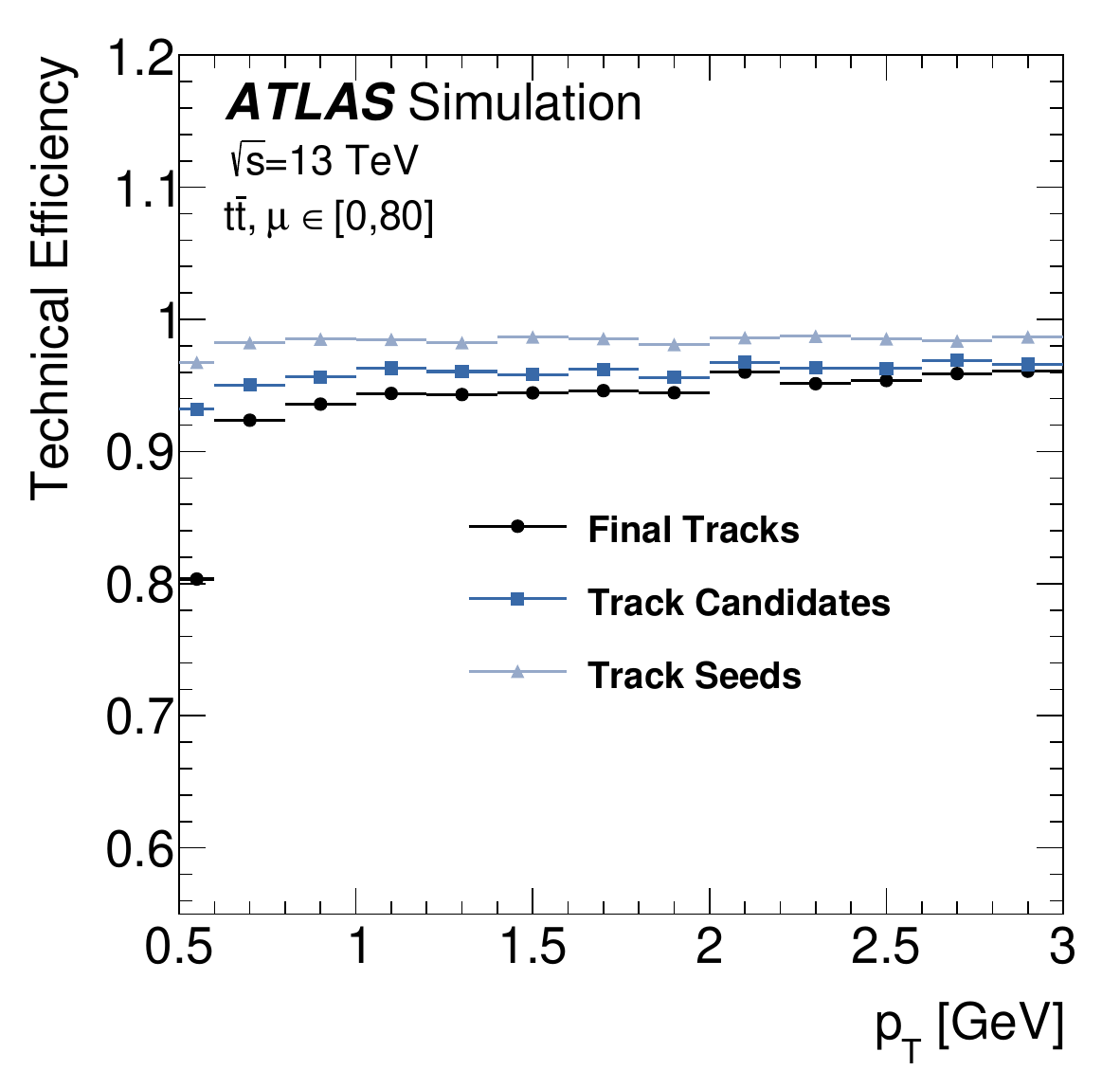}}
\subfloat[]{\includegraphics[width=0.49\textwidth]{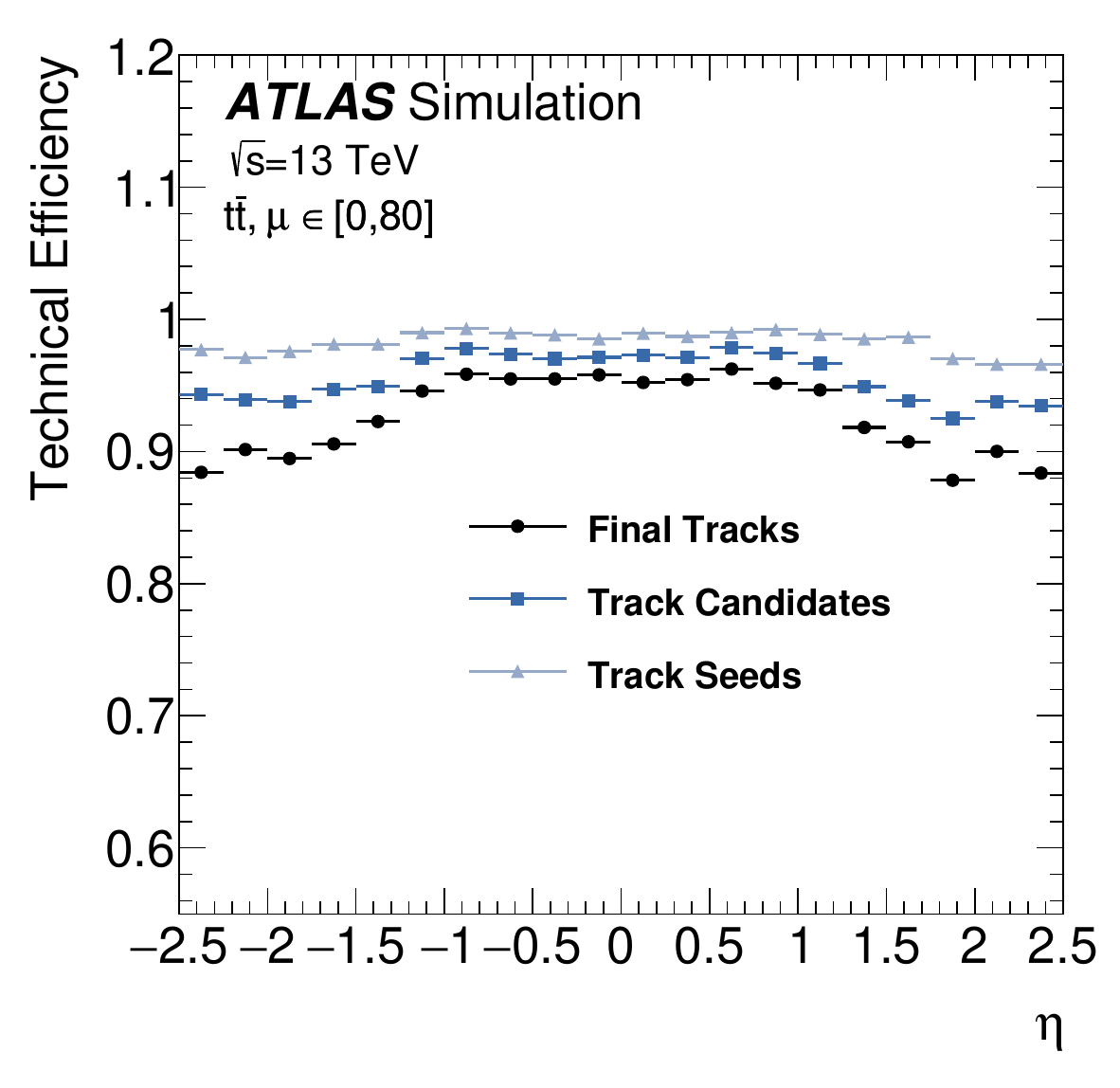}}
\caption{Technical efficiency to reconstruct a track seed, a track candidate and a final silicon track as a function of the true particle (a) \pT and (b) $\eta$.}
\label{fig:pattern:seedcandTechEff}
\end{center}
\end{figure}

\begin{figure}[htbp]
\begin{center}
\subfloat[]{\includegraphics[width=0.49\textwidth]{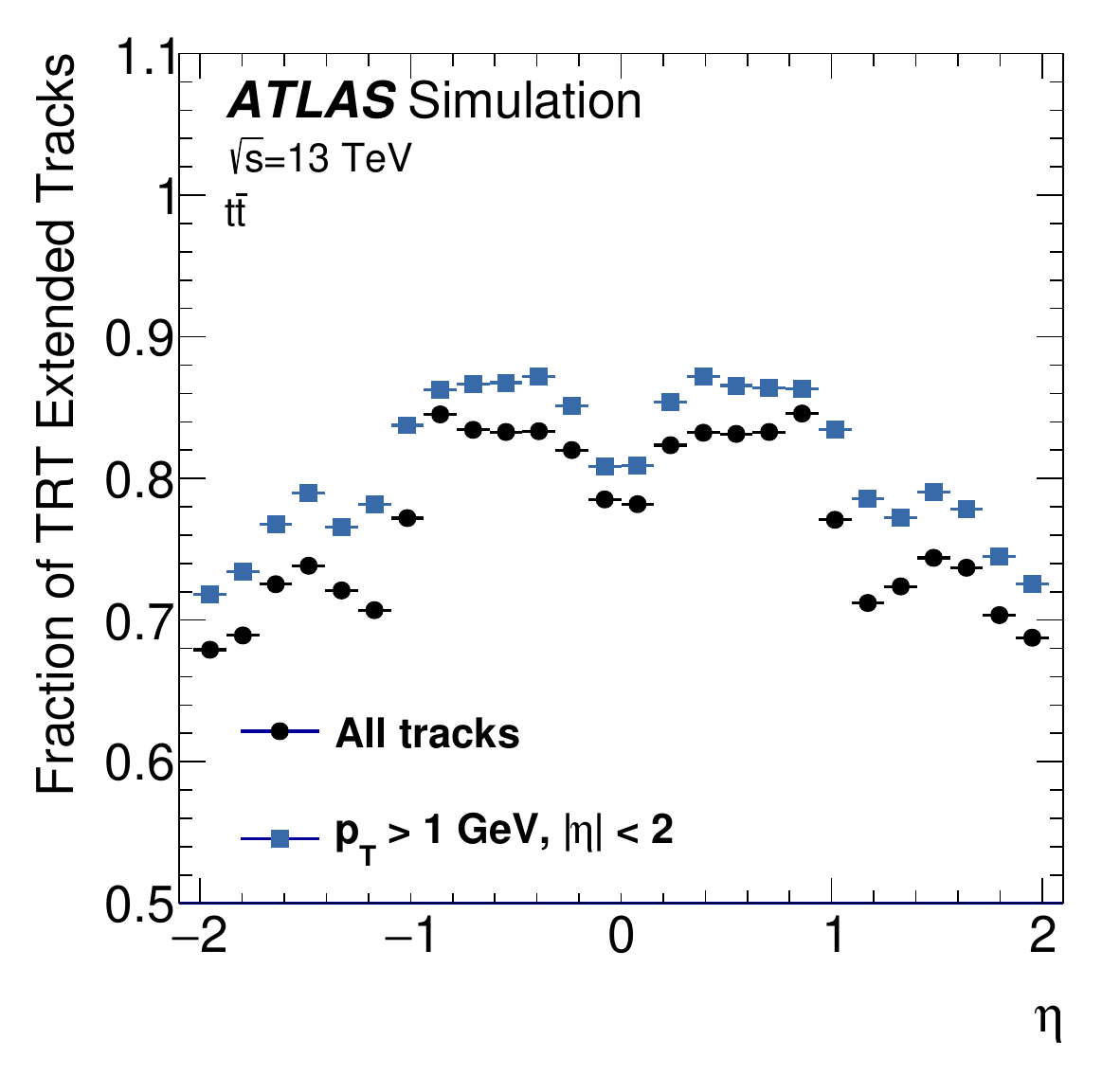}}
\subfloat[]{\includegraphics[width=0.49\textwidth]{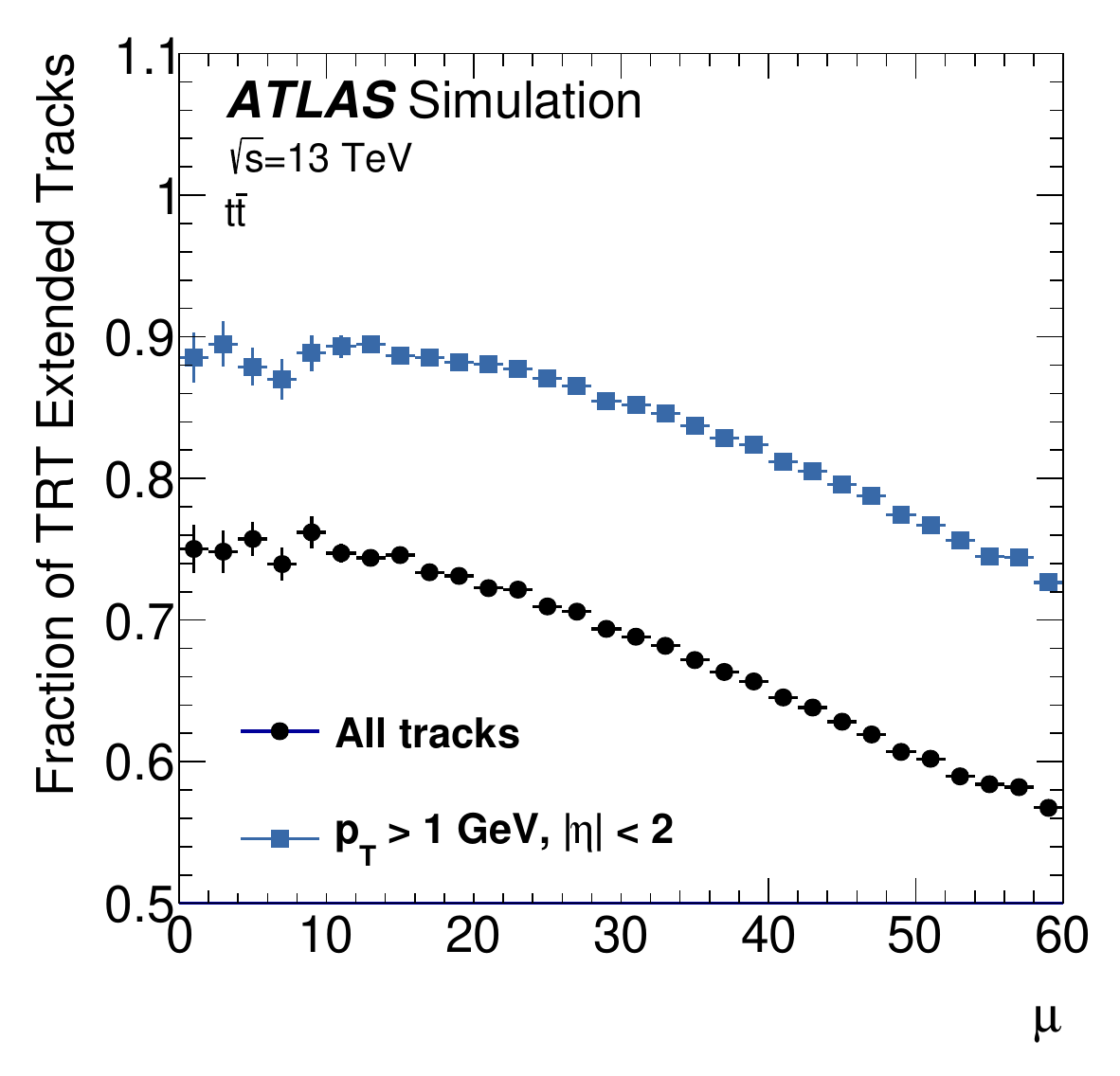}}
\caption{Fraction of TRT-extended tracks as a function of (a) $\eta$ and (b) $\mu$. The fraction is shown for all reconstructed tracks, and for all tracks with $\pt > \SI{1}{\GeV}$ and $|\eta| < 2$, which corresponds approximately to the acceptance of the TRT detector. The features in (a) at $\eta=0$ and $\eta=1$ correspond to the gap between the two halves of the TRT, and the barrel-endcap transition respectively.}
\label{fig:pattern:trtextension}
\end{center}
\end{figure}

Truth-matched seeds and track candidates are defined, analogously to the final tracks, as having a value of $R_\textrm{match} > 0.5$~(see Section~\ref{sec:perf:truthmatching}) to be considered correctly matched to a truth particle\footnote{For short objects such as seeds, especially in the PPP case where only three clusters are present, the $R_{\text{match}} > 0.5$ requirement corresponds to just two matching clusters.}.
Figure~\ref{fig:pattern:seedcandEff} shows the reconstruction efficiency separately for track seeds, track candidates, and the output silicon tracks as a function of \pT\ and $\eta$. Most of the efficiency loss occurs in the transition from seeds to track candidates, primarily due to particles not crossing a sufficient number of measurement surfaces before undergoing a hadronic interaction. This effect is more pronounced at high $|\eta|$, where charged particles traverse a greater amount of material (see also Figure~\ref{fig:detector:materialbudget}).
To separate this effect, which arises from detector material and design choices, from the impact of the reconstruction procedure, Figure~\ref{fig:pattern:seedcandTechEff} presents the \textit{technical} efficiency. Here, the denominator is restricted to particles that leave at least eight measurements in the silicon tracker, a threshold chosen to align with the reconstruction-level selection criteria (see Section~\ref{sec:tracking:finding}). The technical efficiency demonstrates that when sufficient measurements are available, the reconstruction algorithms are highly efficient in reconstructing seeds, track candidates, and the final silicon tracks.
After a silicon track is reconstructed, Figure~\ref{fig:pattern:trtextension} shows the fraction of primary-sequence tracks that are assigned a TRT extension. This fraction is approximately 85\% in the barrel region and 70--80\% in the endcaps up to the TRT acceptance of $|\eta| < 2$. The same figure illustrates that as \pileup increases, the average TRT extension rate\footnote{It should be noted that the TRT extension rate does not represent a true efficiency, since criteria for accepting an extension does not aim to maximise the number of TRT extensions, but rather the overall quality of the track, as explained in Section~\ref{sec:trk:trt}} decreases due to the increased detector occupancy.

\begin{figure}[htbp]
\begin{center}
\subfloat[]{\includegraphics[width=0.49\textwidth]{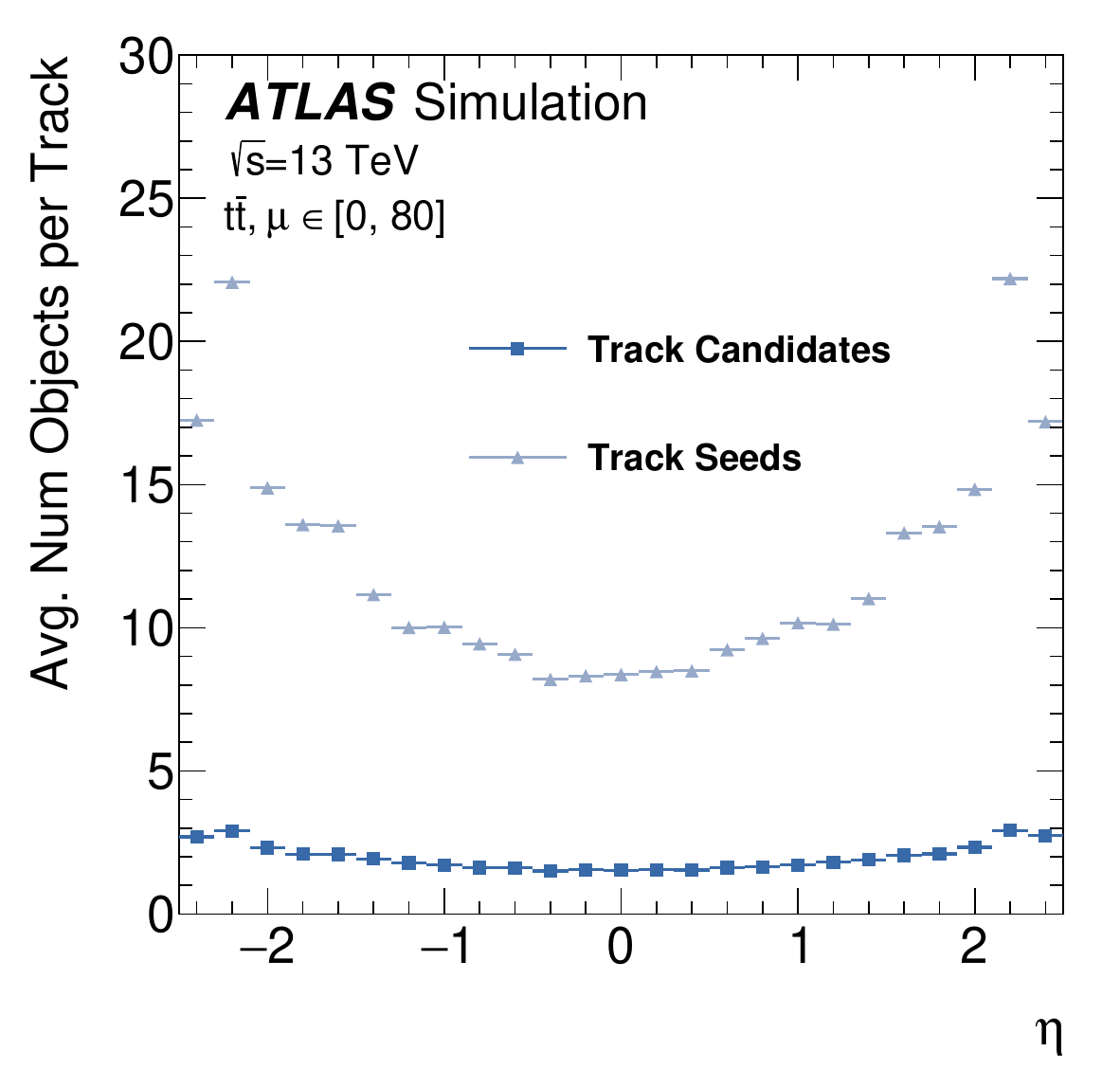} \label{fig:pattern:n_seedDup}}
\subfloat[]{\includegraphics[width=0.49\textwidth]{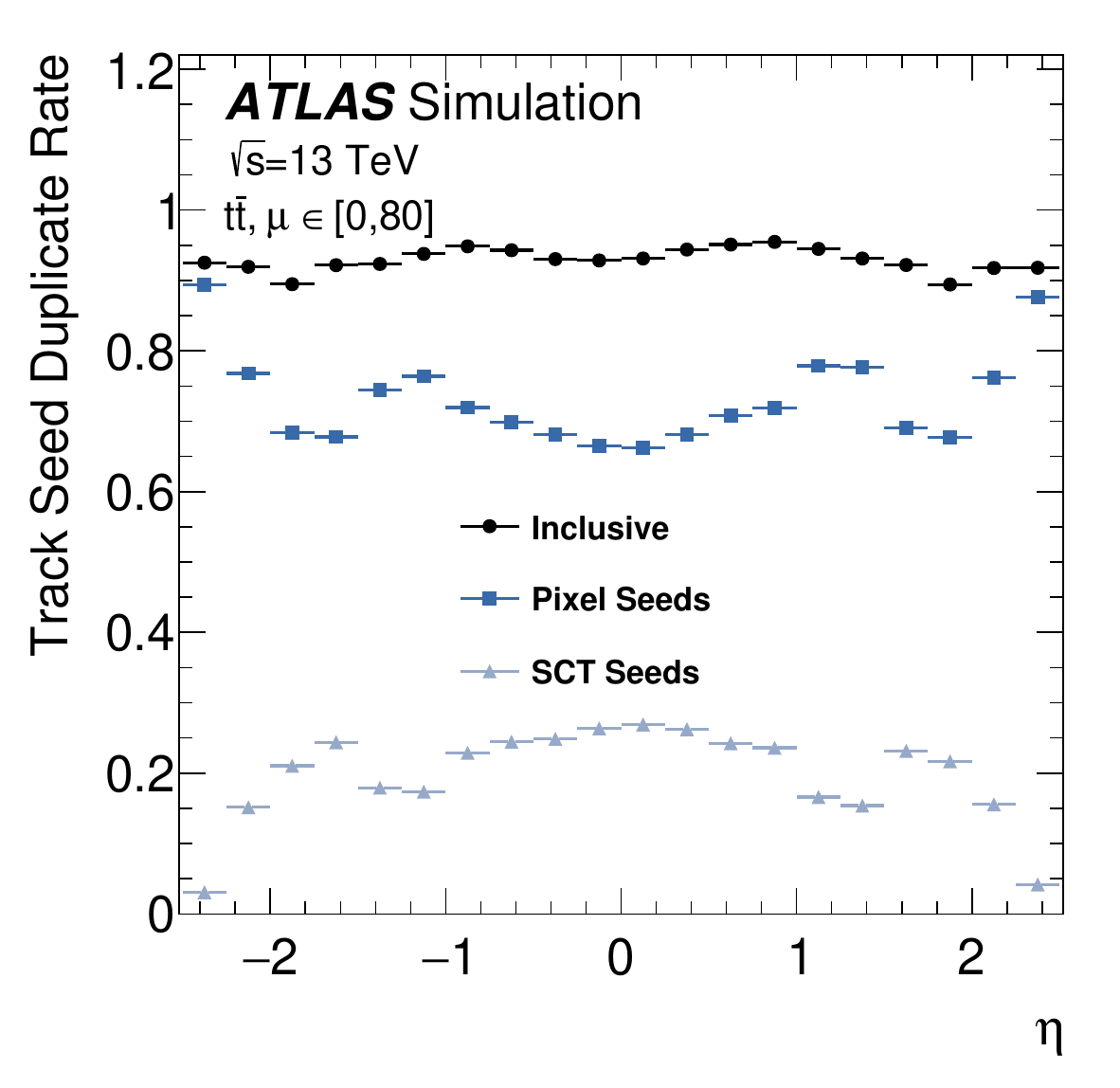} \label{fig:pattern:seedDup}}
\caption{(a) Number of track seeds and candidates, divided by the number of reconstructed tracks, as a function of $\eta$. (b) Seed duplicate rate as a function of $\eta$ for pixel seeds, SCT seeds and inclusively.}
\end{center}
\end{figure}

As mentioned previously, multiple seeds can be created from measurements that were deposited by a single particle.
The number of track seeds and track candidates, normalised to the number of reconstructed tracks, is shown as a function of $\eta$ in Figure~\ref{fig:pattern:n_seedDup}.
If multiple seeds are truth-matched to the same truth particle, they are considered duplicates. The duplicate rate is defined as the fraction of seeds truth-matched to a truth particle that has already been matched to another seed. It provides insight into the behaviour of the seed-finding algorithm, but given the loose requirements for seeding, a high rate of duplicate seeds is expected. The challenge lies in efficiently identifying and eliminating redundant seeds early in the reconstruction process. Figure~\ref{fig:pattern:seedDup} shows the seed duplicate rate as a function of $\eta$. The separation of the duplicate rate into Pixel and SCT is determined by which subdetector contributes the majority of seeds for the particle. A significantly lower duplicate rate can be observed for SCT seeds, which are processed first and used to reduce the $z(r=0)$ range of Pixel seeds considered as discussed in Section~\ref{sec:seeding}.

\begin{figure}[htbp]
\begin{center}
\subfloat[]{\includegraphics[width=0.49\textwidth]{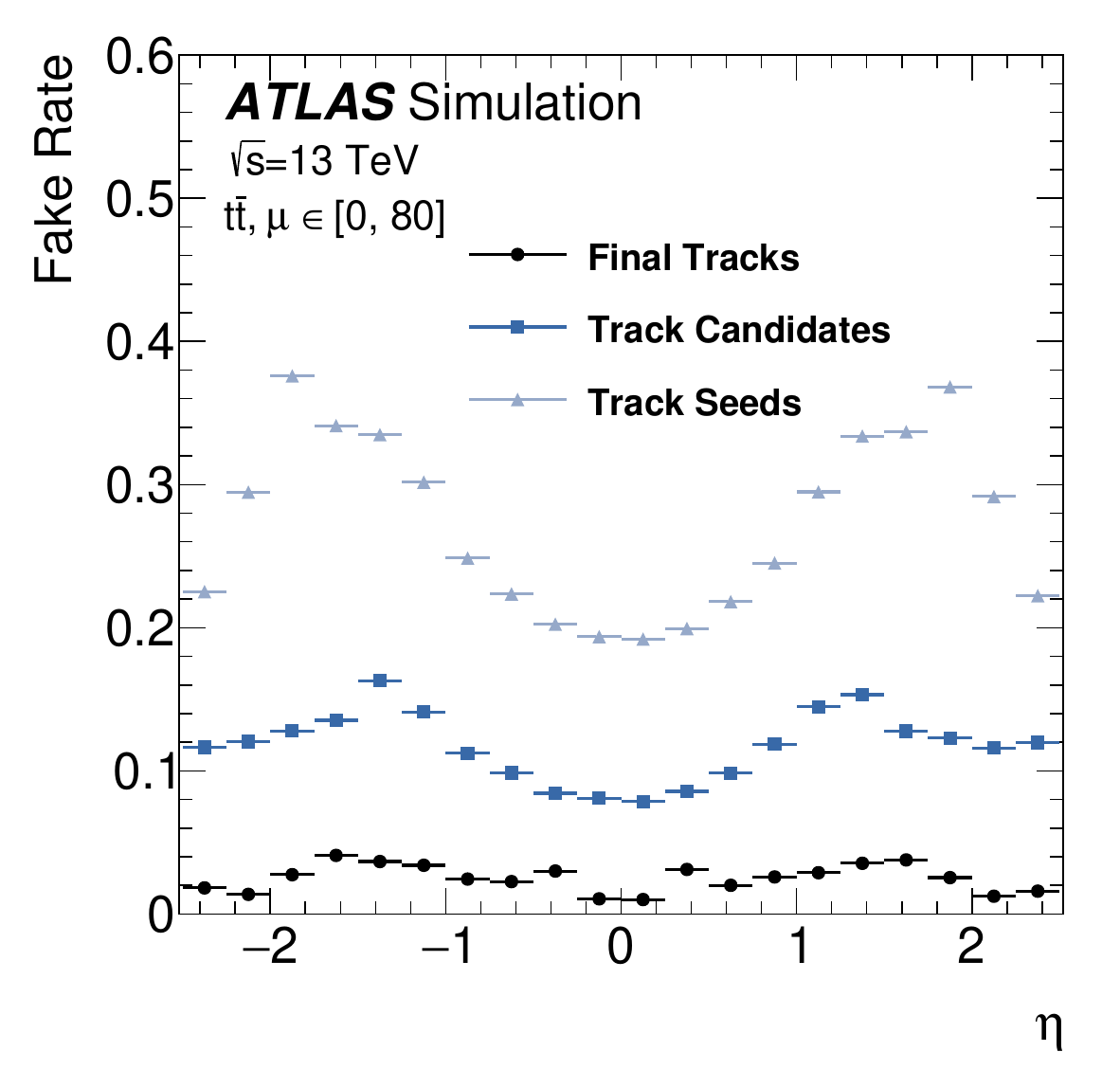}}
\subfloat[]{\includegraphics[width=0.49\textwidth]{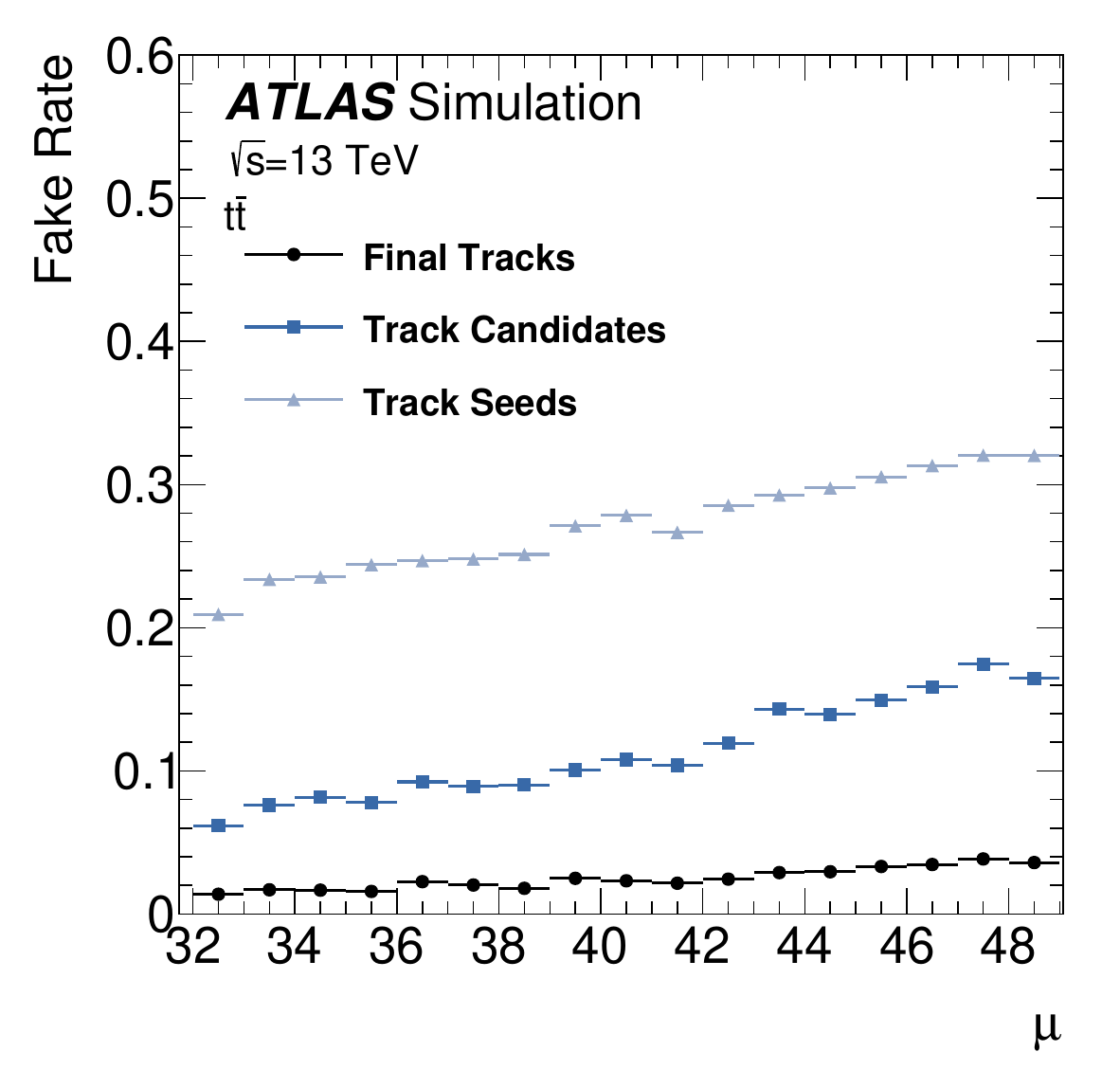}\label{fig:pattern:fakerate-detailed_b}}
\caption{Fake rate as a function of (a) $\eta$ and (b) $\mu$ for track seeds, track candidates, and final tracks.}
\label{fig:pattern:fakerate-detailed}
\end{center}
\end{figure}

The seed efficiency remains stable within the \pileup range of interest, varying by less than 1\%~\cite{ATL-PHYS-PUB-2023-034}. This stability is partly due to the presence of duplicate seeds, which enhance robustness. However, the combinatorial nature of the tracking algorithm leads to a super-linear increase in the number of seeds and track candidates as \pileup increases, as a higher number of available measurements results in more incorrect combinations.
Figure~\ref{fig:pattern:fakerate-detailed} shows the fake rate, defined as the ratio of fake tracks to all tracks, as a function of $\eta$ and $\mu$ for track seeds, track candidates and final tracks. As expected, the rate of non-truth-matched instances steeply decreases when going from seeds to track candidates and finally to the output tracks.
The strong $\eta$ dependence is connected to the material distribution in the detector as well as the average extrapolation distance between layers. The increase as a function of $\mu$ has direct consequences on the computing resources needed to execute the tracking algorithms, as shown later in Section~\ref{sec:perf:timing}.

The resolution on track parameters estimated at early stages of the track reconstruction sequence is expected to be significantly worse than for the final tracks.
In particular, track seeds contain only three measurements and have not undergone a precision fit stage, resulting in significantly worse resolution, especially in the endcap region.
Furthermore, the resolution depends on the radial position of the seed and therefore differs between PPP and SSS seeds~\cite{ATL-PHYS-PUB-2023-034}.

\begin{figure}[htbp]
\begin{center}
\subfloat[]{\includegraphics[width=0.49\textwidth]{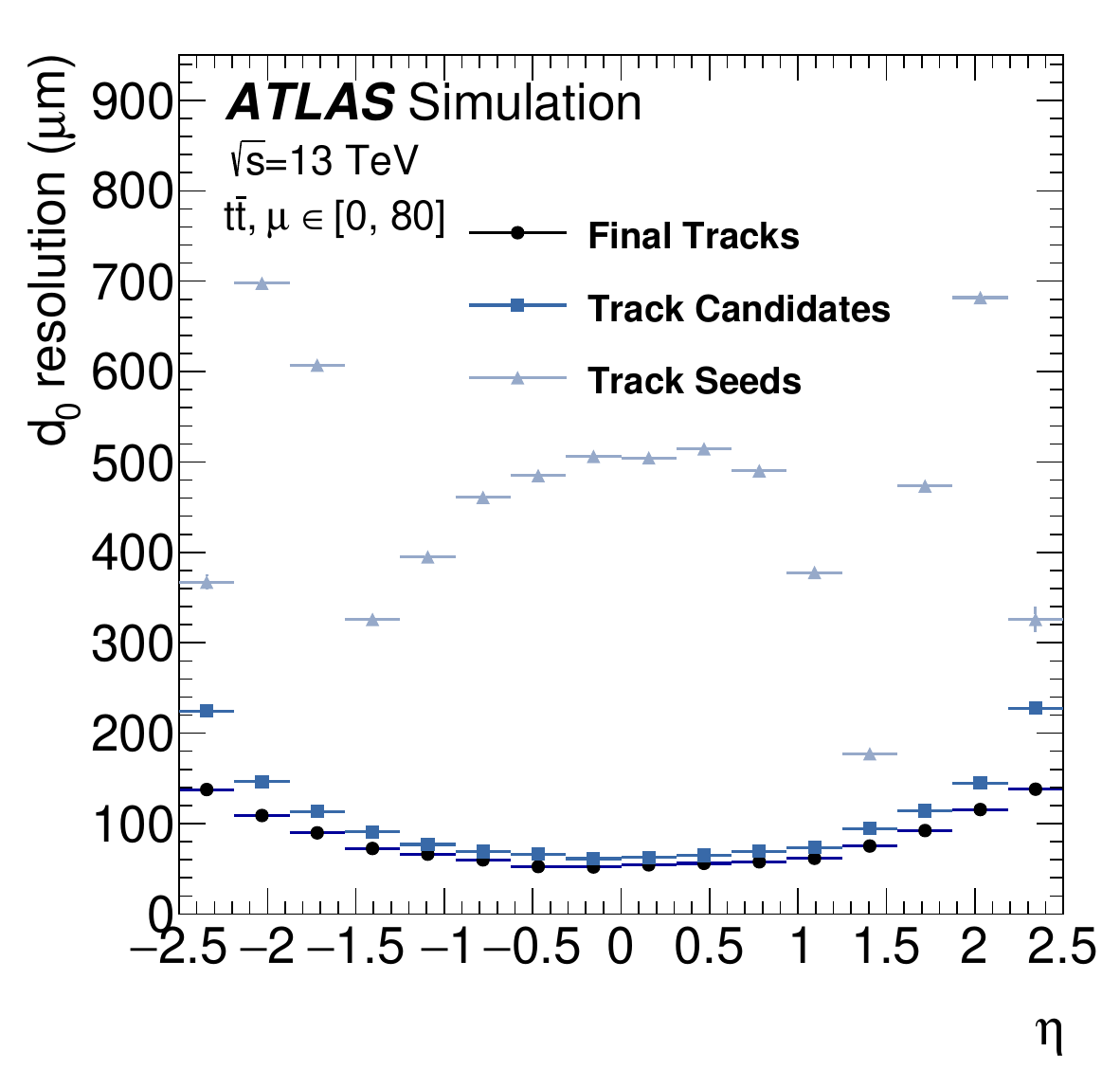}}
\subfloat[]{\includegraphics[width=0.49\textwidth]{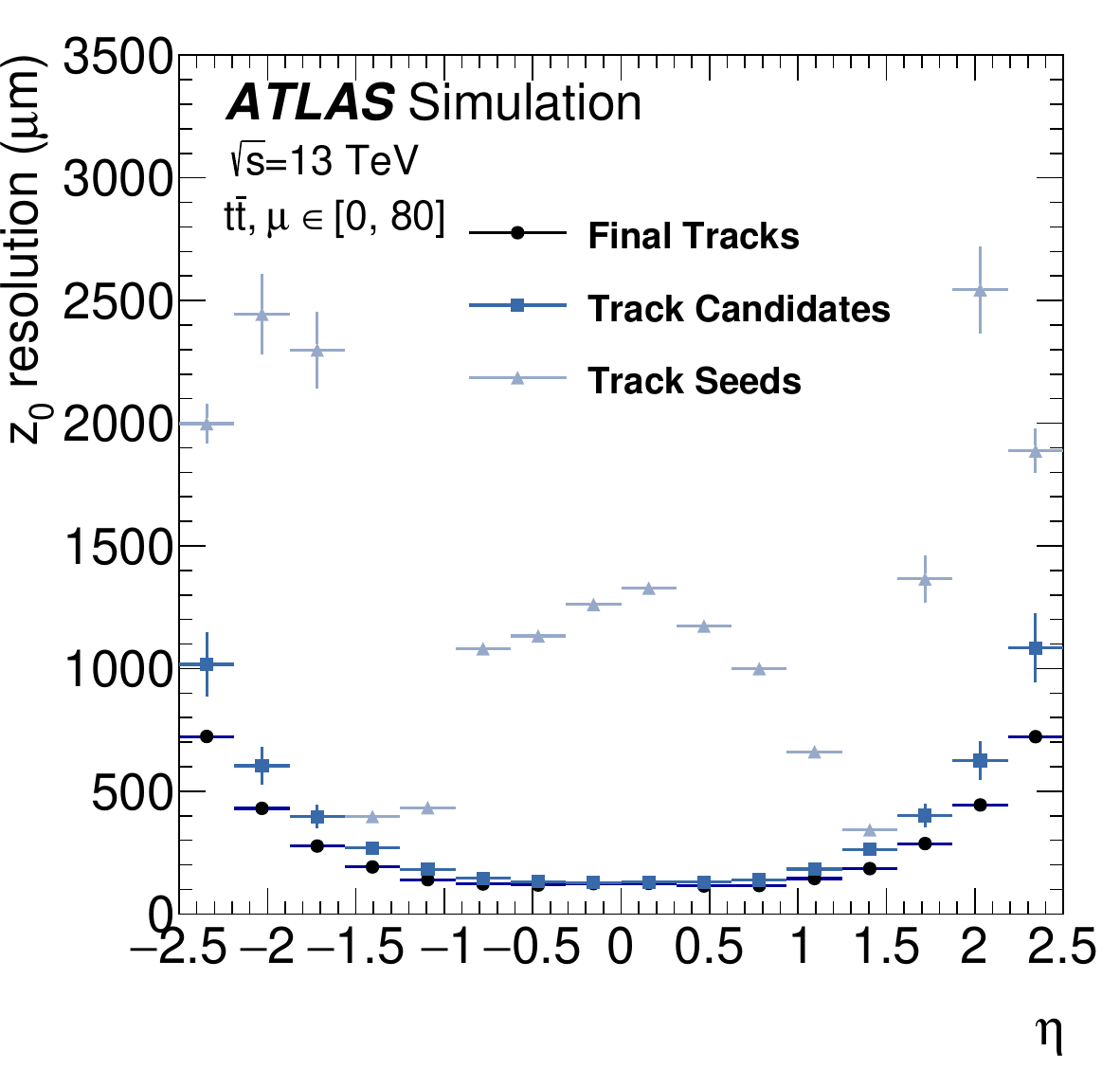}}\\
\subfloat[]{\includegraphics[width=0.49\textwidth]{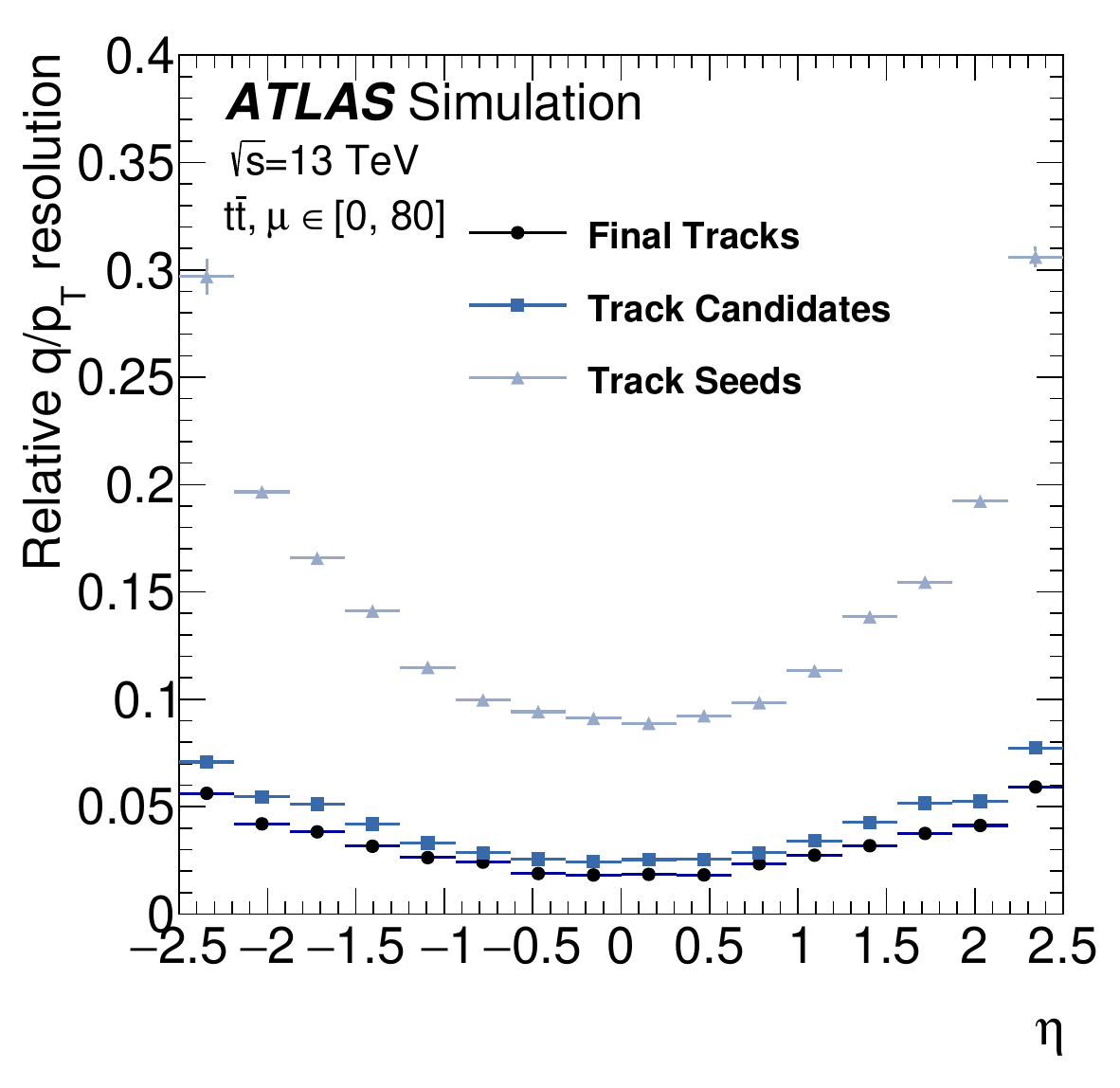}}
\caption{(a) $\dZero$, (b) $z_0$, and (c) relative $q/\pt$ resolution for track seeds, track candidates, and selected tracks as a function of the $\eta$ of the generator-level particles to which they are matched. The resolutions represent the average over the full $\pt$ spectrum of the sample. The observed asymmetries in the distributions for seeds are due to the inactive modules included in the simulation, to reflect data taking conditions during Run 2.}
\label{fig:pattern:SeedCandRes}
\end{center}
\end{figure}

As an example, Figure~\ref{fig:pattern:SeedCandRes} shows the resolution for $\dZero$, $z_0$, and the relative resolution on $q/\pT$ of all truth-matched seeds, track candidates, and final silicon tracks as a function of $\eta$. In all cases, multiple matches per truth particle are allowed, meaning that for seeds several entries per truth particle are included, while for selected tracks there is typically a single entry.
The limited number of measurements included in a given track seed also implies that an estimate of the seed charge, extracted from the direction in which the seed track ``curves'' in the transverse plane, has significant chance to be mis-reconstructed. A rate of about $\sim10\%$ for assigning the wrong charge is found for track seeds, except in the SCT forward region, where a charge mis-reconstruction rate of about 30\% is observed.

For track candidates and final silicon tracks, the degradation of the $\dZero$, $z_0$ and $q/\pT$ resolutions with increasing $|\eta|$ follows directly from the detector geometry and material budget. At large polar angles the bending power of the magnet in the $r$--$\phi$ plane shrinks due to the reduced solenoid field in the forward region, while tracks also traverse more silicon and support material in the forward pixel disks and SCT layers, amplifying multiple‑scattering and worsening the $z$‑coordinate precision.  Consequently, the curvature (hence $q/\pT$) measurement is dominated by a reduced sagitta and an enhanced multiple‑scattering term, leading to a resolution that scales roughly as $\cosh^{2}(\eta)$.

\subsection{Track reconstruction efficiency and mis-reconstruction rates}
%

%
%
%

As discussed previously, the efficiency with which tracks are provided for use in subsequent event reconstruction steps and/or direct analysis is one of the key drivers of the overall performance of the ATLAS experiment. Maximising the efficiency must be balanced against allowing for an acceptable level of mis-reconstructed tracks for the intended purpose, with the working points discussed in Section~\ref{sec:perf:quality} providing different options when selecting tracks for a given purpose.

ATLAS track reconstruction has been developed and validated over many years, with extensive comparisons of track-dependent observables between data and simulation. While these do not directly measure tracking efficiency in data, they demonstrate consistent performance across a wide range of conditions. Early studies with hadronic resonances~\cite{ATLAS-CONF-2010-034,ATLAS-CONF-2010-032,ATLAS-CONF-2011-146} confirmed the ability to reconstruct these signatures and reproduce mass distributions in data, though they lacked the statistical power and techniques, such as \emph{tag-and-probe}, needed to extract absolute efficiencies from collision data.

Several higher-level reconstruction algorithms, such as those for electrons~\cite{EGAM-2018-01} and muons~\cite{MUON-2018-03}, rely on tracks as input and their multi-detector aspect makes them suitable for performing tag-and-probe measurements, in which pairs of particles from specific decays (such as those of $Z$ bosons) are selected. One of the pair (the \emph{tag} particle) is subject to strict identification criteria, and functions as a proxy for the identification of both particles, allowing the other (the \emph{probe} particle) to be used for measurements of the identification efficiency. By using a muon-based tag-and-probe method~\cite{PERF-2014-05,PERF-2015-10}, the muon spectrometer can serve as a reference for the reconstruction of muons in the ID; this measurement is presented in Section~\ref{sec:track-reco-muon}. For hadrons, indirect probes of tracking performance include flavour tagging scale factors~\cite{FTAG-2018-01,FTAG-2020-08} and studies of tracking efficiency in dense environments~\cite{PERF-2015-08}, where detector granularity allows relative efficiency to be assessed.

Although direct measurements for hadronic tracks in data are limited, this does not constrain the physics programme. Physics-object-level comparisons are sensitive to possible mismodelling, and the dominant uncertainties in tracking performance arise from the simulation of detector material and alignment, not from reconstruction itself. Since charged particle interactions with material are well understood, uncertainties primarily reflect how well the detector material is represented in simulation.

\subsubsection{Working point efficiencies and efficiency uncertainties}

\begin{figure}
\centering
\subfloat[]{\includegraphics[width=0.49\textwidth]{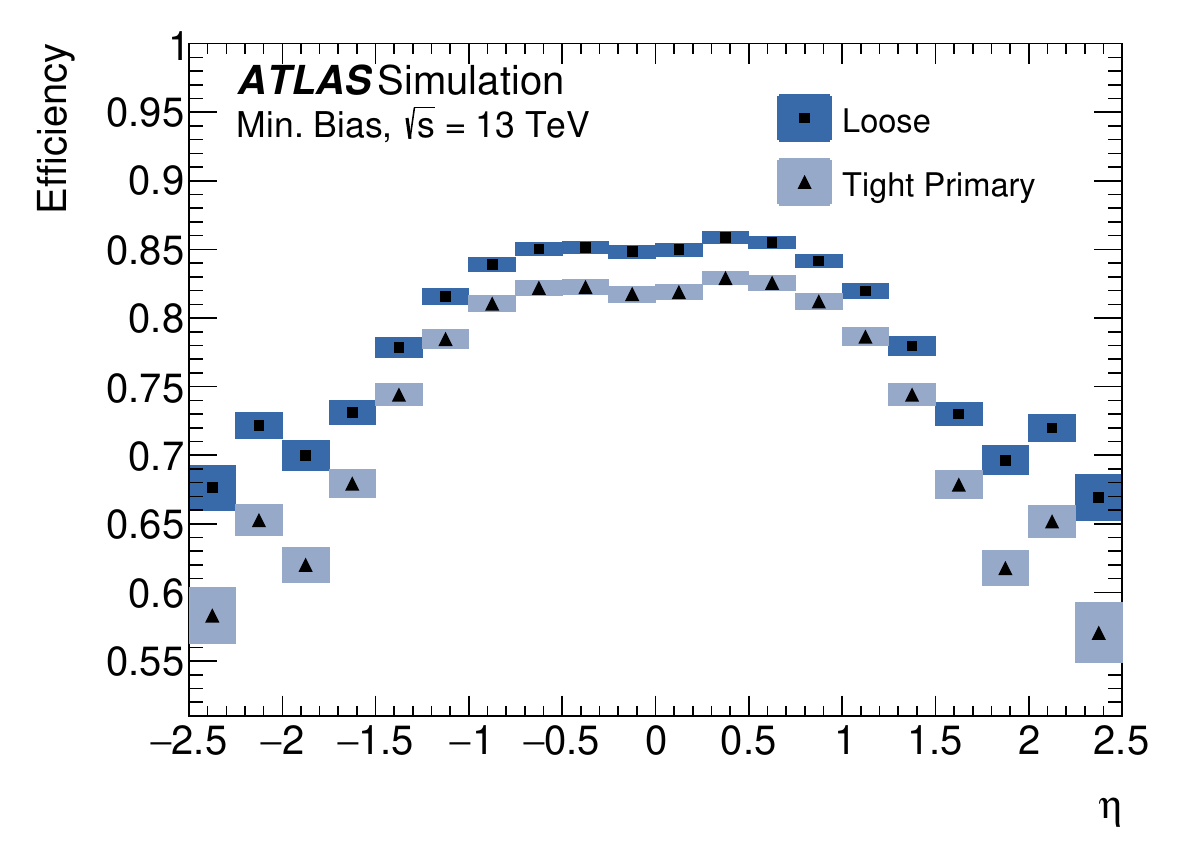}}
\subfloat[]{\includegraphics[width=0.49\textwidth]{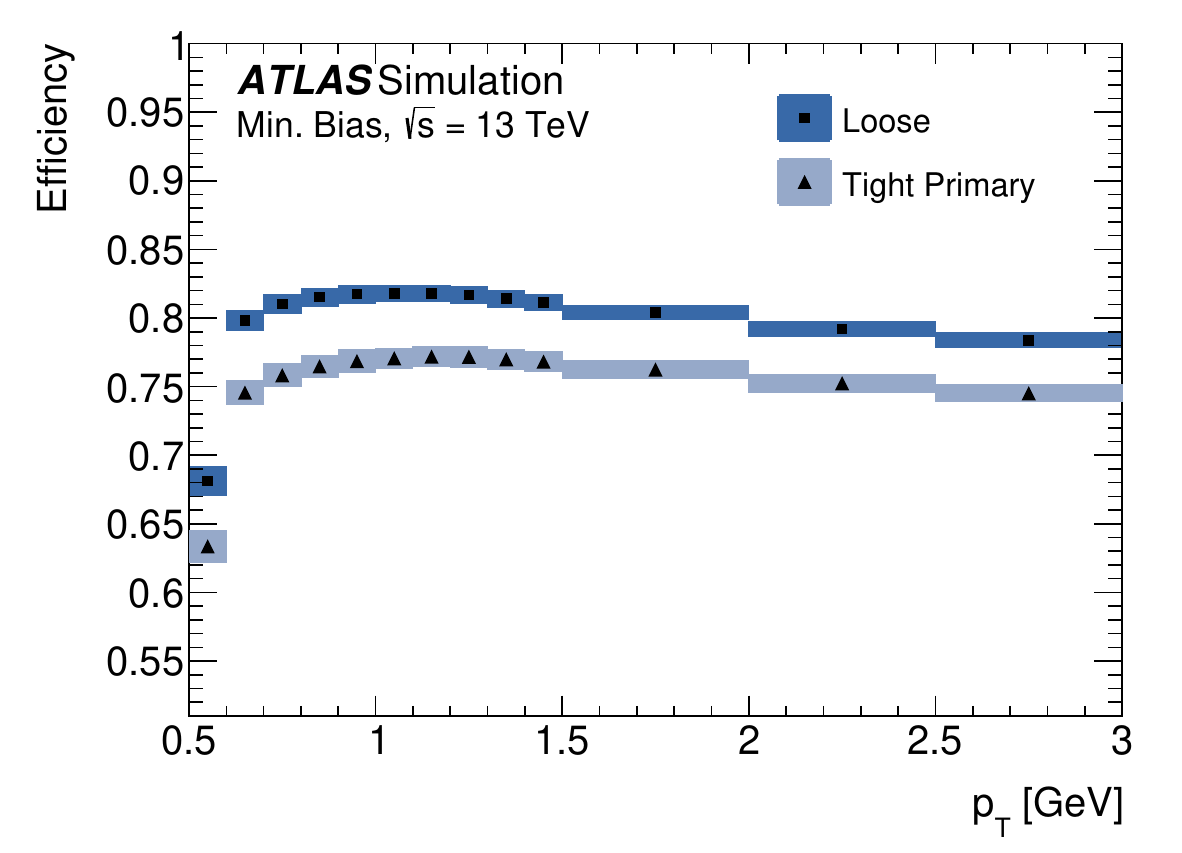}} \\
\caption{Track reconstruction efficiency as a function of the truth particle (a) $\eta$ and (b) $p_{\mathrm{T}}$ for a minimum bias MC sample. The total systematic uncertainty is shown by the error bands and corresponds to the sum in quadrature of the difference between the nominal efficiency and the efficiency determined in simulated samples with material and physics list variations.}
\label{fig:track_efficiency}
\end{figure}

The efficiency, $\epsilon_{\text{trk}}$ defined in \cref{eq:eff}, is parametrised as a function of $\pT$ and $\eta$. While correlations between the efficiency as a function of $\pT$ and $\eta$ could exist due to detector effects, the observed variations suggest they can be treated as largely independent.
The distributions in $\pT$ and $\eta$ are shown in \Fig{\ref{fig:track_efficiency}}, comparing the Loose quality working point to the Tight Primary quality working point.

The Tight Primary working point has 5--10\% lower efficiency, with the difference increasing with $|\eta|$ and 5\% lower efficiency at low \pt with respect to the Loose working point.
The central region $|\eta| < 1.5$ has less material than the forward region $|\eta|>1.5$ which leads to a higher and relatively constant efficiency, ~85\% and 80\% for Loose and Tight Primary respectively.
The slight increase in efficiency around $|\eta| = 2.0$ is due to the geometry of the detector layout that increases the number of active layers traversed by the particle.
In \pt, both working points show the same trend, reaching plateau efficiency around \qty{1.5}{\GeV}.
The Loose working point is around 5\% higher than Tight Primary, with percent level variations.

In addition to the baseline efficiency, the tracking efficiency is computed for variations in the amount of material in the tracker, the largest source of systematic uncertainty for the tracking efficiency in simulation.
This uncertainty is characterised by three components: 5\% overall material mismodelling estimated from the upper limit of the Run~1 material budget~\cite{PERF-2015-06}; 10\% variation in IBL material based on studies of hadronic interactions and photon conversions~\cite{PERF-2015-07}; and 25\% variation in the pixel service material in the Patch Panel 0 (PP0) region to cover discrepancies at forward rapidities~\cite{PERF-2015-07}. %
In addition to uncertainty in the amount of material, another source of systematic uncertainty comes from the choice of \GEANT physics list. These lists set which interaction processes are considered during simulation and at which particle energies they are applied~\cite{Agostinelli:2002hh}.

\begin{figure}
\centering
\subfloat[]{\includegraphics[width=0.49\textwidth]{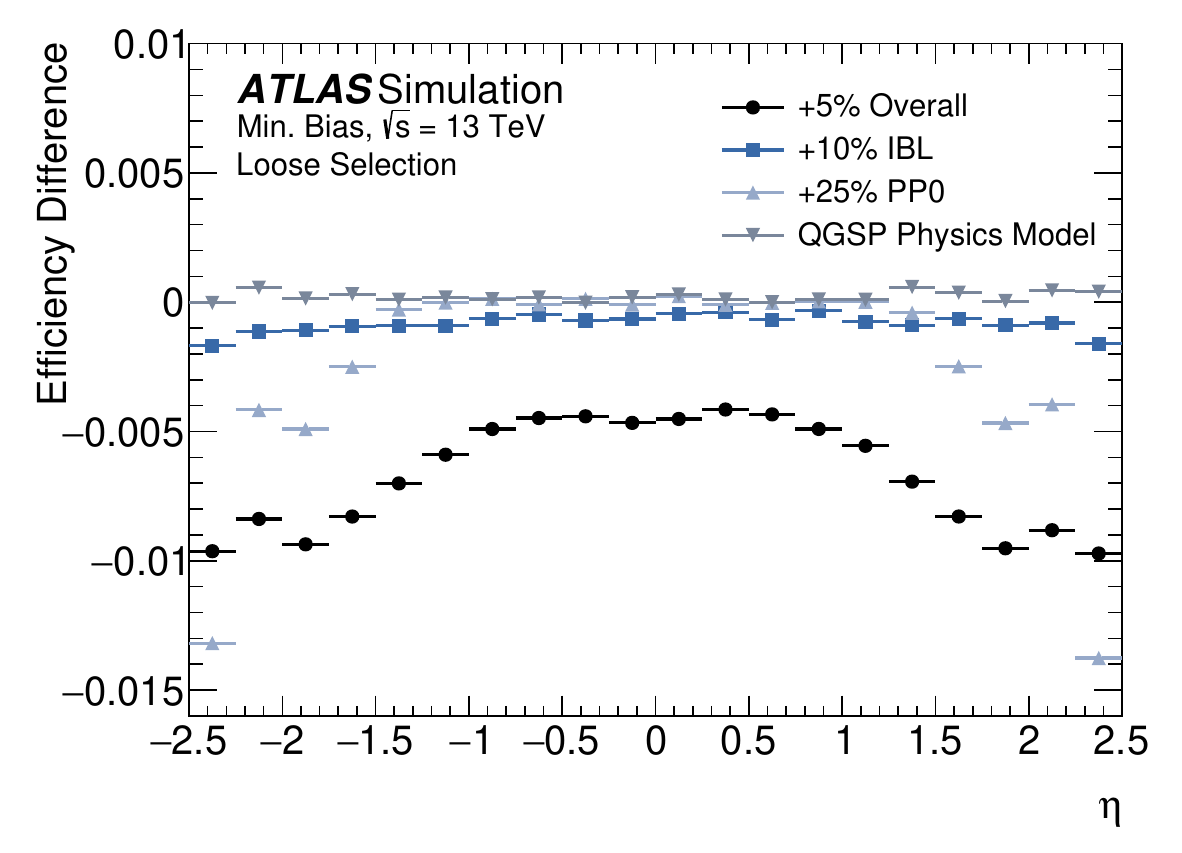}}
\subfloat[]{\includegraphics[width=0.49\textwidth]{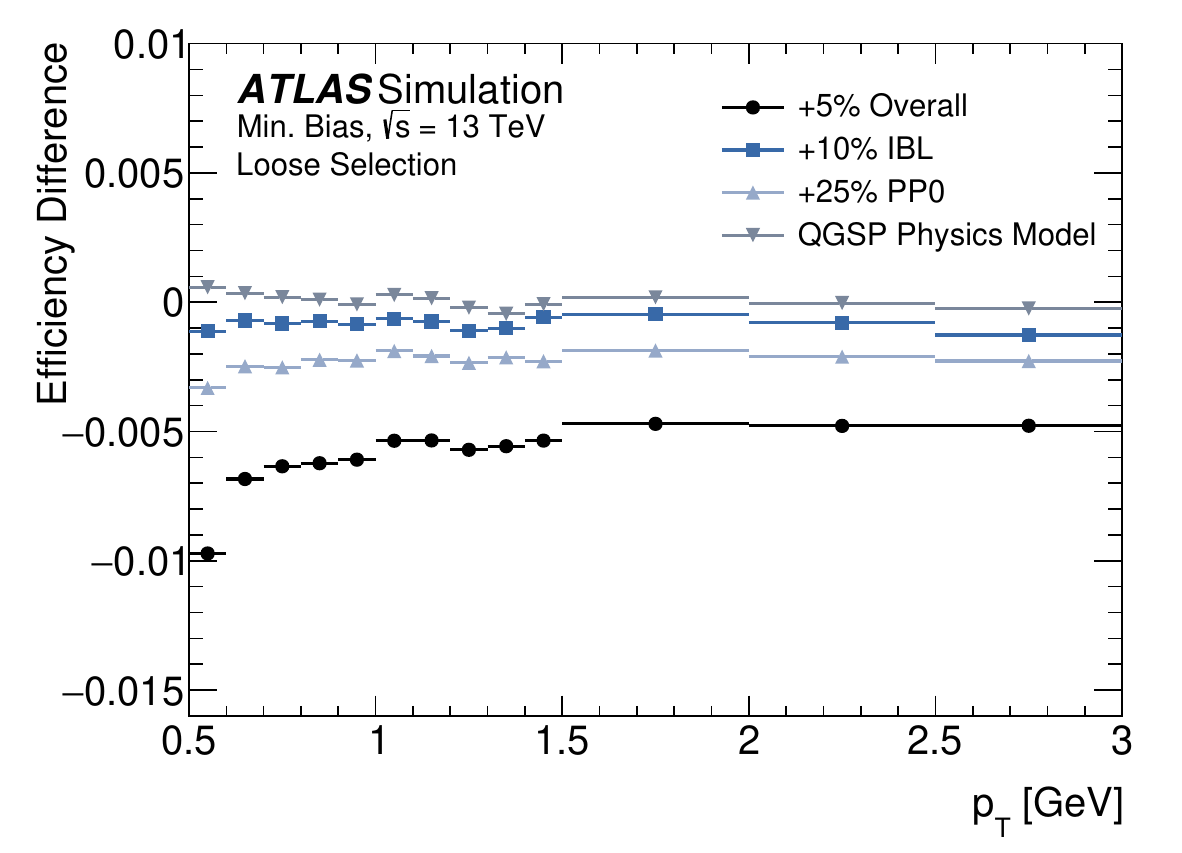}} \\
\caption{Track reconstruction efficiency difference for the Loose working point as a function of the truth particle $\eta$ and \pT\ between a standard MC sample (minimum bias) and samples with systematic variations in the material and \GEANT QGSP physics model.}
\label{fig:track_efficiency_syst}
\end{figure}

Figure~\ref{fig:track_efficiency_syst} shows the effect on the efficiency of the Loose working point from the various systematic variations. In general the 5\% overall variation tends to dominate, with the exception of the very forward region where the 5\% overall and 25\% PP0 variations are comparable.
The total systematic uncertainty is shown by the error bands in Figure~\ref{fig:track_efficiency} and corresponds to the sum in quadrature of the difference between the nominal efficiency and each of the material and physics list variations.
Systematic uncertainties concerning the misalignment of the physical detector are discussed in Ref.~\cite{IDTR-2019-05}.

\FloatBarrier

\subsubsection{Determination of the track reconstruction efficiency with collision data}
\label{sec:track-reco-muon}
For muons, inefficiencies in reconstructing their trajectories in the ID primarily arise from algorithmic limitations; effects such as radiative energy loss have a smaller impact within the momentum range of interest.
A clean sample of $Z$ bosons decaying to muons therefore allows a measurement
of the algorithmic efficiency with a purely data-driven method.

The data set corresponds to an integrated luminosity of \qty{140}{\ifb}
collected in the years from 2015 to 2018. The ID track
efficiencies are measured using a tag-and-probe method applied to
$Z\to \mu\mu$ candidate events. To select $Z\to\mu\mu$
decays, the invariant mass $m_{\mathrm{tag-probe}}$ of the
tag-and-probe pair is required to be between 61 and \qty{121}{\GeV}. One leg
of the decay, the \textit{tag}, is required to satisfy stringent
identification criteria and to have triggered the online event
selection. The second muon candidate in the pair, the \textit{probe},
is a track reconstructed in the muon spectrometer (MS) detectors. The
use of the MS tracking information for event selection allows for an unbiased measurement of
the ID track reconstruction efficiency for muons within the kinematic acceptance of the MS. To enhance the purity in the selection of $\text{Z}\to\mu\mu$
decays the probe tracks have to satisfy $\pt>\qty{10}{\GeV}$ and $|\eta|<2.5$, and
the impact parameters of the probe track must satisfy
$\left|d_{0}/\sigma\left(d_{0}\right)\right|<3$ and
$\left|z_{0}\right|<\qty{10}{\mm}$. Both tag and probe tracks are subject to (different) isolation requirements, to suppress misidentified candidates from jet activity. For the efficiency calculations, the MS probe is required to be
matched to an ID track by using a requirement of
$\Delta R < 0.05$.
No additional track-to-vertex association or track selection working point is applied to the ID tracks used in this matching. More details on the methodology of the efficiency measurement are given in Ref.~\cite{MUON-2018-03}.

\begin{figure}
\centering
\includegraphics[width=0.75\textwidth]{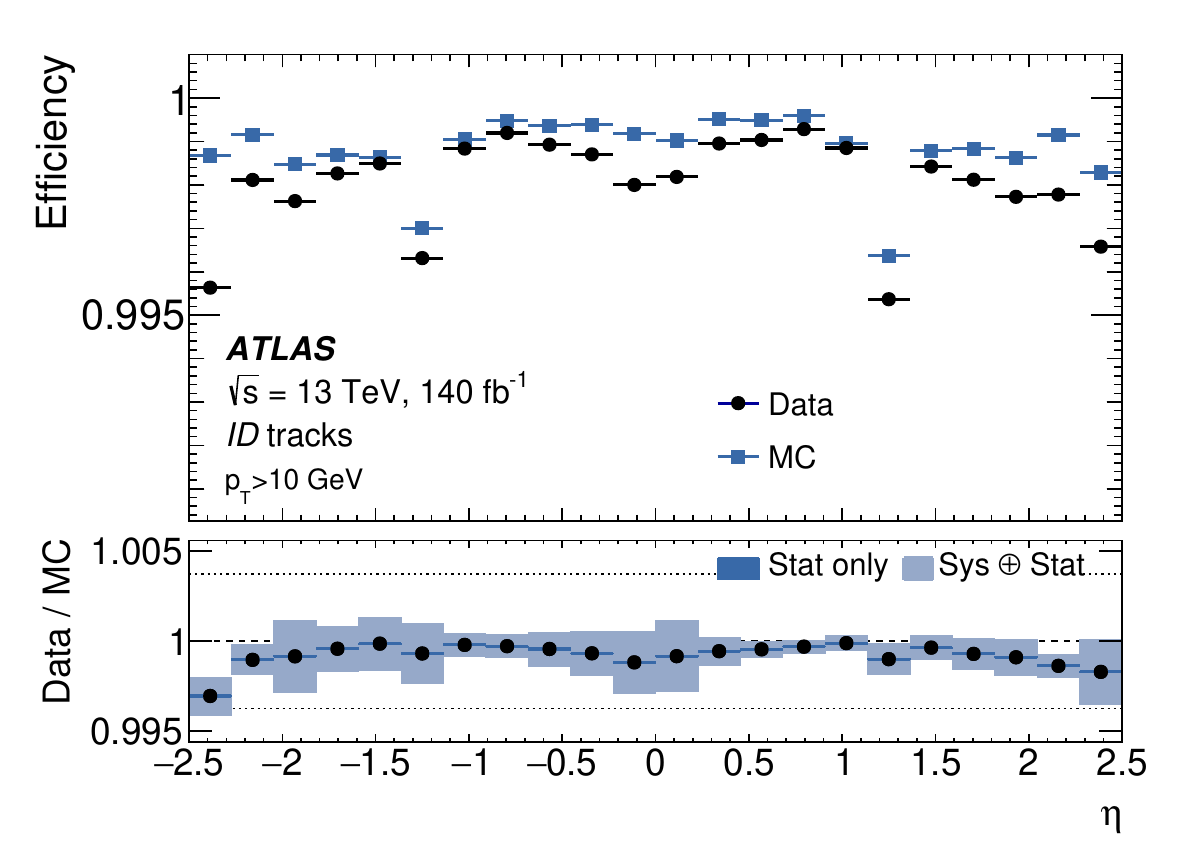}
\caption{Track reconstruction efficiencies in bins of $\eta$ for muons with $\pt >\qty{10}{\GeV}$ in data and Monte Carlo. The panel at the bottom shows the ratio of the measured to predicted efficiencies, with statistical and systematic uncertainties.}
\label{fig:eff_eta_run2}
\end{figure}

Figure~\ref{fig:eff_eta_run2} shows the ID track reconstruction efficiencies measured in $Z\to\mu\mu$ events in bins of $\eta$. The efficiency scale factors are defined as the ratios of the efficiencies measured in data to the efficiencies in simulation. The resulting efficiency is very close to unity, with nearly all measured values of above 99.7\%, and the obtained scale factors are consistent with one. This high efficiency reflects the fact that the measurement probes isolated muons, and therefore primarily tests algorithmic reconstruction performance rather than effects related to material interactions. The systematic uncertainties in the measurement are dominated by biases in the tag-and-probe method. The uncertainty is estimated in simulation by comparing the efficiency measured via the tag-and-probe method with the fraction of truth-level muons that are successfully reconstructed as ID tracks.
Other sources of systematic uncertainties are estimated by studying the modelling of the background parameterisation as well as the effect of the MC normalisation. Detailed descriptions of different sources of systematic uncertainties are given in Ref.~\cite{MUON-2018-03}.

\subsubsection{Large-Radius Tracking}

\begin{figure}[!t]
\centering
\subfloat[]{\includegraphics[width=.47\textwidth]{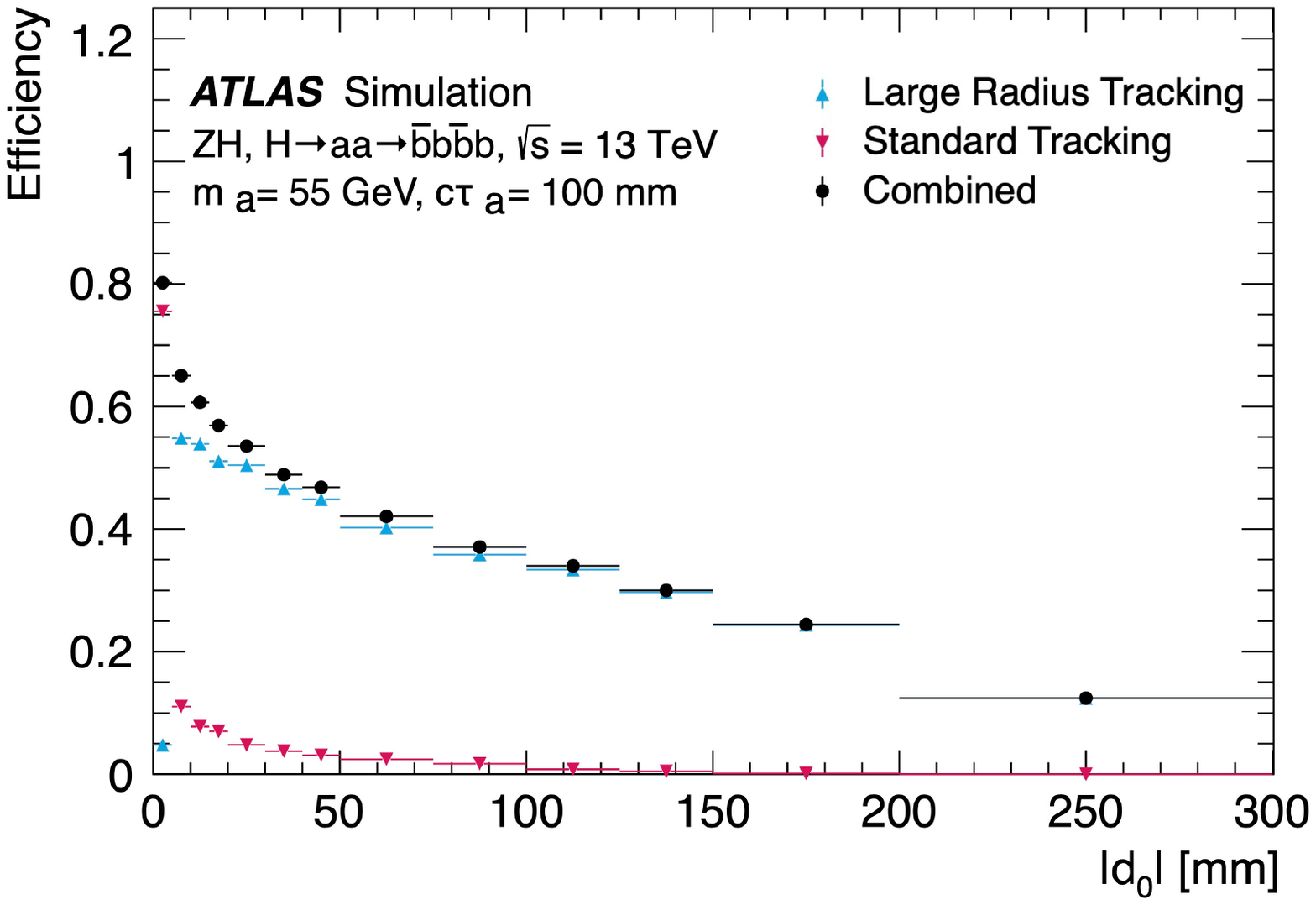} \label{fig:LRT_eff}}
\qquad
\subfloat[]{\includegraphics[width=.47\textwidth]{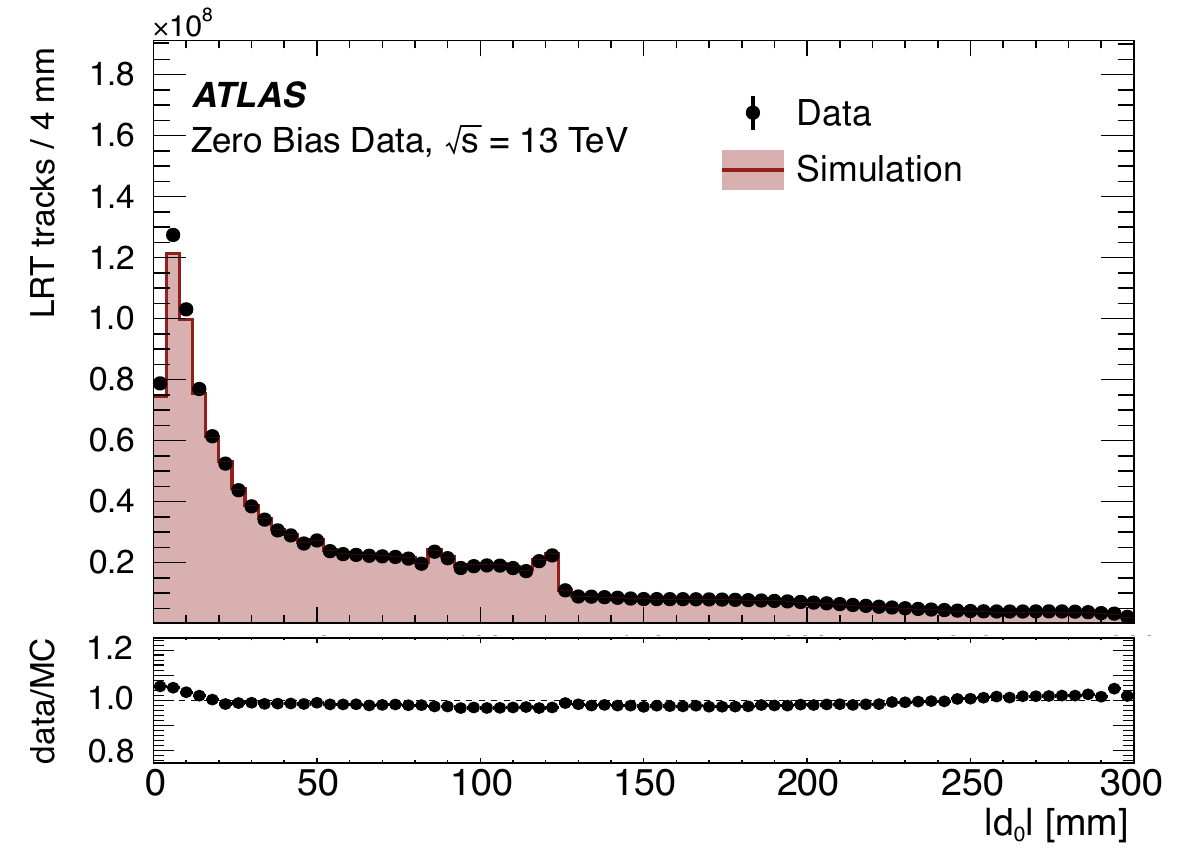} \label{fig:LRT_d0}}
\caption{(a) Track reconstruction efficiencies for displaced charged particles produced by the decay of the LLP in the Higgs portal signal model~\cite{patt2006higgsfieldportalhiddensectors}. The truth particles are subject to a fiducial selection requiring the particle to be from an LLP decay within a transverse distance of \SI{400}{\mm} from the interaction point, stable, charged, with $\pT>1.2~\GeV$, and within tracking acceptance $|\eta|<2.5$. Efficiencies are shown as a function of $|\dZero|$. (b) The distribution of the $|\dZero|$ of reconstructed LRT tracks in 2018 zero bias data (black markers) and simulated inelastic \pp collisions (filled red histogram). The simulation is normalised to the number of tracks in data. The discontinuity at $\qty{125}{\mm}$ reflects a drop in efficiency for tracks originating beyond the outermost pixel layer, where the number of remaining silicon layers is insufficient to meet the cluster requirements. Figures are adapted in style from Ref.~\cite{IDTR-2021-03}.}
\end{figure}

The efficiency for LRT for a Higgs portal LLP benchmark model is shown in Figure~\ref{fig:LRT_eff} as a function of $|\dZero|$. After $|\dZero| > \qty{5}{\mm}$, the standard reconstruction becomes inefficient. LRT recovers the loss in efficiency, with significant efficiencies up to $|\dZero| = \qty{300}{\mm}$. The quality of the LRT sequence modelling is demonstrated in Figure~\ref{fig:LRT_d0} which shows excellent agreement in a comparison between the $|\dZero|$ distribution of LRT tracks in 2018 zero bias data (Section~\ref{sec:samples}) and simulated inelastic events.
Further details are found in Ref.~\cite{IDTR-2021-03}.

\subsubsection{Mis-reconstructed tracks}
\label{sec:track-reconstr-fake}
\begin{figure}
\centering
\subfloat[]{\includegraphics[width=0.49\textwidth]{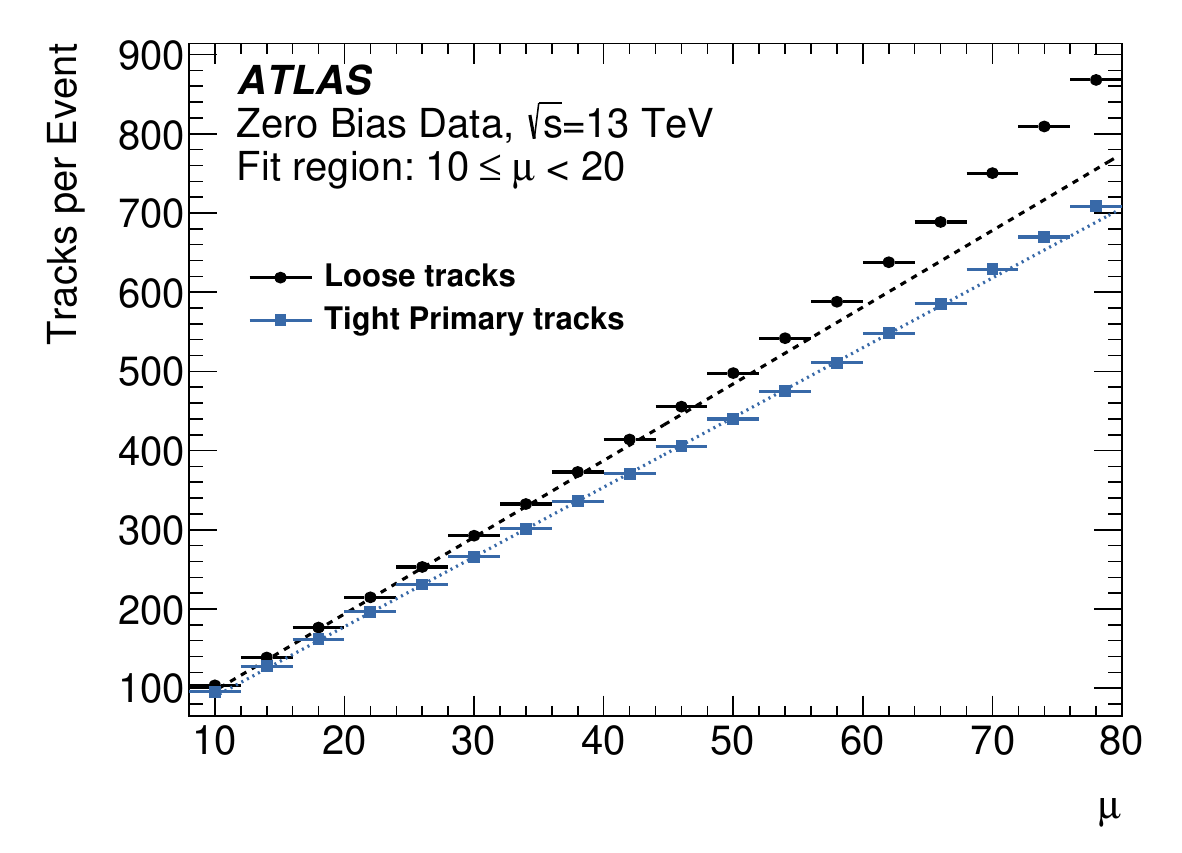}}
\subfloat[]{\includegraphics[width=0.49\textwidth]{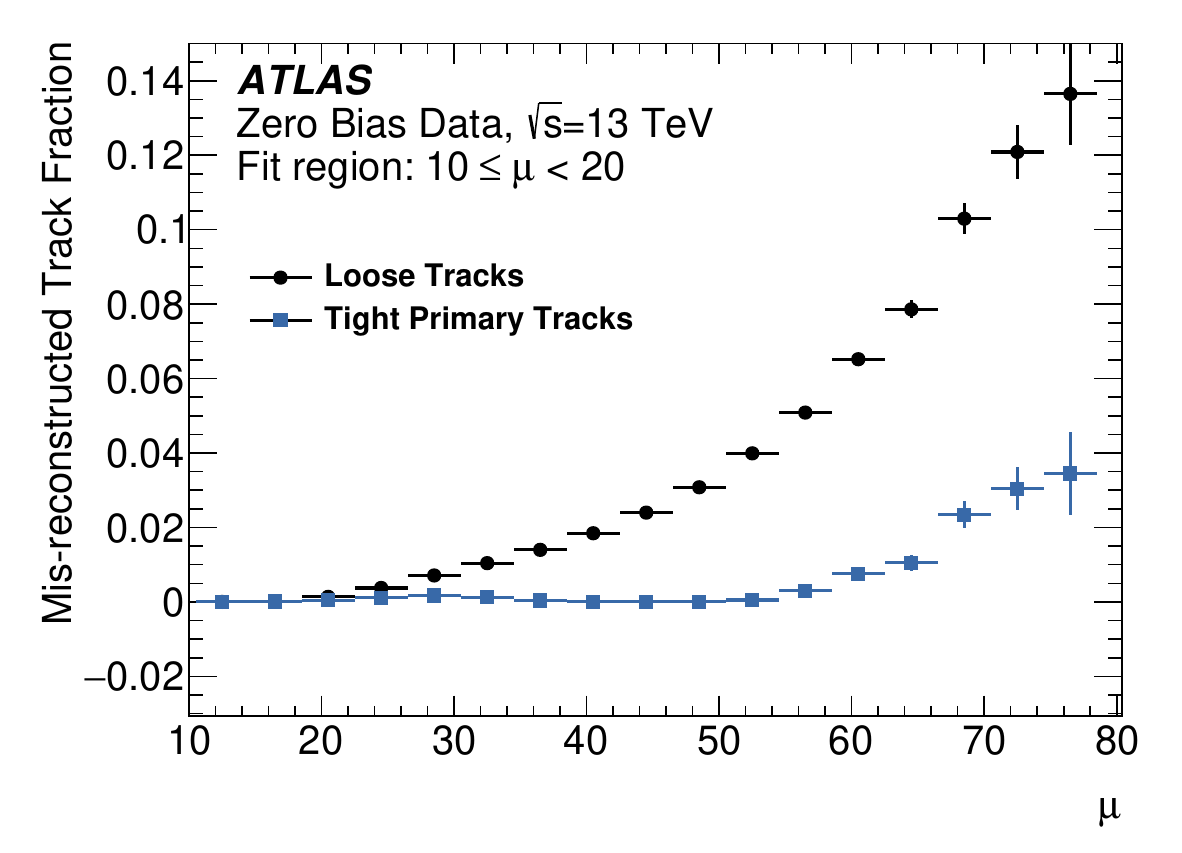}}
\caption{(a) Average number of total reconstructed Loose and
Tight Primary tracks per event ($N_{\mathrm{trk}}$) as a function of
$\mu$ in a 2018 zero bias data sample. The dashed lines illustrate the results of linear fits performed in the
$10 < \mu < 20$ region.
(b) Relative excess of reconstructed tracks after subtracting the expectation from a linear fit to the low-$\mu$ track multiplicity, normalised to the total track multiplicity.}
\label{fig:fakerateData}
\end{figure}

Mis-reconstructed tracks primarily arise from incorrect combinations of measurements from multiple particles. Thus, tracks consistent with the MC-based definition of fakes are expected to be the dominant contribution to mis-reconstruction. Their occurrence is strongly dependent on the measurement density and, consequently, on $\mu$.
The dependence on $\mu$ has been extensively studied throughout the entire lifetime of the detector, from early data analyses~\cite{ATLAS-CONF-2012-042} to ongoing preparations for the HL-LHC era~\cite{IDTR-2023-01}.
Because the efficiency for reconstructing good tracks remains largely stable as a function of $\mu$, the contribution from mis-reconstructed tracks can be estimated by analysing the scaling behaviour of the average number of reconstructed tracks per event as a function of $\mu$.
A linear fit is performed to the average number of reconstructed tracks per event at low $\mu$, where the contribution from mis-reconstructed tracks is assumed to be negligible, and the additional super-linear contribution to the number of reconstructed tracks at higher $\mu$ is then assumed to represent the contribution from mis-reconstructed tracks. These assumptions were tested in Monte Carlo simulations, where the method gave good agreement with truth-based calculations (as is also demonstrated by Figure~\ref{fig:pattern:fakerate-detailed_b}). An estimation of the mis-reconstructed track fraction using this method for both Loose and Tight Primary reconstructed tracks in a 2018 zero bias data~\cite{TRIG-2016-01} sample is shown in Figure~\ref{fig:fakerateData}, with fractions of up to $\sim 0.14$ and $\sim 0.04$, respectively, observed for the two working points at $\mu=80$.
For average Run~2 conditions, with $\mu$ around 30, the mis-reconstructed track fraction is around $9 \times 10^{-3}$ and $7 \times 10^{-4}$ for the Loose and Tight Primary working points, respectively.
For average Run~3 conditions, with $\mu$ around 60, the mis-reconstructed track fraction is around $6.5 \times 10^{-2}$ and $1.6 \times 10^{-3}$.

\FloatBarrier

%
%
%
%


\subsection{Track Impact Parameter Resolution}

The resolution of the measured impact parameters is a key performance quantity, directly influencing vertexing and lifetime-based reconstruction. It is governed by two main effects: multiple scattering effects and the intrinsic detector resolution. Multiple scattering effects depend on the detector material and are proportional to $1/{\pT}$. They are the dominant effect on tracks with \pT\ below approximately 5~\GeV. Intrinsic measurement resolution and alignment residuals dominate the impact parameter resolution at high \pt. The resolution also varies with pseudorapidity due to the detector geometry and the varying distance from the first measurement layer to the beamline. Mis-reconstruction of tracks (in particular, incorrect choices of innermost measurements) are an additional source of resolution degradation.

For primary vertex reconstruction, precise $\dZero$ and $z_{0}$ values mitigate the effects of pile‑up on vertex‑based observables. The enhanced spatial resolution reduces the probability that two nearby primary vertices are reconstructed as a single merged vertex. It also allows a tighter track‑to‑vertex association, which improves the reliability of downstream observables such as lepton isolation, jet‑vertex fraction~\cite{PERF-2014-03} used to distinguish pile‑up jets from hard‑scatter jets, and the soft term of missing transverse energy \met.
High‑quality impact parameters enhance the \met\ calculation by rejecting non‑prompt tracks from secondary interactions, thereby reducing the fake \met\ component and improving its scale and resolution.

\begin{figure}[htbp]
\centering
\subfloat[]{
\includegraphics[width=0.5\textwidth]{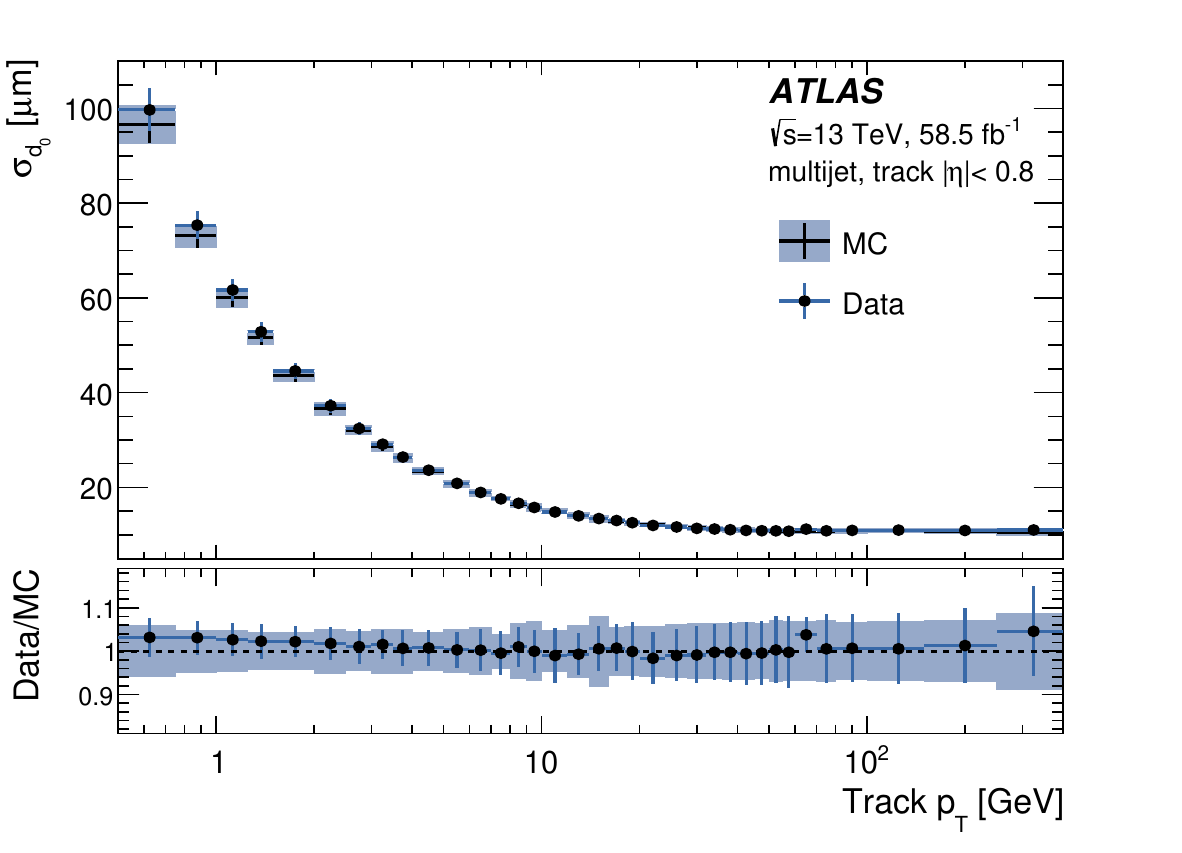}
\label{fig:d0}
}
\centering
\subfloat[]{
\includegraphics[width=0.5\textwidth]{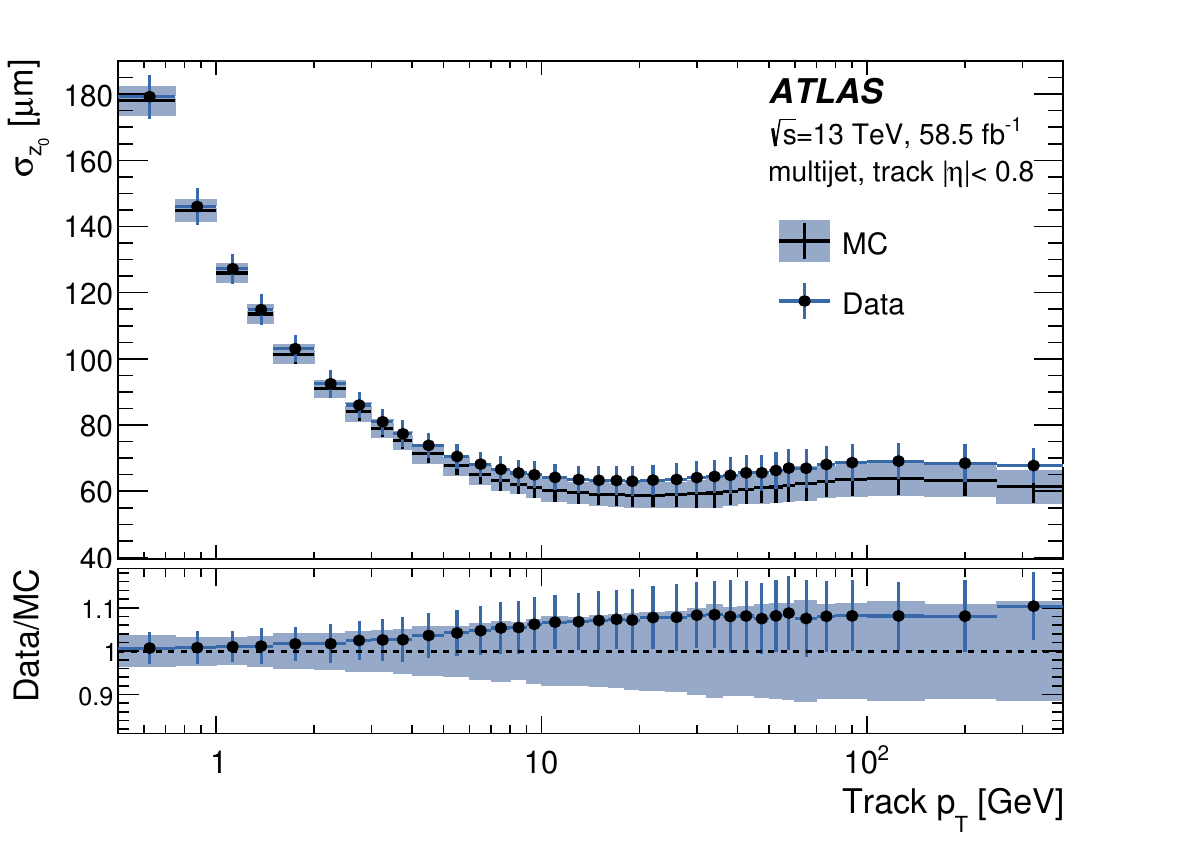}
\label{fig:z0}
}
\caption{Tight primary track resolution for 2018 collision data and MC \protect\subref{fig:d0} in the transverse plane, $d_{0}$, and \protect\subref{fig:z0} along the longitudinal axis, $z_{0}$. For $d_{0}$, the uncertainties reflect variations of the fit window from 1.5$\sigma$ to 2.0$\sigma$, with an additional beam-spot width variation included in quadrature for data. For $z_{0}$, the uncertainties are the quadrature sum of the same fit-window variation and a 20\% increase in the primary-vertex resolution.}
\label{fig:d0z0}
\end{figure}

Figure~\ref{fig:d0z0} shows the measured resolutions in data and MC.
The resolution is measured using central tracks ($|\eta| < 0.8$) passing the Tight Primary selection in jets with $\pt > \qty{300}{\GeV}$. Data from 2018 collisions are compared to simulated multijet events reweighted to match the jet \pt\ and track multiplicity. The beamspot is used as the reference point, and its transverse width and the longitudinal size of the distribution of primary vertices ($\sim\qty{10}{\um}$ and $\sim\qty{35}{\um}$ respectively~\cite{ATLAS-CONF-2010-027}) are subtracted in quadrature from the measured impact parameters in the respective directions to obtain the intrinsic resolution.

Distributions of $d_0$ and $z_0$ are analysed in \pt\ intervals. The core of each distribution, dominated by prompt particles, is fit with a single Gaussian in a $1.5\sigma$ window\footnote{The window is defined by an initial fit to the full distribution, with the subsequent fit performed in a reduced range to suppress outliers.} to extract the resolution ($\sigma$) and bias ($\mu$). For $d_0$, the error bands represent the impact of varying the fit window from $1.5\sigma$ to $2.0\sigma$. In data, an additional uncertainty from varying the beamspot width according to the observed range within the data period is added in quadrature. At low \pt below \qty{2}{\GeV} the resolution in data is slightly broader than simulation, potentially indicating differences in material description~\cite{PERF-2015-07}. For $z_0$, the error bands include the same fit window variation as $d_0$ and add in quadrature the impact of a 20\% increase in the primary vertex resolution. The resolution degrades slightly above \qty{20}{\GeV} due to an increased fraction of lower-quality tracks. This includes tracks missing IBL clusters or containing a higher proportion of P2/P3 clusters due to the longer pixel pitch along $\abseta$. The differences between relative rate of lower-quality tracks is potentially responsible for the increasing divergence between data and simulation, as well as residual issues in beamspot modelling.

Biases in $d_0$ and $z_0$ remain below \qty{1}{\um} in data, while MC is nearly unbiased. The resolution is also affected by the angular separation ($\Delta R$) between tracks. For $\Delta R < 0.02$, $d_0$ resolution worsens by up to 20\% due to merged clusters and cluster confusion, but stabilises at larger separations. The effect of \pileup is small: the resolution remains stable as a function of \pileup, with excellent agreement between data and MC.


\FloatBarrier
\subsection{Tracking in dense environments}
\label{sec:perf:tide}

%
The density of tracks in high-\pt\ hadronic jets increases with momentum and as discussed above, the charged particle separation in these environments is comparable to the ID sensor granularity.
Therefore, single clusters are often constructed from the energy deposited by multiple particles and must be shared between tracks.
As shared clusters are usually a sign of mis-reconstructed tracks, it is challenging to maintain efficient track reconstruction in dense environments.

Jets used in the following studies are reconstructed from topological clusters calibrated at the electromagnetic scale (EMTopo)~\cite{PERF-2014-02}.
The anti-\kt algorithm~\cite{Cacciari:2008gp} using a radius parameter of $R=0.4$ is used to reconstruct jets, followed by a calibration procedure~\cite{JETM-2018-05} to determine their kinematic properties such as the transverse momentum \pT.
Unless otherwise stated, jets are required to have $\pT>500~\GeV$ and $\vert\eta\vert<2.5$.
Tracks are matched to the jets using ghost association~\cite{Cacciari:2007fd}.

\subsubsection{Track Reconstruction Efficiency and Fake Rate}
\begin{figure}[th!]
\centering
\includegraphics[width=0.49\textwidth]{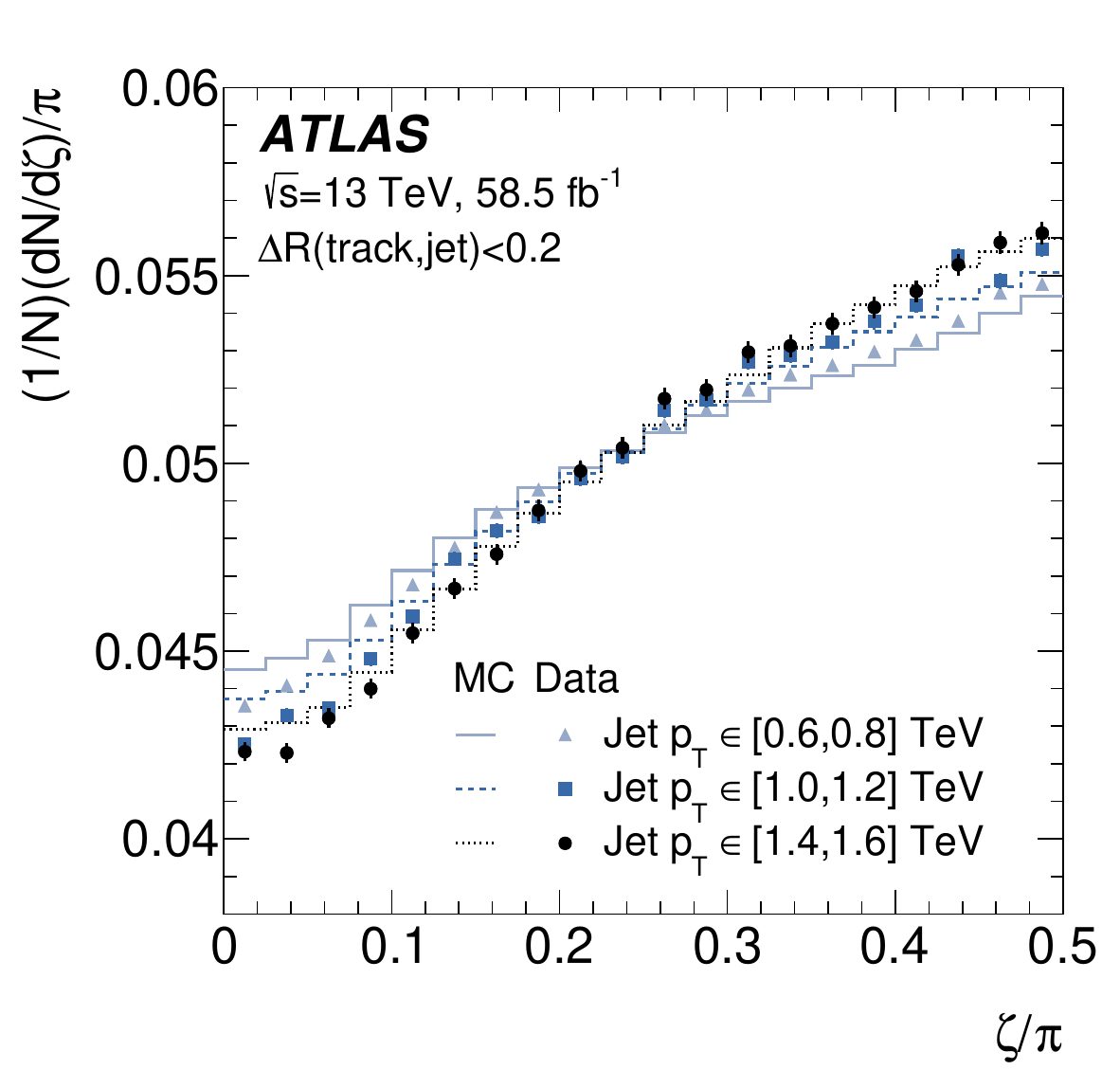}
\caption{
Distribution of $\zeta$ for several jet \pt\ intervals calculated for pairs of tracks that are inside the jet core (within $\dR < 0.02$ to the jet axis) compared between collision data from 2018 and simulated events.
}
\label{tide:qualitative}
\end{figure}

\cref{tide:qualitative} demonstrates the loss of track reconstruction efficiency in these challenging environments~\cite{ATL-PHYS-PUB-2017-016}.
The quantity $\zeta$ is defined as $|\tan^{-1}(\dphi/\deta)|$, where $\dphi$ and $\deta$ are the angular distances between pairs of tracks. Since the two sensor axes of a pixel module point along the $\eta$ and $\phi$ directions respectively, this observable therefore encodes the relative orientation of pairs of close-by tracks with respect to the local pixel sensor axes.
In the absence of reconstruction effects, a uniform distribution between 0 and $\pi/2$ is expected, as the relative orientation of track pairs is isotropic in the plane transverse to the jet axis.
The (innermost) pixel layer sensors have dimensions of \qtyproduct[product-units = power]{50 x 400}{\um} (\qtyproduct[product-units = power]{50 x 250}{\um}) in the $r$--$\phi$ and $z$ directions, respectively.
Due to the larger pitch in the $z$ direction, clusters from close-by tracks in the barrel region are more likely to merge along $z$ (track $\eta$) than along $r$--$\phi$.
As reconstruction efficiency is reduced in proportion to the rate of cluster merging, track pairs with small $\deta$ separations are reconstructed with lower efficiency compared to pairs with similar separation in $\dphi$.
This anisotropy manifests itself as a slope in the $\zeta$ distribution, which becomes more pronounced at high jet \pT, where cluster merging effects are enhanced.

\begin{figure}[th!]
\centering
\includegraphics[width=0.47\textwidth]{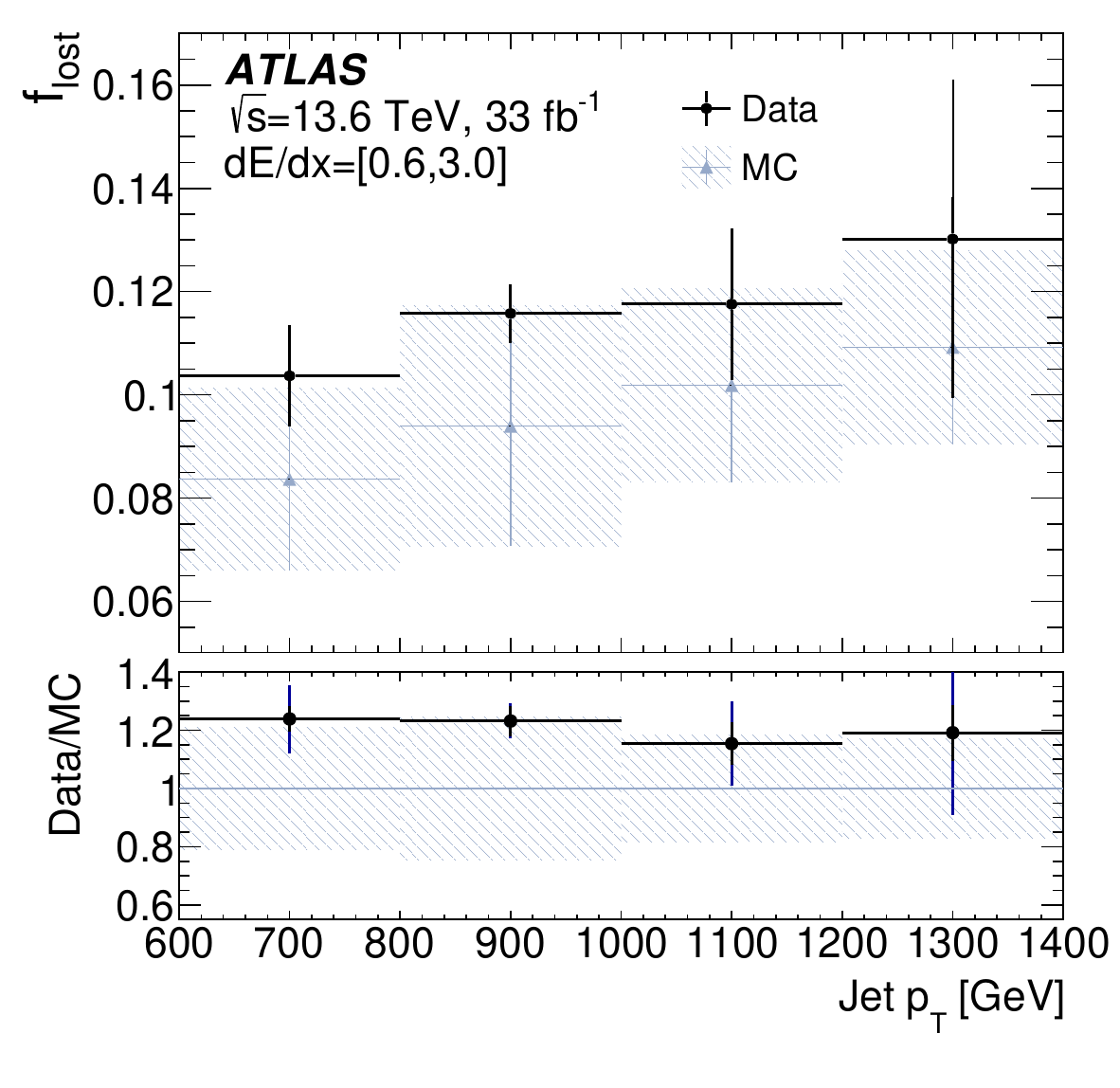}
\caption{Fraction of lost tracks ($f_\text{lost}$) as a function of jet \pT in Run~3 data collected in 2022 and simulation. The lost-track fraction is estimated using a fit to the \dedx\ distribution of B-Layer pixel clusters, using templates for single- and multi-particle ionisation. The range $0.6<$ \dedx\ $<3.0$ $\MeV~g^{-1}~cm^{2}$ was considered in the fit. Error bands include both statistical and systematic uncertainties.
}
\label{tide:flost}
\end{figure}

To probe the efficiency of track reconstruction in dense environments, a method based on the measurement of the energy loss (\dedx) in the pixel detector was employed, as originally described in Ref.~\cite{PERF-2015-08}. This data-driven technique estimates the fraction of lost tracks by fitting the \dedx\ distribution in B-Layer pixel clusters with templates corresponding to single-particle and multi-particle ionisation. The method has been updated with an improved multi-track template constructed as a convolution of single-track templates, producing a more realistic model of merged energy deposits. Systematic uncertainties on the mean \dedx\ of the single- and two-particle cluster templates used in the fit, as well as the effect of changing the range of \dedx\ to be fit over were considered, with the latter being the dominant contribution overall.
Run~3 data and simulation were utilised, in order to take advantage of the radiation damage modelling included in the simulation~\cite{IDET-2017-10,IDET-2020-01}. Jets with $\pT > 600~\GeV$ and $|\eta| < 2.5$ were used, along with tracks passing the Tight Primary selection. Figure~\ref{tide:flost} displays the fraction of lost tracks as a function of jet \pT, showing increasing inefficiency in denser jets, both for data and MC. The differences of up to 25\% observed between data and MC likely come from the accuracy with which the simulation models the \dedx\ distributions used in the fit. In particular, the three-particle cluster \dedx\ distributions in MC exhibit a higher level of secondary-peak contamination at low \dedx\ compared to those in data.

\begin{figure}[th!]
\centering
\includegraphics[width=0.45\textwidth]{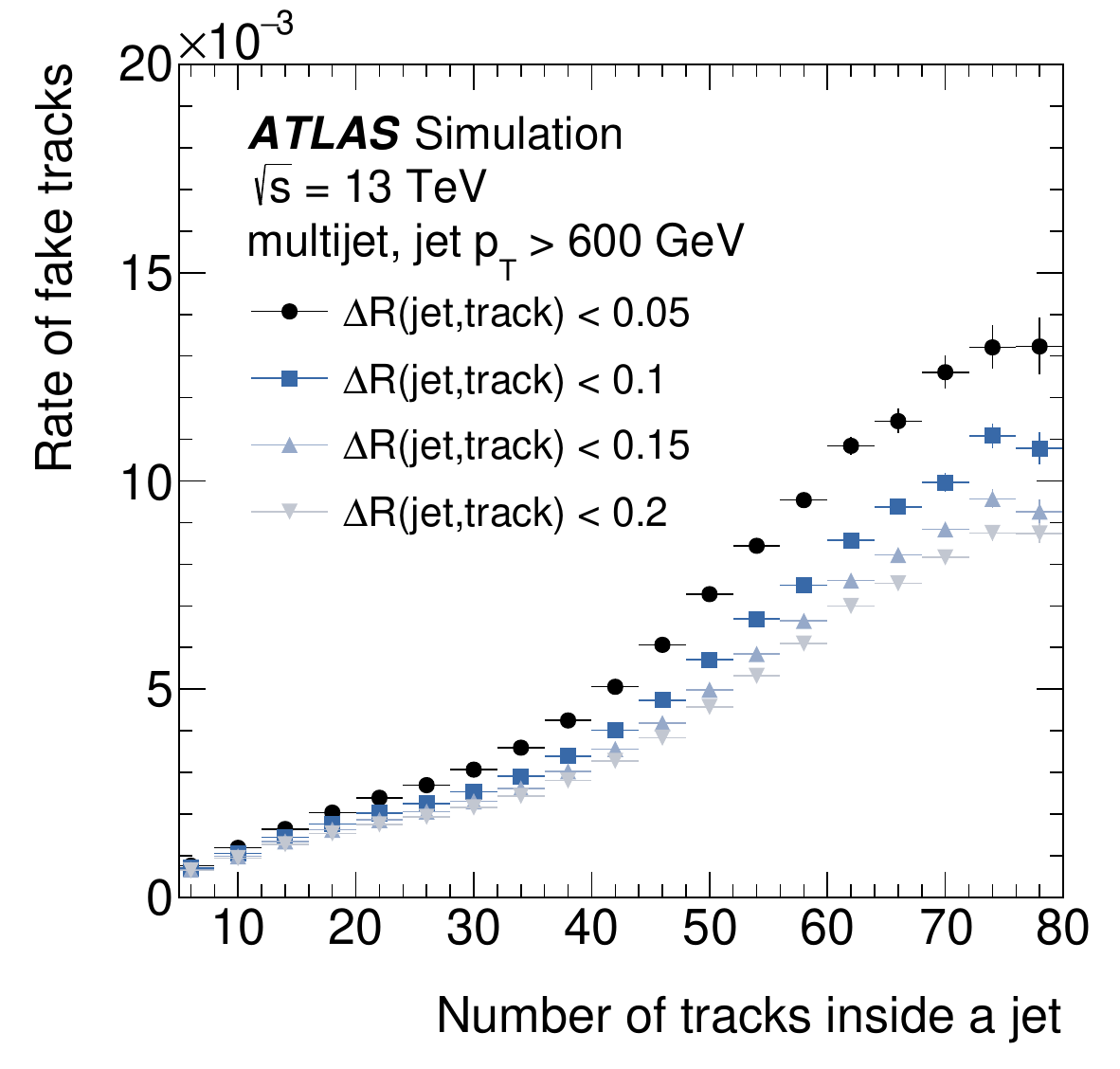}
\caption{
Rate of fake tracks in the jet core as a function of number of tracks found in the jet.
Distributions obtained in MC simulation for several $\DRjt$ intervals representing the angular distance between tracks and the jet core.
Only the statistical uncertainties are shown.
}
\label{fake:fakeTrks_Origin}
\end{figure}

In addition to the fake tracks occurring due to incorrect combinations of measurements from multiple pile-up tracks, covered in Sections~\ref{sec:perf:lowlevel} and~\ref{sec:track-reconstr-fake}, confusion can also happen between close-by tracks in dense jet cores, leading to the production of fakes. As the combinatorial complexity of the pattern-recognition problem increases rapidly with track density, the rate at which fakes occur is strongly dependent on the number of tracks in the jet (and therefore also dependent on \pt) as shown in Figure~\ref{fake:fakeTrks_Origin}. Since these fakes arise from a different source, this can be considered an additional contribution on top of the inclusive rates discussed elsewhere. Figure~\ref{fake:fakeTrks_Origin} shows the fake rate due to tracks in jets for various definitions of the core of a jet, which ranges from 0.1\% to 1.25\% depending on the track multiplicity. It can be observed that there is a strong dependence on the jet core definition, with differences of up to 50\% in the fake rate between the widest and narrowest $\DRjt$ intervals. For a typical high-\pt jet core, the additional fake contribution will be 0.2\% to 0.3\% absolute.
Pile-up has a second order impact on the rate of fake tracks in a jet~\cite{ATL-PHYS-PUB-2017-016}.

A template fit to variables in a fake-enhanced region (FER) of events containing high-\pt\ ($\pT > 600~\GeV$ and $|\eta| < 2.5$) jets, following the method described in Ref.~\cite{ATL-PHYS-PUB-2017-016}, allows for a comparison of the fake rate in data and simulated events which can be extrapolated to the full phase space. The FER is enriched with fake tracks by applying a selection of criteria designed to isolate tracks with characteristics typical of fakes, such as large impact parameters, poor fit quality, and a low number of associated clusters.

To account for potential mismodelling in the FER, four variables are fit to extract the fake rate in data: the fit quality $\chi^2/\mathrm{DoF}$, the number of TRT measurements inclusively and within $|\eta|<2$, and $d_0$. The variation in the results, indicating the degree of mismodelling, is less than 35\%. Extrapolated to the full phase space, the fake track fraction ranges from 0.19–0.29\% in data depending on the variable used, which can be compared to 0.25\% in simulation.


\subsection{Primary vertex reconstruction}
%

The performance of vertex reconstruction is primarily assessed using simulation. Previous studies of the primary vertex reconstruction efficiency have been performed using data from minimum-bias events at \(\sqrt{s} = 8\,\TeV\)~\cite{STDM-2014-19} and \(\sqrt{s} = 13\,\TeV\)~\cite{STDM-2015-02}. While these measurements validate the overall vertex reconstruction efficiency in data, they do not allow the study of the dependence of the performance on the level of \pileup.

Reconstructed primary vertices are classified into four types based on the truth-matching of reconstructed tracks and the associated weight from vertex fitting~\cite{PERF-2015-01}:
\begin{itemize}
\item \textit{Matched}: At least 70\% of the total track weight in the reconstructed vertex originates from a single
simulated \pp interaction.
\item \textit{Merged}: Less than 70\% of the total track weight in the reconstructed vertex originates from any
single simulated \pp interaction.
\item \textit{Split}: If a single simulated \pp interaction contributes the largest fraction of track weights to two or more reconstructed vertices, the reconstructed vertex with the largest track $\sum \pT^2$ is classified as either matched or merged, whilst the other(s) are labelled split.
\item \textit{Fake}: Fake tracks contribute more weight to the reconstructed vertex than any simulated \pp interaction.
\end{itemize}

\begin{figure}[htbp]
\begin{center}
\subfloat[][]{
\includegraphics[width=0.49\textwidth]{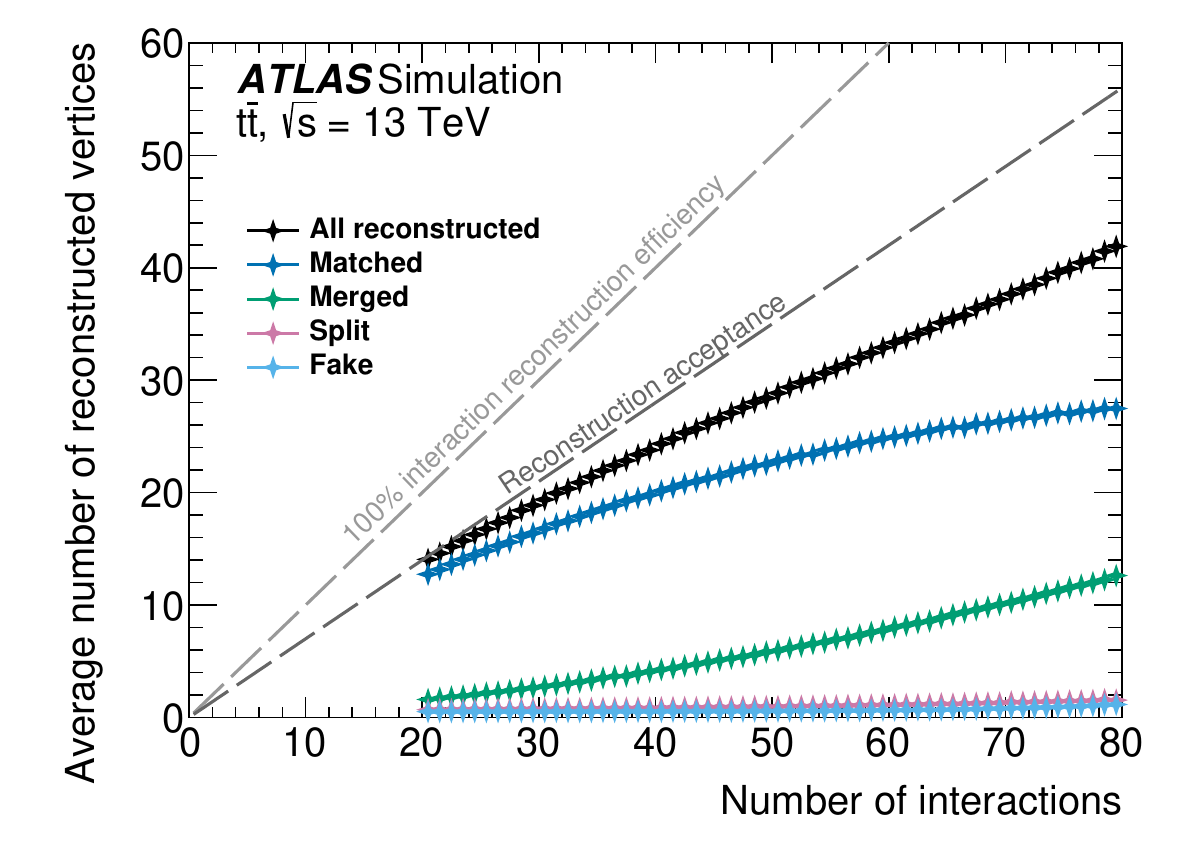}
\label{fig:perf-vertex:nPV_vs_Pileup}
}
\subfloat[][]{
\includegraphics[width=0.49\textwidth]{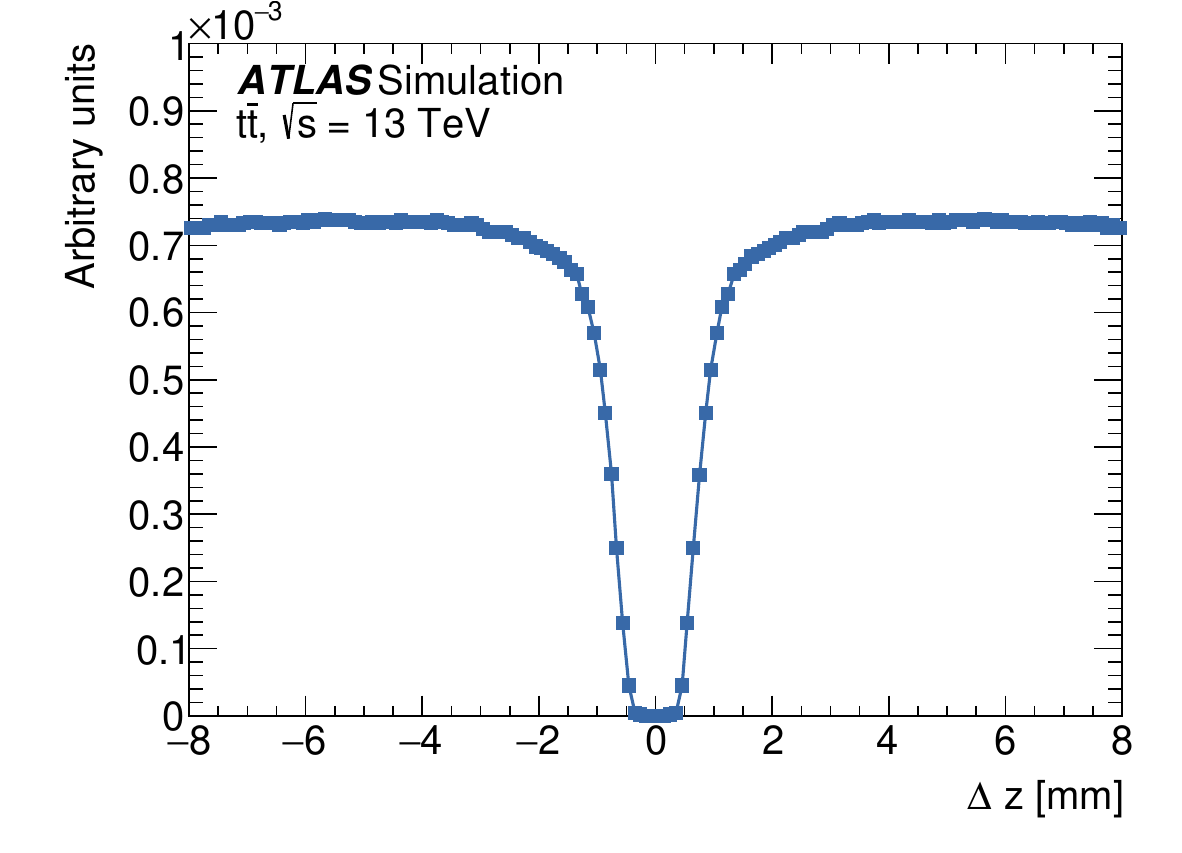}
\label{fig:perf-vertex:vertex_separation}
}
\caption{(a) Number of reconstructed primary vertices as a function of the number of simulated \pp interactions per bunch crossing and (b) longitudinal separation between reconstructed primary vertices, in simulated $t\overline{t}$ events.}
\end{center}
\end{figure}

Figure~\ref{fig:perf-vertex:nPV_vs_Pileup} shows the average number of reconstructed vertices per event in a sample of $t\bar{t}$ interactions with accompanying \pileup, as a function of the number of simulated \pp interactions per event.
The line labelled \textit{reconstruction acceptance} shows the number of interactions having at least two reconstructed tracks in the detector, which is considered the minimum requirement for having a successful reconstructed primary vertex.
The fraction for the split and fake vertices is small compared to matched and merged vertices, which indicates a robust primary vertex reconstruction.
The main effect contributing to a non-linear behaviour in the number of reconstructed vertices is the merging of two nearby interactions into a single vertex.
Figure~\ref{fig:perf-vertex:vertex_separation} shows the distance between nearby vertices,  where a clear dip around zero is observed caused by two interactions merging into one reconstructed vertex. In the absence of merging, the distribution of vertices along the $z$-axis would be expected to be Gaussian with a standard deviation equal to $\sqrt{2}$ times the beamspot size\footnote{The $\sqrt{2}$ arises from the convolution of two independent vertex position uncertainties.}, i.e.\ about \qty{60}{\mm} for this simulated sample.

\begin{figure}[htbp]
\begin{center}
\subfloat[][]{
\includegraphics[width=0.49\textwidth]{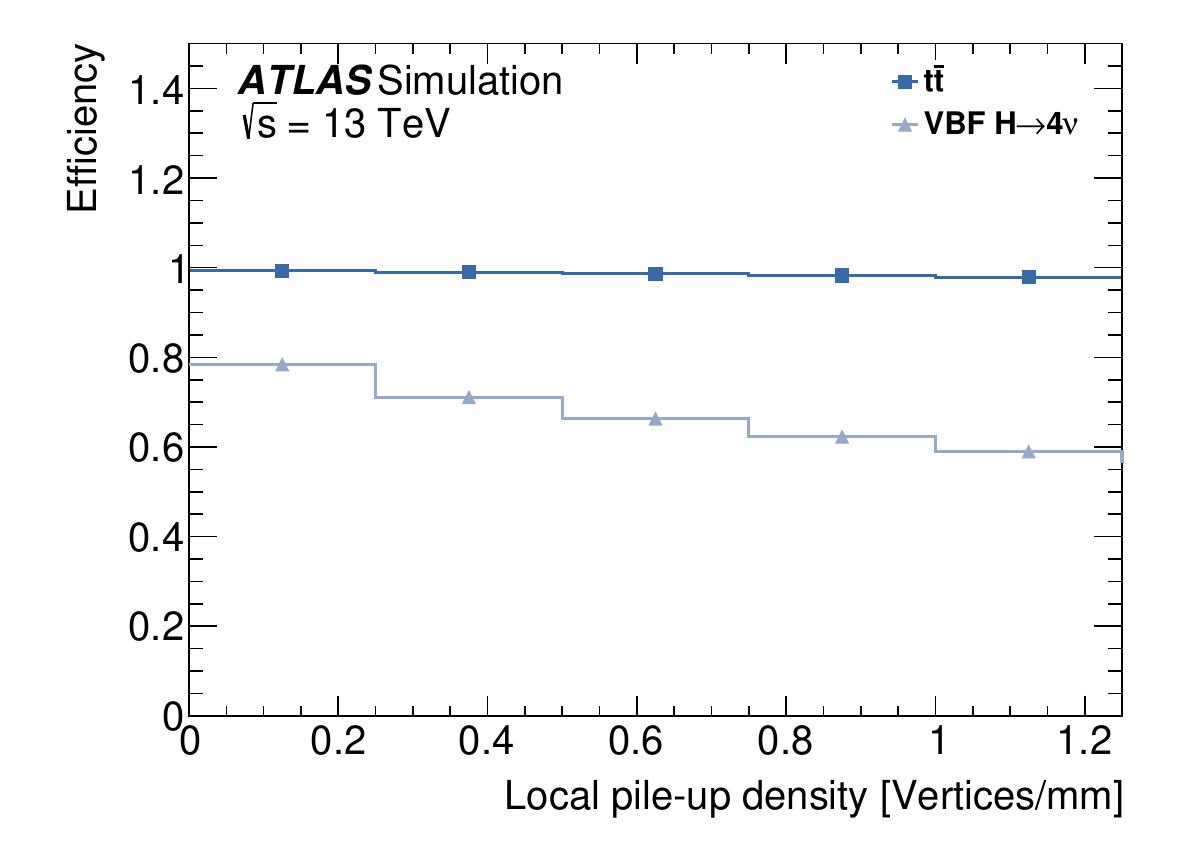}
\label{fig:perf-vertex:sel_eff}
}
\subfloat[][]{
\includegraphics[width=0.49\textwidth]{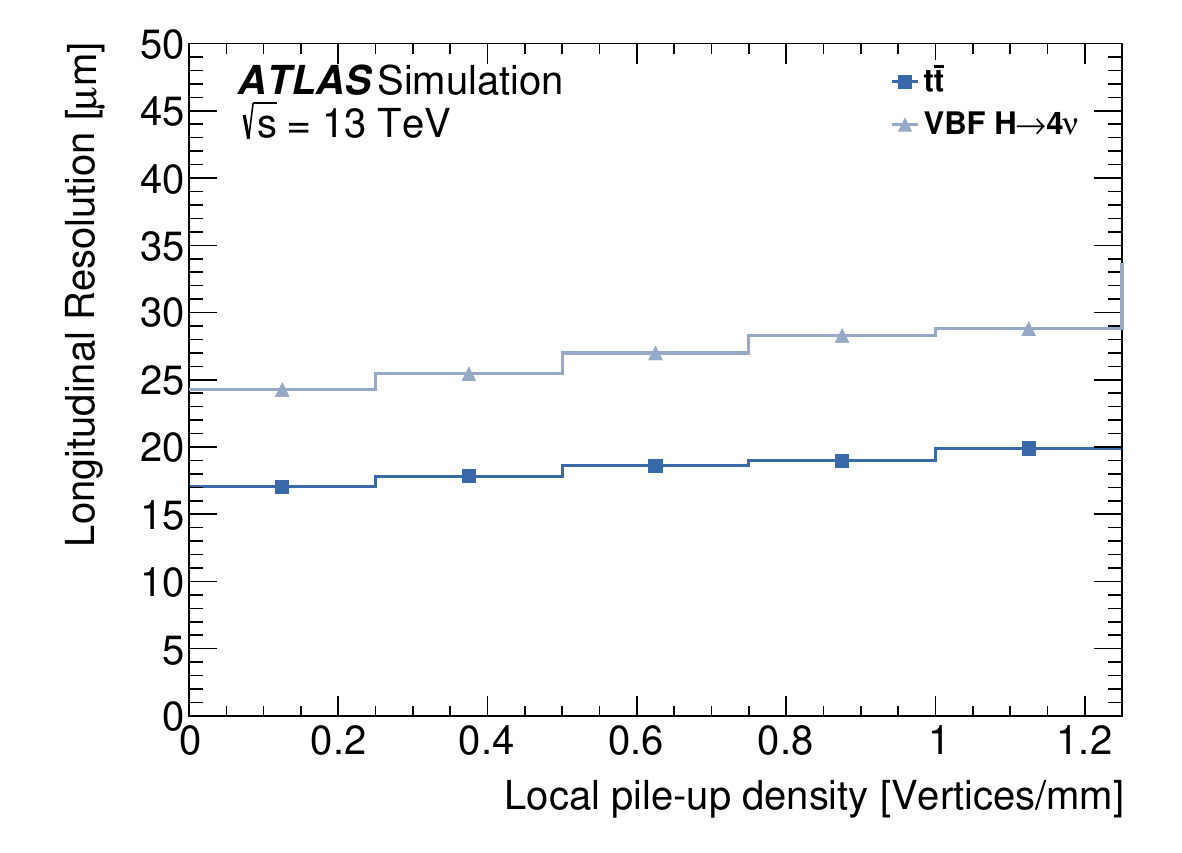}
\label{fig:perf-vertex:hs_pos_res}
}
\caption{(a) Hard-scatter vertex selection efficiency as a function of \pileup density and (b) longitudinal resolution for the hard-scatter vertex as a function of \pileup density, for simulated $t\overline{t}$ and VBF $H\to 4\nu$.}
\end{center}
\end{figure}

There are several algorithms that can be deployed to select the reconstructed vertex best associated with the hard-scatter interaction, depending on the specific final state of interest (for example the neural network based approach taken for $H\to\gamma\gamma$ decays~\cite{HIGG-2019-16}). For inclusive events the best generic choice is to take the reconstructed vertex with the highest sum of the squared transverse momenta of associated tracks, which is the selection used here.
The selection efficiency is defined as the fraction of the events in which the selected hard-scatter vertex is matched to the truth hard-scatter interaction. This efficiency quantifies effects from \pileup interactions. Figure~\ref{fig:perf-vertex:sel_eff} shows the distribution of hard-scatter selection efficiency for $t\bar{t}$ and VBF $H\to 4\nu$. This simple algorithm can achieve almost 100\% efficiency for $t\bar{t}$ events, while some inefficiency can be noted in the case of VBF $H\to 4\nu$.
Figure~\ref{fig:perf-vertex:hs_pos_res} shows the position resolution for the hard-scatter vertex in the longitudinal direction as a function of \pileup density; only minimal degradation is found even for very large \pileup density values.

\FloatBarrier


%
%
%

\subsection{Computation time}
\label{sec:perf:timing}

%
%
%
The initial reconstruction of ATLAS data is performed at the CERN Tier-0 computing centre~\cite{Stewart:1457501,SOFT-2022-02}, where events must be processed within the available bandwidth in order to avoid delays in delivering outputs critical for ATLAS operations. Tier-1 and Tier-2 sites distributed worldwide are responsible for large-scale data reprocessing and the reconstruction of simulated events. Across all computing tiers, it is important to make efficient use of the available resources, both to maximise throughput and to ensure that data processing is carried out in a sustainable manner by minimising the energy consumption associated with computation.

\begin{figure}[htbp]
\begin{center}
\includegraphics[width=0.75\textwidth]{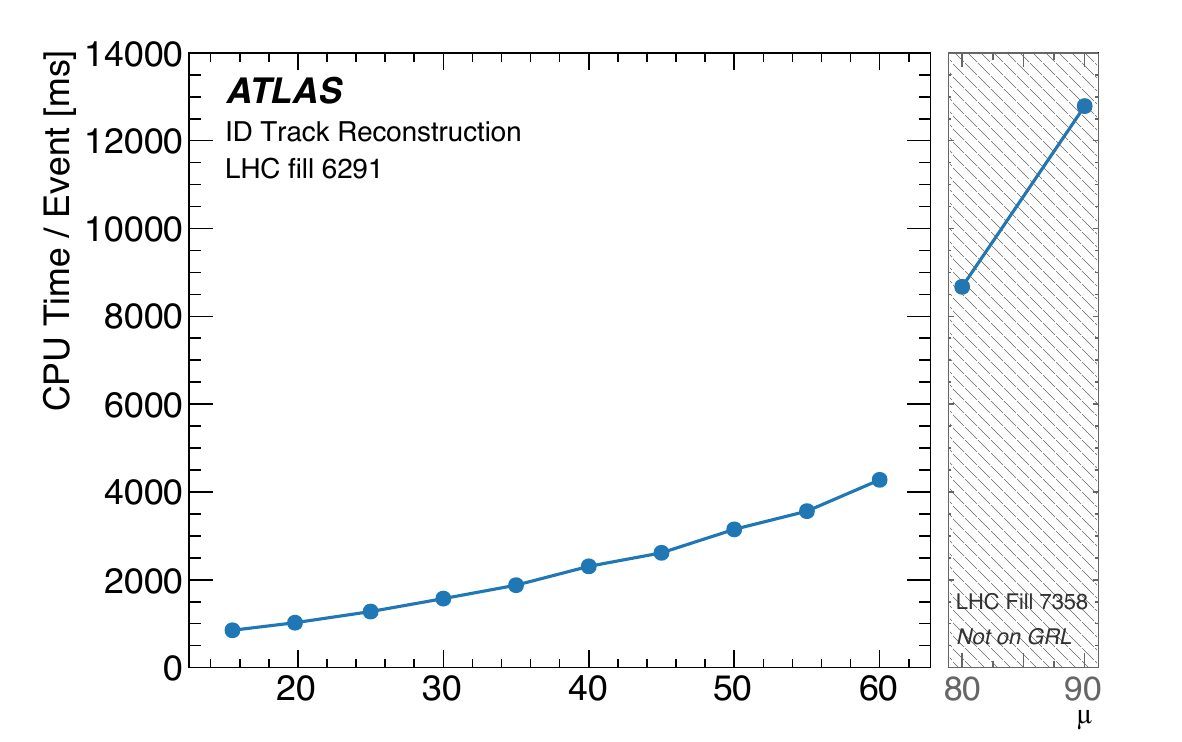}
\caption{Processing time taken per event as a function of the average number of \pp interactions per bunch crossing, $\mu$, for the full ID reconstruction chain.}
\label{fig:perf-timing:CPU_vs_mu}
\end{center}
\end{figure}

Since ID track reconstruction is one of the dominant contributions to overall reconstruction time~\cite{IDTR-2022-04}, it is of critical importance to ensure that its CPU usage is kept at an acceptable level. Given the combinatorial nature of track reconstruction, it is also critical to ensure that this performance scales well with increasing $\mu$ and does not exhibit significant super-linear behaviour. Figure~\ref{fig:perf-timing:CPU_vs_mu} demonstrates the average CPU time (in ms) taken to process an event on a typical computing node used for offline reconstruction\footnote{A dedicated single-user node with a 16-core processor and a HS06 score of 22.6~\cite{HS06} per core, run in single-thread mode with multiple simultaneous jobs leading to the node running at 50\% capacity.} as a function of $\mu$. Up to a $\mu$ of 60, the scaling behaviour remains stable, and the CPU usage has been well within the available bandwidth for Tier-0 processing. The higher range (shown in grey) demonstrates a significant increase in the required time for processing events at higher $\mu$. The LHC machine development run used for producing this estimate was not part of the standard physics programme and does not meet criteria to be part of the good-run list (GRL)~\cite{DAPR-2018-01}. While it therefore may not be completely representative of standard ATLAS running, it nonetheless demonstrates the scale of challenge faced for future track reconstruction at higher $\langle \mu \rangle$.

\FloatBarrier

\section{Conclusion}
\label{sec:conclusions}

The ATLAS experiment has developed a highly robust procedure for the reconstruction of charged particle trajectories and interaction vertices within the Inner Detector. The reconstruction algorithms demonstrate excellent performance, both in simulation and real data, for the levels of \pileup experienced during Run~3 of the Large Hadron Collider. Using a set of reconstruction algorithms based around a combinatorial Kalman filter, tracks are reconstructed with high efficiency and purity, achieving efficiencies of over $75\%$ on average for charged particles above \qty{500}{\MeV} within the detector acceptance when applying the most stringent (\emph{Tight Primary}) selection criteria, with the rate of spuriously-reconstructed tracks below 0.1\% (0.2\%) for typical Run~2 (Run~3) $\mu$ values. Following a precision fit using a global $\chi^2$ approach, precise determinations of the track parameters, including the transverse and longitudinal impact parameters $d_0$ and $z_0$, are achieved for high-\pT particles. Specially-tailored optimisations for electrons, particles within high-\pT jets, and long-lived particles allow the mitigation of performance losses for these challenging objects. Thanks to the strong, early rejection of low-quality candidates during the pattern recognition stage, there is sufficient CPU time available in the processing budget for the additional tracking chains targeting converted photons and LLPs to run as standard on all events.

Vertex reconstruction using an adaptive method that simultaneously finds and fits multiple vertices allows for clean and efficient reconstruction of the primary interaction locations within the detector. Very high efficiencies are achieved in particular for reconstructing and selecting events which are of interest for physics analyses.

Looking ahead, the decades of knowledge encapsulated in the reconstruction strategy used for the Run~3 Inner Detector forms the basis of the upgraded reconstruction pipeline for the all-silicon Inner Tracker (ITk) that will replace the Inner Detector for Run~4 and the High-Luminosity (HL-LHC) era where $\langle\mu\rangle$ is expected to reach values in the range of 140--200~\cite{IDTR-2023-01}.
The ITk, composed of a pixel and a strip subdetector~\cite{ATLAS-TDR-25,ATLAS-TDR-30}, has been designed to maintain or improve upon the track reconstruction performance of the ID. The current ITk reconstruction implementation is described in Ref.~\cite{IDTR-2023-01}.
For Run~4, ATLAS plans to make extensive use of the ACTS toolkit~\cite{ACTS}, replacing its own implementations of the various tracking algorithms with those from ACTS~\cite{CERN-LHCC-2020-015} as already done for the AMVF vertex reconstruction in the Run~3 software. This will allow the track reconstruction to benefit from the modernised implementations and broader developer base of ACTS, ensuring the code remains maintainable throughout the lifetime of the ITk. In parallel, machine learning techniques are actively under development~\cite{ATL-PHYS-PUB-2024-018,ATL-PHYS-PUB-2025-045} in search of more accurate and faster methods for tracking.
Research and development efforts are ongoing to extend tracking and vertexing performance under the challenging conditions of HL-LHC triggering, including both conventional and ML-based approaches~\cite{ATLAS-TDR-29-ADD-1}.
These developments, like those discussed in this paper, are made possible by the high-level performance and operational stability of the current ATLAS Inner Detector. Ongoing detector alignment~\cite{IDTR-2019-05}, radiation damage monitoring~\cite{IDET-2017-10,IDET-2020-01,ATL-PHYS-PUB-2022-033,PIXE-2023-02}, and optimisation of detector operating conditions~\cite{SCTD-2019-01} ensure the continued delivery of high-quality data and provide a solid foundation for the success of the ITk upgrade.


%
\section*{Acknowledgements}
%

%
%

%
%

We thank CERN for the very successful operation of the LHC and its injectors, as well as the support staff at
CERN and at our institutions worldwide without whom ATLAS could not be operated efficiently.

The crucial computing support from all WLCG partners is acknowledged gratefully, in particular from CERN, the ATLAS Tier-1 facilities at TRIUMF/SFU (Canada), NDGF (Denmark, Norway, Sweden), CC-IN2P3 (France), KIT/GridKA (Germany), INFN-CNAF (Italy), NL-T1 (Netherlands), PIC (Spain), RAL (UK) and BNL (USA), the Tier-2 facilities worldwide and large non-WLCG resource providers. Major contributors of computing resources are listed in Ref.~\cite{ATL-SOFT-PUB-2026-001}.

We gratefully acknowledge the support of ANPCyT, Argentina; YerPhI, Armenia; ARC, Australia; BMWFW and FWF, Austria; ANAS, Azerbaijan; CNPq and FAPESP, Brazil; NSERC, NRC and CFI, Canada; CERN; ANID, Chile; CAS, MOST and NSFC, China; Minciencias, Colombia; MEYS CR, Czech Republic; DNRF and DNSRC, Denmark; IN2P3-CNRS and CEA-DRF/IRFU, France; SRNSFG, Georgia; BMFTR, HGF and MPG, Germany; GSRI, Greece; RGC and Hong Kong SAR, China; ICHEP and Academy of Sciences and Humanities, Israel; INFN, Italy; MEXT and JSPS, Japan; CNRST, Morocco; NWO, Netherlands; RCN, Norway; MNiSW, Poland; FCT, Portugal; MNE/IFA, Romania; MSTDI, Serbia; MSSR, Slovakia; ARIS and MVZI, Slovenia; DSI/NRF, South Africa; MICIU/AEI, Spain; SRC and Wallenberg Foundation, Sweden; SERI, SNSF and Cantons of Bern and Geneva, Switzerland; NSTC, Taipei; TENMAK, T\"urkiye; STFC/UKRI, United Kingdom; DOE and NSF, United States of America.

Individual groups and members have received support from BCKDF, CANARIE, CRC and DRAC, Canada; CERN-CZ, FORTE and PRIMUS, Czech Republic; COST, ERC, ERDF, Horizon 2020 and Marie Sk{\l}odowska-Curie Actions, European Union; Investissements d'Avenir Labex, Investissements d'Avenir Idex and ANR, France; DFG and AvH Foundation, Germany; Herakleitos, Thales and Aristeia programmes co-financed by EU-ESF and the Greek NSRF, Greece; BSF-NSF and MINERVA, Israel; NCN and NAWA, Poland; La Caixa Banking Foundation, CERCA and AGAUR programs from Generalitat de Catalunya and PROMETEO and GenT Programmes Generalitat Valenciana, Spain; G\"{o}ran Gustafssons Stiftelse, Sweden; The Royal Society and Leverhulme Trust, United Kingdom; Eric and Wendy Schmidt Fund for Strategic Innovation, United States of America.

In addition, individual members wish to acknowledge support from Chile: Agencia Nacional de Investigaci\'on y Desarrollo (ANID FONDECYT reg. 1230987, FONDECYT 1230812, FONDECYT 1240864, Fondecyt 3240661, Fondecyt Regular 1240721); China: Chinese Ministry of Science and Technology (MOST-2023YFA1605700, MOST-2023YFA1609300), National Natural Science Foundation of China (NSFC 12275265, NSFC-W2543005); Czech Republic: Czech Science Foundation (GACR - 24-11373S), Ministry of Education Youth and Sports (ERC-CZ-LL2327, FORTE CZ.02.01.01/00/22\_008/0004632), PRIMUS Research Programme (PRIMUS/21/SCI/017); EU: H2020 European Research Council (ERC - 101002463); European Union: European Research Council (BARD No. 101116429, ERC - 948254, ERC 101089007), European Regional Development Fund (HE COFUND GA No.101081355, ERDF), Marie Sklodowska-Curie Actions (GAP-101168829); France: Agence Nationale de la Recherche (ANR-21-CE31-0013, ANR-22-EDIR-0002, ANR-24-CE31-0504-01); Germany: Deutsche Forschungsgemeinschaft (DFG - 469666862); China: Research Grants Council (GRF); Italy: Ministero dell'Università e della Ricerca (NextGenEU 153D23001490006 M4C2.1.1, NextGenEU I53D23000820006 M4C2.1.1, NextGenEU I53D23001490006 M4C2.1.1, SOE2024\_0000023); Japan: Japan Society for the Promotion of Science (JSPS KAKENHI  JP25H0063, JSPS KAKENHI JP22H01227, JSPS KAKENHI JP22H04944, JSPS KAKENHI JP22KK0227, JSPS KAKENHI JP24K23939, JSPS KAKENHI JP24KK0251, JSPS KAKENHI JP25H00650, JSPS KAKENHI JP25H01291, JSPS KAKENHI JP25K01011, JSPS KAKENHI JP25K01023); Norway: Research Council of Norway (RCN-314472); Poland: Polish National Science Centre (NCN 2021/42/E/ST2/00350, NCN OPUS 2023/51/B/ST2/02507, NCN OPUS nr 2022/47/B/ST2/03059, NCN UMO-2019/34/E/ST2/00393, UMO-2022/47/O/ST2/00148, UMO-2023/49/B/ST2/04085, UMO-2023/51/B/ST2/00920, UMO-2024/53/N/ST2/00869); Spain: Agència de Gestió d'Ajuts Universitaris i de Recerca. (AGAUR - 2023 BP 00141), Ministry of Science and Innovation (RYC2019-028510-I, RYC2020-030254-I, RYC2021-031273-I, RYC2022-038164-I), Ministerio de Ciencia, Innovación y Universidades/Agencia Estatal de Investigaci\'on (EU NextGenerationEU (PRTR-C17.I1), PID2022-142604OB-C22); Sweden: Carl Trygger Foundation (Carl Trygger Foundation CTS 22:2312), Swedish Research Council (Swedish Research Council 2023-04654, VR 2021-03651, VR 2022-03845, VR 2022-04683, VR 2023-03403, VR 2024-05451, VR 2025-05940), Knut and Alice Wallenberg Foundation (KAW 2023.0366); Switzerland: Swiss National Science Foundation (SNSF - PCEFP2\_194658); United Kingdom: The Binks Trust, Royal Society (NIF-R1-231091); United States of America: U.S. Department of Energy (ECA DE-AC02-76SF00515), John Templeton Foundation (John Templeton Foundation 63206), Neubauer Family Foundation.

%
%


%
%
%
%
%
%

%
%
%
%
%
%

%
%
%
\printbibliography

\clearpage
 
\begin{flushleft}
\hypersetup{urlcolor=black}
{\Large The ATLAS Collaboration}

\bigskip

\AtlasOrcid[0000-0002-6665-4934]{G.~Aad}$^\textrm{\scriptsize 102}$,
\AtlasOrcid[0000-0001-7616-1554]{E.~Aakvaag}$^\textrm{\scriptsize 17}$,
\AtlasOrcid[0000-0002-5888-2734]{B.~Abbott}$^\textrm{\scriptsize 121}$,
\AtlasOrcid[0000-0002-0287-5869]{S.~Abdelhameed}$^\textrm{\scriptsize 83b}$,
\AtlasOrcid[0000-0002-1002-1652]{K.~Abeling}$^\textrm{\scriptsize 54}$,
\AtlasOrcid[0000-0001-5763-2760]{N.J.~Abicht}$^\textrm{\scriptsize 48}$,
\AtlasOrcid[0000-0002-8496-9294]{S.H.~Abidi}$^\textrm{\scriptsize 30}$,
\AtlasOrcid[0009-0003-6578-220X]{M.~Aboelela}$^\textrm{\scriptsize 44}$,
\AtlasOrcid[0000-0002-9987-2292]{A.~Aboulhorma}$^\textrm{\scriptsize 36e}$,
\AtlasOrcid[0000-0001-5329-6640]{H.~Abramowicz}$^\textrm{\scriptsize 154}$
% ...
\bigskip
\\

% ...
$^{17}$Department for Physics and Technology, University of Bergen, Bergen; Norway.\\
% ...
$^{30}$Physics Department, Brookhaven National Laboratory, Upton NY; United States of America.\\
% ...
$^{36}$$^{(a)}$Facult\'e des Sciences Ain Chock, Universit\'e Hassan II de Casablanca;$^{(b)}$Facult\'{e} des Sciences, Universit\'{e} Ibn-Tofail, K\'{e}nitra;$^{(c)}$Facult\'e des Sciences Semlalia, Universit\'e Cadi Ayyad, LPHEA-Marrakech;$^{(d)}$LPMR, Facult\'e des Sciences, Universit\'e Mohamed Premier, Oujda;$^{(e)}$Facult\'e des sciences, Universit\'e Mohammed V, Rabat;$^{(f)}$Institute of Applied Physics, Mohammed VI Polytechnic University, Ben Guerir; Morocco.\\
% ...
$^{44}$Physics Department, Southern Methodist University, Dallas TX; United States of America.\\
% ...
$^{48}$Fakult\"{a}t Physik, Technische Universit{\"a}t Dortmund, Dortmund; Germany.\\
% ...
$^{54}$II. Physikalisches Institut, Georg-August-Universit\"{a}t G\"ottingen, G\"ottingen; Germany.\\
% ...
$^{83}$$^{(a)}$Khalifa University of Science and Technology, Abu Dhabi;$^{(b)}$New York University Abu Dhabi, Abu Dhabi;$^{(c)}$United Arab Emirates University, Al Ain;$^{(d)}$University of Sharjah, Sharjah; United Arab Emirates.\\
% ...
$^{102}$CPPM, Aix-Marseille Universit\'e, CNRS/IN2P3, Marseille; France.\\
% ...
$^{121}$Homer L. Dodge Department of Physics and Astronomy, University of Oklahoma, Norman OK; United States of America.\\
% ...
$^{154}$Raymond and Beverly Sackler School of Physics and Astronomy, Tel Aviv University, Tel Aviv; Israel.\\
% ...
$^{a}$ Also at Affiliated with an institute formerly covered by a cooperation agreement with CERN.\\
$^{b}$ Also at An-Najah National University, Nablus; Palestine.\\
$^{c}$ Also at Borough of Manhattan Community College, City University of New York, New York NY; United States of America.\\
$^{d}$ Also at Center for Interdisciplinary Research and Innovation (CIRI-AUTH), Thessaloniki; Greece.\\
$^{e}$ Also at Centre of Physics of the Universities of Minho and Porto (CF-UM-UP); Portugal.\\
$^{f}$ Also at CERN, Geneva; Switzerland.\\
% ...

\end{flushleft}


%
%

%

%
%
%
%

\end{document}